\documentclass[useAMS,usedcolumn,usenatbib,usegraphicx]{mn2e}

\usepackage{epsfig,graphicx,natbib}
\usepackage{cite}
\usepackage{amssymb}
\usepackage{amsfonts}
\usepackage{amsmath}
\usepackage{color}
 
\newcommand{\ud}{\,\mathrm{d}}

\begin{document}
\title[The Angular Power Spectra of Photometric SDSS LRGs]{The Angular Power Spectra of Photometric SDSS LRGs}

\author[Thomas et al.]
{Shaun A. Thomas,
Filipe B. Abdalla,
Ofer Lahav \\
Department of Physics and Astronomy, University College London,
Gower Street, London, WC1E 6BT, UK.\\
\\
Email: sat@star.ucl.ac.uk
}

\maketitle

\begin{abstract}
We construct new galaxy angular power spectra $C_{\ell}$ based on the extended, updated and final SDSS II Luminous Red Galaxy (LRG) \emph{photometric} redshift survey--MegaZ (DR7). Encapsulating $7746$ $\mathrm{deg}^{2}$ we utilise 723,556 photometrically determined LRGs between $0.45<z<0.65$ in a 3.3 (Gpc $h^{-1}$)$^3$ spherical harmonic analysis of the galaxy distribution. By combining four photometric redshift bins we find preliminary parameter constraints of $f_{b} \equiv \Omega_{b}/\Omega_{m} = 0.173 \pm 0.046$ and $\Omega_{m} = 0.260 \pm 0.035$ assuming $H_{0} = 75$ km $s^{-1}$ Mpc$^{-1}$, $n_{s}=1$ and $\Omega_{k} = 0$. These limits are consistent with the CMB and the previous data release (DR4). The $C_{\ell}$ are sensitive to redshift space distortions and therefore we also recast our constraints into a measurement of $\beta \approx \Omega_{m}^{0.55}/b$ in different redshift shells. The robustness of these power spectra with respect to a number of potential systematics such as extinction, photometric redshift and ANNz training set extrapolation are examined. The latter includes a cosmological comparison of available photometric redshift estimation codes where we find excellent agreement between template and empirical estimation methods. MegaZ DR7 represents a methodological prototype to next generation surveys such as the Dark Energy Survey (DES) and, furthermore, is a photometric precursor to the spectroscopic BOSS survey. Our galaxy catalogue and all power spectra data can be found at http://zuserver2.star.ucl.ac.uk/$\sim$sat/MegaZ/MegaZDR7.tar.gz.
\end{abstract}


\section{Introduction}

The analysis of the statistical distribution of fluctuations in the Universe is a potent method for constraining theories or components within Cosmology. In fact, the power spectrum will fully describe these variations, which are predicted by theory, if they are given by a Gaussian random field. The Cosmic Microwave Background (CMB) has been a great example of this principle in action with recent high precision measurements \citep{Komatsu10} confirming that a clear and consistent picture of cosmology is emerging. It is desirable however to test this picture with additional and independent data that explores a contrasting epoch of cosmic evolution and breaks the parameter degeneracies that exist from a single probe of the early Universe. A galaxy redshift survey is therefore a powerful tool in Cosmology \citep{Peebles73}. In addition, this late-time galaxy distribution is sensitive to the emergence of dark energy (\citealt{Riess98} and \citealt{Perlmutter99}) and arising through the growth of structure enables a test of gravity (\citealt{Jain07}, \citealt{HutererLinder07}, \citealt{Thomas08}) and the mass of the neutrino (\citealt{Hu98} and \citealt{Thomas09}).

The structure and aim of this paper is as follows: To construct and present the angular power spectra $C_{\ell}$ of the new SDSS Luminous Red Galaxy (LRG) photometric survey, along with the associated error and individual cosmological constraints. Specifically, we determine the colour, redshift and angular selection functions that define the survey in Section~\ref{sec:LRGs}. The spherical harmonic analysis is described in Section~\ref{sec:measurement} and Section~\ref{sec:powerspectrum}. The cosmological constraints inferred and the potential systematics of the data set are discussed in Section~\ref{sec:result} and Section~\ref{sec:systematics1}, respectively. This last section also includes a cosmological comparison of different photometric redshift methods. Extended and combined \emph{cosmological implications} are to be presented in a companion paper with the likelihood.

\section{Data} \label{sec:LRGs}

The development of galaxy surveys over the past few years reflects the balance between observational technology and gains in cosmological parameter estimation. This has at present culminated in the impressive 2-degree Field Galaxy Redshift Survey (2dFGRS - \citealt{Colless01}) and the Sloan Digital Sky Survey (SDSS - \citealt{York00}). However, the acquisition of a vast number of precise redshifts through spectroscopy is an expensive, challenging and time consuming task. An alternative method is to use photometric redshifts (E.g. \citealt{Csabai03}) resulting from observations of broadband galaxy colours through a series of filters. The motivation is that a decrease in redshift accuracy is outweighed by measurements of a vast number of galaxies over a wide area of the sky, therefore encompassing a large cosmic volume. Photometric redshift surveys have been shown to be competitive by \citealt{Blake07} and \citealt{Padmanabhan07} and here we follow these papers. Upcoming surveys, such as the Dark Energy Survey (\citealt{DES05}), are heavily based on this efficiency principle.

We therefore aim to analyse the clustering of the latest and final SDSS II photometry given by Data Release 7 (DR7) - \citet{Abazajian09}. The $\approx 1.5$ million LRG catalogue (MegaZ-LRG DR7) is produced as an updated version of MegaZ-LRG \citep{Collister07}. These LRGs are old red elliptical galaxies that provide a clean and consistent galaxy sample. With a stable spectral energy distribution (SED) and a sharp $4000 \AA$ break they provide good photometric redshift estimates. Furthermore, they are known to strongly trace the underlying mass density; a distribution we are striving to quantify. Also, being among the brightest galaxies in the Universe they allow detailed studies over a large cosmic volume. This is highly desirable for a cosmological study given that it diminishes the effect of sample variance.

\subsection{Redshift Selection} \label{sec:redshiftselection}

The redshift estimates for this above sample were constructed by using the redshift output as given by ANNz \citep{Collister04} an Artificial Neural Network code. This empirical photometric redshift estimator learns an effective parameterisation of redshift with varying galaxy magnitudes (here in $u$, $g$, $r$, $i$ and $z$ bands) by working on a representative training set. For our training set we use a subset of the $\sim 13,000$ spectroscopic redshifts from the 2dF-SDSS LRG and Quasar survey (2SLAQ - \citealt{Cannon06}), a $\delta \approx 0\,^{\circ}$ (declination) stripe within the DR7 imaging area. Specifically, we use $5,482$ of these objects as the training set with the rest utilised for testing. For this reason and for this specific galaxy sample over the range of redshifts of interest ($0.45<z<0.65$) \citealt{Abdalla08} found the ANNz training method to have the best performance on an evaluation LRG sample compared with other redshift estimation codes, with average scatter $\sigma_{z} = 0.0575$ and $\sigma_{z}$ defined by,
\begin{equation} \label{eq:photoz}
\sigma_{z} = <(z_{\mathrm{phot}} - z_{\mathrm{spec}})^{2}>^{\frac{1}{2}}.
\end{equation}
\noindent
The performance of this procedure and the representative photometric-spectroscopic scatter can be seen in Figure 2 of \citealt{Blake07} and Figures 2-5 in \citealt{Abdalla08}. The reliability of the neural network training procedure depends on the training set being completely representative of the target galaxy sample. It is noted that by applying this 2SLAQ stripe to the wider photometric LRGs there is an extrapolation of the training set with sky position. The discussion of this potential systematic, however, is left to Section~\ref{sec:systematiccodes}. 

\subsection{The Colour Selection} \label{sec:colourselection}

Our SDSS pre-selection and secondary colour selection of galaxies is based on and described in \citealt{Collister07}, \citealt{Blake07} and \citealt{Cannon06}. For example, at the start of the 2SLAQ survey there was an alteration in the selection criteria used to extract the homogeneous LRG sample from the overall galaxy and object population. This is associated with the de Vaucouleurs model magnitude $i_{\mathrm{deV}}$ and also $d_{\mathrm{perp}}$, a colour cut which is related to the $g$, $r$ and $i$ model magnitudes via,
\begin{equation} \label{eq:d_perp}
d_{\mathrm{perp}} \equiv (r - i) - (g - r)/8.0.
\end{equation}
\noindent
We prefer to act cautiously in order to analyse a galaxy sample that most represents the training set used to infer its properties. Therefore, we use the colour cuts $i_{\mathrm{deV}} \le 19.8$ and $d_{\mathrm{perp}} \geq 0.55$ to select and extract the LRG population given that these were the selection criteria used in the strict majority of 2SLAQ. Again, these cuts were also used in the earlier MegaZ-LRG analysis \citep{Blake07}.

\subsection{M-star Contamination} \label{sec:Mstars}

Comparison with the spectroscopic 2SLAQ survey verifies that the pre-selection of SDSS galaxies and further colour cuts from \citealt{Collister07} and \citealt{Blake07} are accurate. Star-galaxy separation was also ensured using the criteria \citep{Collister07},
\begin{equation} \label{eq:psf}
i_{\mathrm{psf}}-i_{\mathrm{model}} > 0.2 \, (21.0 - i_{\mathrm{deV}})
\end{equation}
\begin{equation} \label{eq:radius}
i\mathrm{-band \, de \, Vaucouleurs \, radius} \, > 0.2 ''.
\end{equation}
\noindent
However, the presence of M-stars still persist and represent the main source of object contamination ($\approx 5\%$) within the remaining sample owing to similar broadband colours. Generally, an uncorrelated sample of stars will act to suppress the power of fluctuations \citep{Huterer01}. However, one would expect a slightly correlated variation of stellar material through the Galactic plane and hence our survey area. We therefore remove a large proportion of these contaminants with an extra cut in star-galaxy separation. The ANNz code has a star-galaxy parameter $\delta_{sg}$ as an additional optional output \citep{Collister04}. This uses a variety of further inputs, including angular size and gauges of the light profile throughout the object. These extra parameters are detailed in Figure 3 of \citealt{Collister07}. The $\delta_{sg}$ output parameter varies continuously from `guaranteed' star $\delta_{sg}=0$ to `certain' galaxy $\delta_{sg}=1$. We remove all objects with $\delta_{sg}<0.2$, in the processes decreasing the contamination fraction to $\approx 1.5\%$ with minimum loss of real LRGs (\citealt{Collister07} and \citealt{Blake07}).

\subsection{The Angular Selection Function} \label{sec:angularselection}

\begin{figure*}
  \centering
      \includegraphics[width=6.0in,height=3.6in]{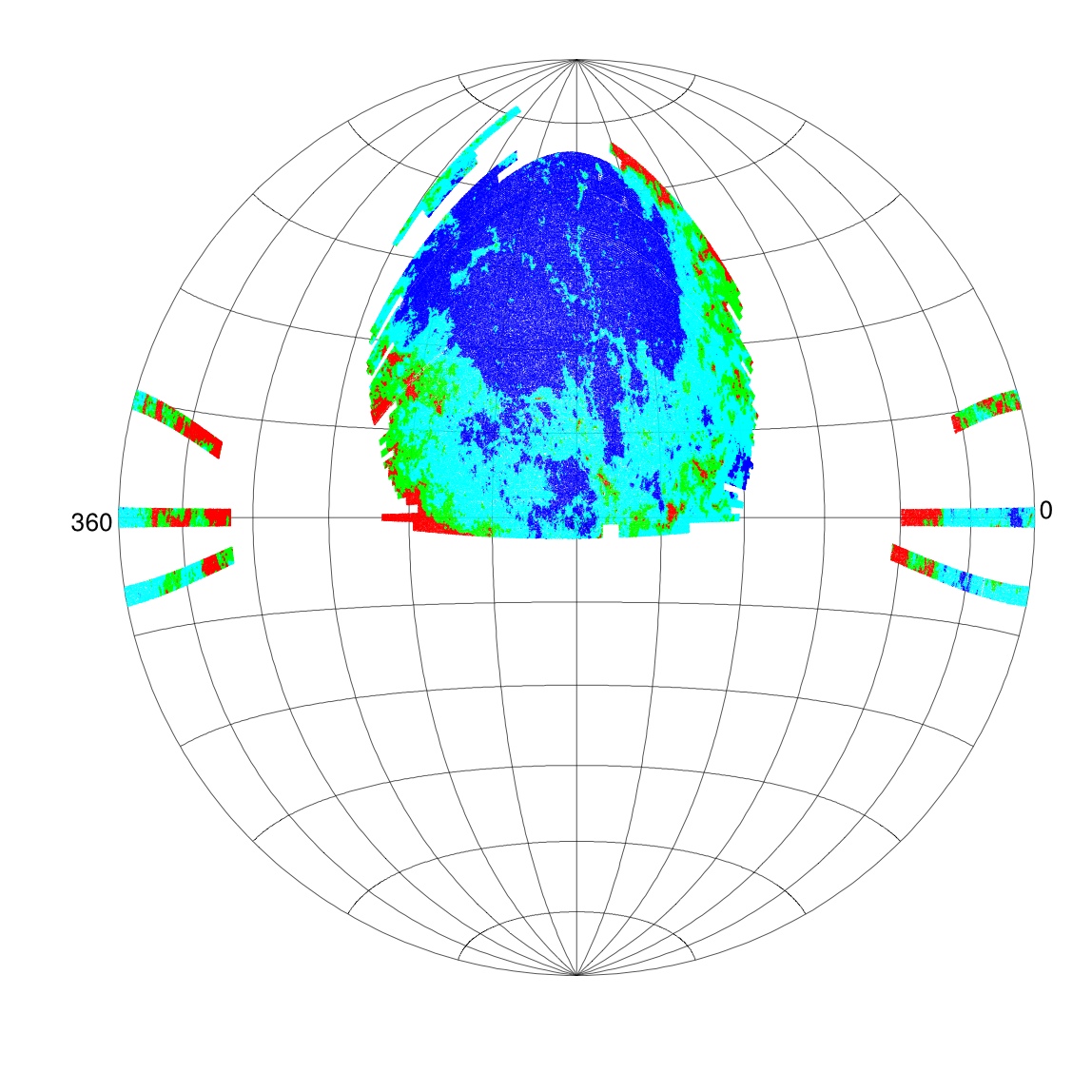}
    \caption{\small{The SDSS Data Release 7 (DR7) photometric LRG coverage. At $7746$ $\mathrm{deg^{2}}$ it covers $723,556$ galaxies over a redshift $0.4<z<0.7$. The three {\it excluded} stripes (76, 82 and 86) are visible towards the boundary of the plot. The 2dF SDSS LRG and Quasar (2SLAQ) survey and training set constitutes a narrow stripe ($\delta \approx 0\,^{\circ}$) that passes approximately through the middle of the equatorial coordinate system and the bottom of the defined survey. The fluctuations in Galactic extinction are also shown across the survey area. The magnitude values are represented by dark blue ($0.0-0.05$ mag), light blue ($0.05-0.1$ mag), green ($0.1-0.15$ mag) and red ($> 0.15$ mag). The dust is particularly abundant near the edges of the survey indicating the outer boundaries of the Galaxy. We test the effects of this extinction by removing regions with $> 0.1$ mag ($15\%$ of the area) in Section~\ref{sec:systematicextinction}.}}
    \label{fig:SDSSmask}
\end{figure*}

The angular selection function, which is used to determine the observable boundaries of the survey, was obtained from \texttt{tsChunk.dr7.best.par} downloaded at \texttt{www.sdss.org/ dr7/coverage}. We converted the provided great circle coordinates ($\mu$,$\nu$) and the survey's stripe numbers to declination and right ascension before undergoing a HEALPix pixelisation on a sphere \citep{Gorski05}. We used a total of 3,145,728 pixels ($12 \times \mathrm{n_{side}} \times \mathrm{n_{side}}$ where $\mathrm{n_{side}} = 512$) over the entire sky, placing a zero in pixels corresponding to holes, gaps or regions not surveyed and a one in genuinely surveyed pixels. This discrete survey mask was then overlaid with the aforementioned LRG catalogue to leave the final galaxy map. We further tested this with $\mathrm{n_{side}} = 1024$ to examine the effects of a pixelised space. After appropriately adjusting the estimated $C_{\ell}$ (found in Section~\ref{sec:measurement}), by dividing by the square of the HEALPix window function $w^{2}_{\ell}$, the pixelisation effect was found to be negligible.

We imposed an additional constraint on the mask/map by excluding the survey stripes 76, 82 and 86, which are widely separated from the rest of the contiguous region. These segments act to increase the complexity of the survey window function and contribute relatively little extra galaxies. The resulting survey used for the primary angular power spectrum analysis spans $7746$ $\mathrm{deg^{2}}$ and $723,556$ galaxies over a redshift $0.4<z<0.7$. This is a $30\%$ larger area for analysis than the first and previous MegaZ-LRG survey (\citealt{Blake07} and \citealt{Collister07}). Likewise, it is significantly more expansive than the earlier \citealt{Padmanabhan05}, which covering $3,528$ $\mathrm{deg^{2}}$ and $0.2<z<0.6$ represents a slightly different LRG population and analysis method. The final sky coverage is shown in Figure~\ref{fig:SDSSmask}.

\section{The Power Spectrum Measurement} \label{sec:measurement}

The measurement of the angular power spectrum is performed by undertaking a spherical harmonic analysis \citep{Peebles73}. By explicitly summing the discrete galaxies over the incomplete sky we follow the derivation, methodology and/or notation of \citealt{Peebles73}, \citealt{Wright94}, \citealt{Blake04} and \citealt{Blake07}. 

One connects the underlying density field in a redshift band to the relevant statistical entities by first projecting the mass distribution $\sigma(\theta, \phi)$. This distribution is then decomposed into a series of spherical harmonics $Y_{l,m}$ and their corresponding coefficients $a_{l,m}$,

\begin{equation} \label{eq:density field}
\sigma(\theta, \phi) = \sum_{l=0}^{\infty} \sum_{m=-l}^{l} a_{l,m} Y_{l,m} (\theta, \phi).
\end{equation}
\noindent

The statistical distribution--the angular power spectrum $C_{\ell}$--is then given by the multi-realisation expectation of these $a_{l,m}$ coefficients, $<|a_{l,m}|^{2}>$. For a full sky survey these coefficients represent an orthogonal and normalised basis and are thus found by a summation of the spherical harmonic conjugate over the galaxy catalogue,

\begin{equation} \label{eq:alm}
A_{l,m} = \sum_{i=1}^{N} Y^{*}_{l,m} (\theta_{i},\phi_{i}).
\end{equation}
\noindent

However, in reality one will observe a masked and therefore incomplete sky. This effectively correlates the spherical harmonic coefficients and induces the correction and adjustment for loss of power given by,

\begin{equation} \label{eq:incompleteskyCl}
C_{l,m}^{\mathrm{psky}} = \frac{|A_{l,m} - \frac{N}{\Delta \Omega}I_{l,m}|^{2} }{J_{l,m}} - \frac{\Delta \Omega}{N}
\end{equation}
\noindent
where $N$ is the number of galaxies, $\Delta \Omega$ is the area of the sky and the $I_{l,m}$ and $J_{l,m}$ integrals in Equations~\ref{eq:ilm} and \ref{eq:jlm} are evaluated over the geometry of the discrete survey area. I.e., $\delta \Omega = 1$ for a surveyed pixel and $\delta \Omega = 0$ for an unsurveyed pixel. The last subtracted term is a correction for the statistical distribution of shot noise and is equivalent to the expectation of the corresponding harmonic coefficient for a random unclustered sample. 
\begin{equation} \label{eq:ilm}
I_{l,m} = \int_{\Delta \Omega} Y^{*}_{l,m} \ud\Omega
\end{equation}
\noindent

\begin{equation} \label{eq:jlm}
J_{l,m} = \int_{\Delta \Omega} |Y_{l,m}|^{2} \ud\Omega
\end{equation}
\noindent
One can then obtain the resulting angular power spectrum for a given multipole $\ell$ via an averaging of $C_{l,m}$ over the ($2\ell+1$) $a_{l,m}$ values,
\begin{equation} \label{eq:averagedCl}
C_{\ell}^{\mathrm{obs}} = \frac{\sum_{m=-l}^{l} C_{l,m}^{\mathrm{psky}}}{2l+1}.
\end{equation}
\noindent
The angular power spectrum is independent of $m$ for statistical isotropy. The $C_{\ell}$ values are further averaged into bins of width $\Delta \ell = 10$. As seen later this has the effect of decorrelating the measurements and provides a more Gaussian likelihood. We weight this average by the corresponding number of $a_{l,m}\mathrm{s}$, 
\begin{equation} \label{eq:weightedCls}
C_{\ell}^{\Delta \ell} = \frac{\sum_{\ell'}^{\ell'+\Delta \ell} (2\ell+1) C_{\ell}^{\mathrm{obs}}}{\sum_{\ell'}^{\ell'+\Delta \ell} (2\ell+1)}.
\end{equation}
\noindent
The angular power spectrum in these $\Delta \ell$ bands is measured up to $\ell = 500$. One can therefore use these statistics for each redshift band within the survey volume. We measure the clustering distribution in four such photometric redshift bins, each having width $\Delta z = 0.05$ from $z=0.45$ to $z=0.65$. It is this expression in Equation~\ref{eq:weightedCls} that is presented in Figure~\ref{fig:Cl_redshift_bin_1} and Section~\ref{sec:observedspectra}. These procedures are in line with \citealt{Blake07} and therefore a direct MegaZ-LRG consistency check can be made. 

The aforementioned redshift bins are correlated, however, as photometric errors scatter galaxies between the bins. A small modification to the angular power spectra,
\begin{equation} \label{eq:crosspowerspectrumdata}
C_{\ell}^{i,j} = \frac{1}{2\ell+1} \sum_{m=-\ell}^{\ell} (A_{l,m}^{i})^{*} A_{l,m}^{j}
\end{equation}
\noindent
enables a measurement where the harmonic coefficients in bin $i$ and bin $j$ have been adjusted for incomplete sky coverage as detailed above. The results are listed in Section~\ref{sec:observedspectra}.

Note there exist other analogous procedures for the analysis of galaxy clustering including, for example, quadratic estimators, maximum likelihood methods and explicit reconstructions of the power spectrum (E.g. \citealt{Huterer01}, \citealt{Tegmark02}, \citealt{Seo03}, \citealt{Tegmark04}, \citealt{Blake05}, \citealt{Tegmark06}, \citealt{Padmanabhan07}, \citealt{Blake07} and \citealt{Reid09}).

\subsection{Simulations and Gaussian error} \label{sec:covariance}

\begin{figure}
  \centering
  \includegraphics[width=3.45in,height=3.45in]{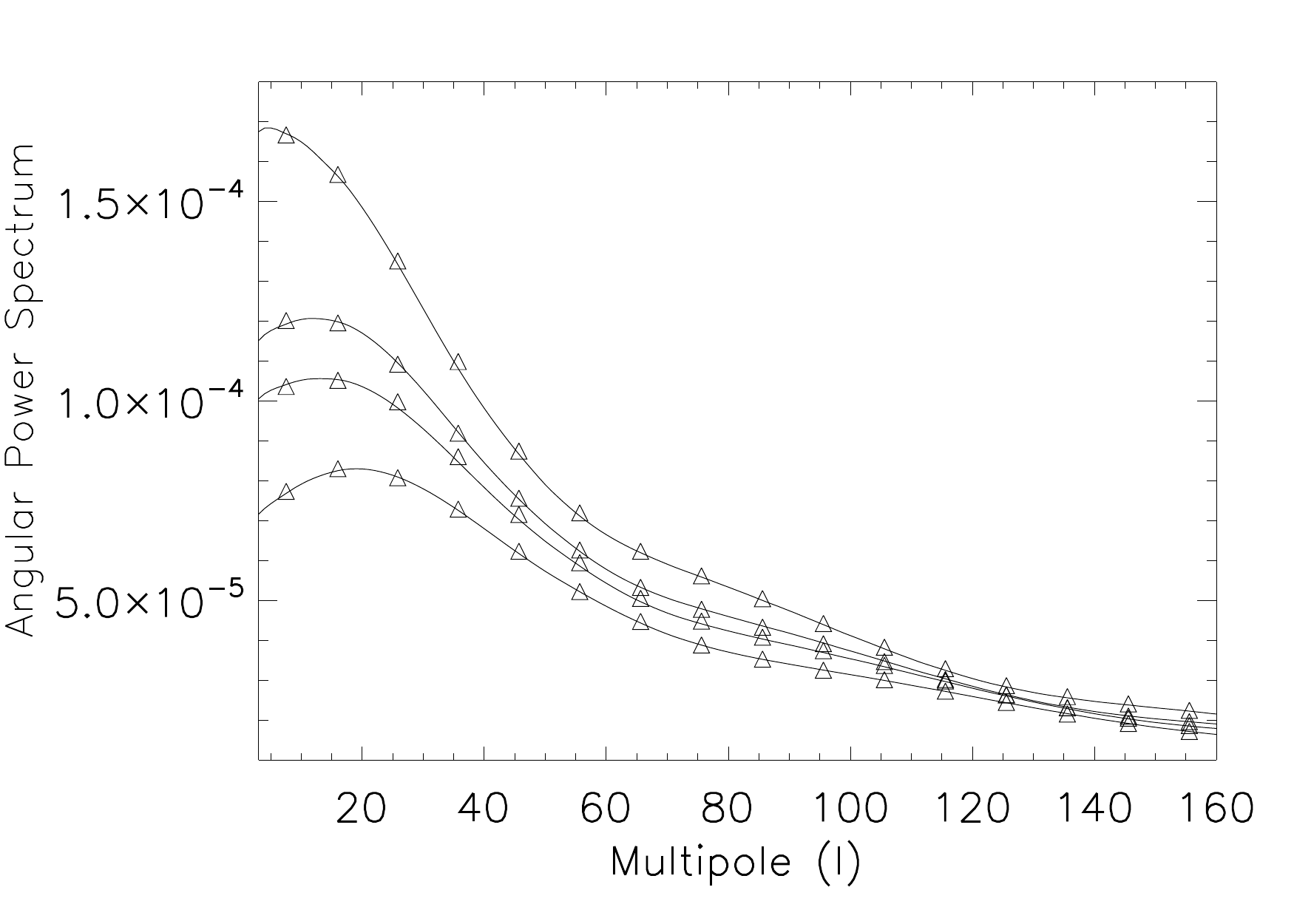}
  \caption{\small{The averaged reconstruction of the input $C_{\ell}$ field for 1000 simulations. The solid lines represent the input cosmology for the four equally spaced redshift bins ($\Delta z = 0.05$) between $z = 0.45$ (top profile) and $z = 0.65$ (bottom profile). The triangles are the binned measured values from Equation~\ref{eq:weightedCls}. The plot has been truncated at $\ell=160$ as a visual aid; the behaviour beyond this point continues in an identical fashion. The accuracy and consistency of the code and measurement procedure is clear.}}
  \label{fig:simulation}
\end{figure}
\noindent

\begin{figure*}
  \begin{flushleft}
    \centering
    \begin{tabular}{ll}
      \includegraphics[width=3.2in,height=3.2in]{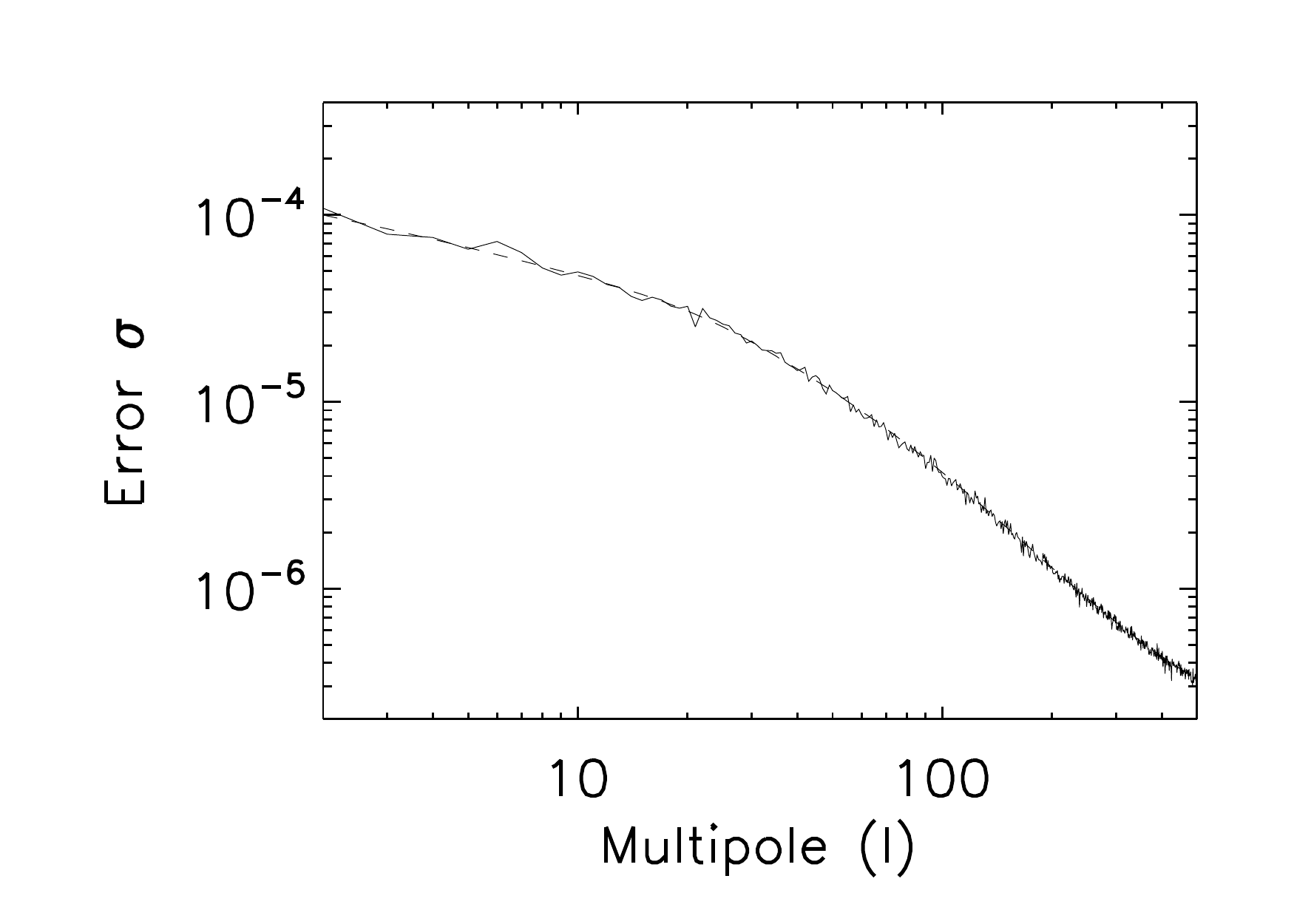} &
      \includegraphics[width=3.2in,height=3.2in]{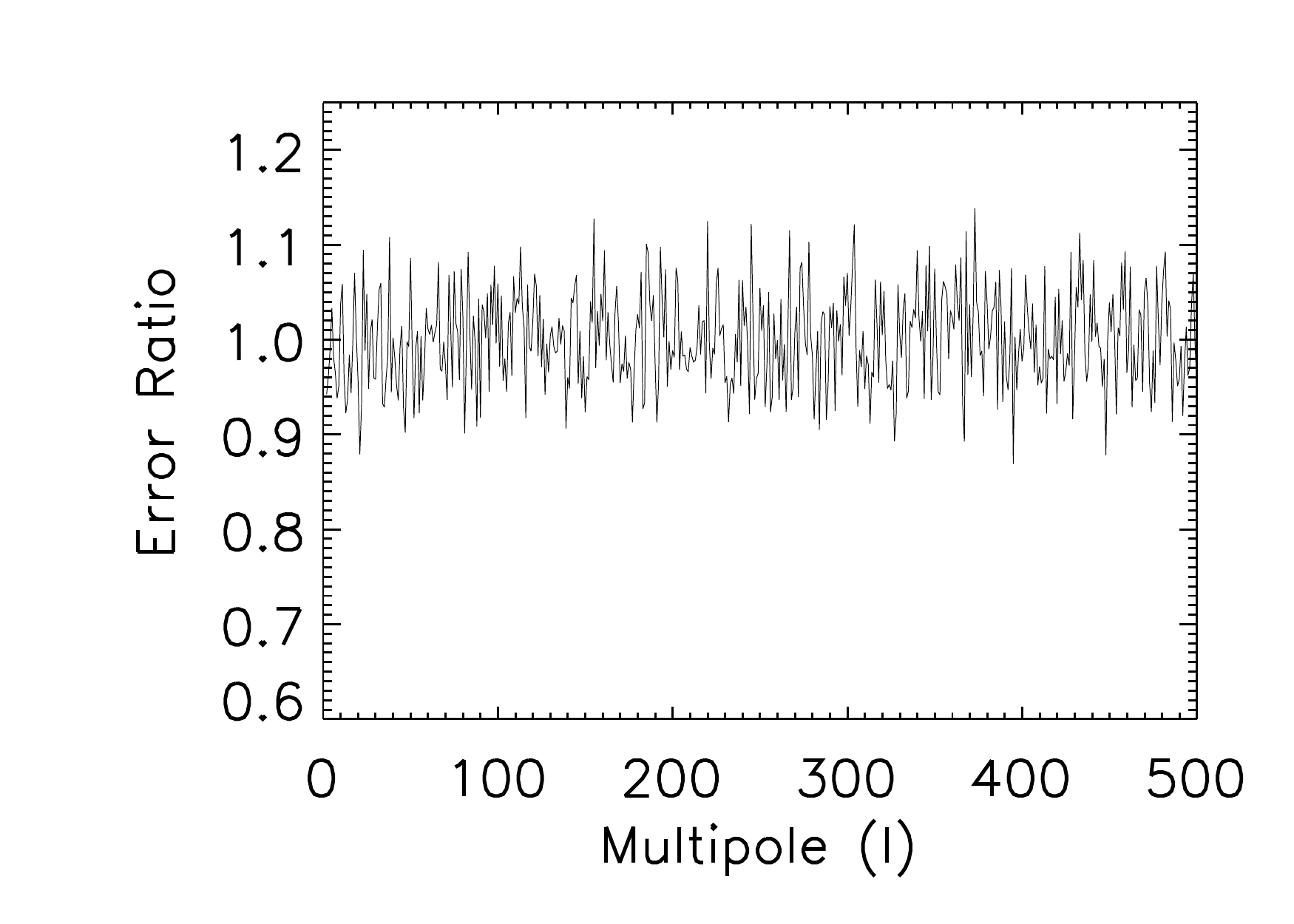}
    \end{tabular}
    \caption{\small{Left Panel: The analytic Gaussian expression (Equation~\ref{eq:gaussianerror}; dashed line) is accurately traced by the 1000 realisation simulated error in a redshift band (solid line), shown here for bin 1 ($0.45 \le z \le 0.5$). This demonstrates the approximate validity of the Gaussian expression. Right Panel: The agreement is further highlighted by the ratio of the analytic and numerical estimations of the statistical error, where the overall behaviour is consistent with unity. The two panels are shown for the first bin only but are representative of all other bin combinations. }}
    \label{fig:errorratio}
  \end{flushleft}
\end{figure*}
\noindent
The methodology described above, for the measurement of the angular power spectra, was applied to simulated data in order to test the procedure and the code. This was performed by first constructing a Gaussian random field for some input cosmology, using best fit WMAP parameters \citep{Larson10}, and subsequently reconstructing this cosmology for each of the four galaxy clustering redshift bins to be measured. We \emph{randomly} selected the full set of spherical harmonic coefficients $a_{\ell',m}$ from Gaussian distributions with widths given by the underlying known cosmology [$(C_{\ell'})^{\frac{1}{2}}$]. The relation between the underlying matter power spectrum and the theoretical angular power spectrum is described in Section~\ref{sec:powerspectrum}. Then, using the {\sc HEALPix} function \texttt{alm2map} \citep{Gorski05} we simulated a pixelised galaxy map from these quantities and sampled objects as a Poisson realisation of the field. The full angular selection function of the survey (Section~\ref{sec:angularselection}; Figure~\ref{fig:SDSSmask}) was imposed on the simulated map and the number of galaxies sampled in each bin were matched to those present in the observed catalogue. This mock data was then analysed with the measurement pipeline in the same manner as the real data and averaged over 1000 simulated realisations. The accuracy and reliability of the code and the power spectrum measurement procedure is evident in Figure~\ref{fig:simulation}.

One can also use these simulations to check the analytic statistical error in the galaxy clustering measurements $\sigma(C_{\ell})$, which we assume to be Gaussian. This is extracted from the standard deviation over the $1000$ mock realisations at each $\ell$. The analytic expression (E.g. \citealt{Dodelson03} and \citealt{Blake07}) is given by,
\begin{equation} \label{eq:gaussianerror}
\sigma(C_{l}) = \sqrt{\frac{2}{f_{\mathrm{sky}}(2l+1) } } \Big( C_{l} + \frac{\Delta \Omega}{N} \Big)
\end{equation}
\noindent
where $\mathrm{f_{sky}}$ is the fraction of sky surveyed, $\Delta \Omega$ is the area, N is the measured number of galaxies in the bin and $C_{\ell}$ is the observed or theoretical angular power spectrum. The first and second terms in Equation~\ref{eq:gaussianerror} include the necessary error contributions from both cosmic variance and shot noise, respectively. It also accounts for the reduced error given the combination of $2\ell+1$ $C_{\ell,m}$ values into the determination of each $C_{\ell}$. For the statistical error in the cross power spectrum this generalises to,
\begin{equation} \label{eq:crossgaussianerror}
\sigma^{2}(C_{l}^{i,j}) = \frac{2}{f_{\mathrm{sky}}(2\ell+1) } \Big( C_{\ell}^{i} + \frac{1}{N_{i}/\Delta \Omega} \Big)\Big( C_{\ell}^{j} + \frac{1}{N_{j}/\Delta \Omega} \Big).
\end{equation}
\noindent
We find the expression reconstructs the simulated error accurately in each of the four redshift bins and across the entire range of $\ell$. This is easily seen in the left panel of Figure~\ref{fig:errorratio}. The error ratio, typified by the first redshift bin, is displayed in the right panel. Therefore, for our cosmological analyses we use the analytic expression only in the covariance matrix, evaluated with the {\it model} $C_{\ell}$s.

\subsection{Results} \label{sec:observedspectra}

\begin{figure*}
  \begin{flushleft}
    \centering
    \begin{minipage}[c]{1.00\textwidth}
      \centering
      \includegraphics[width=2.87in,height=2.87in]{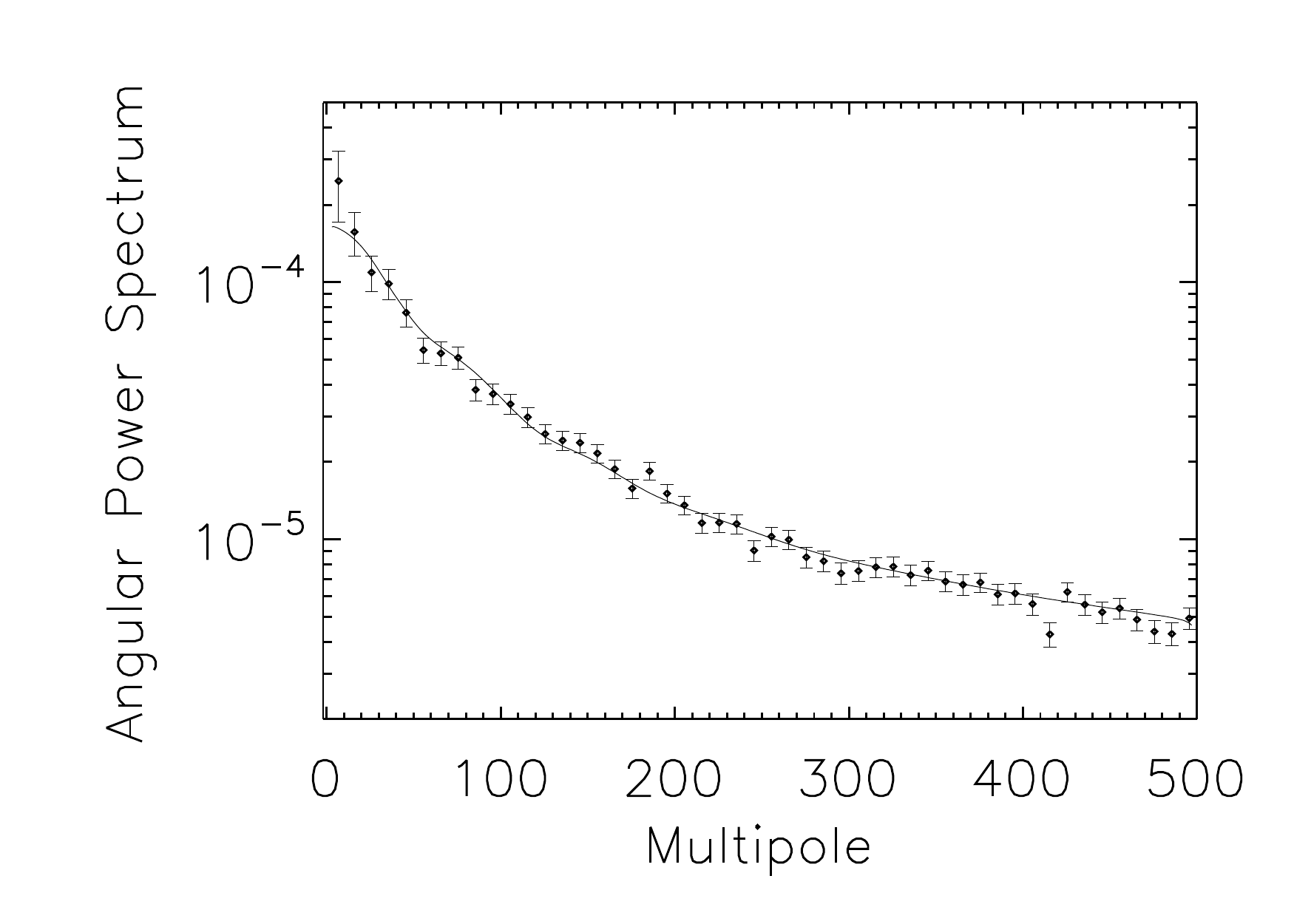} 
      \includegraphics[width=2.87in,height=2.87in]{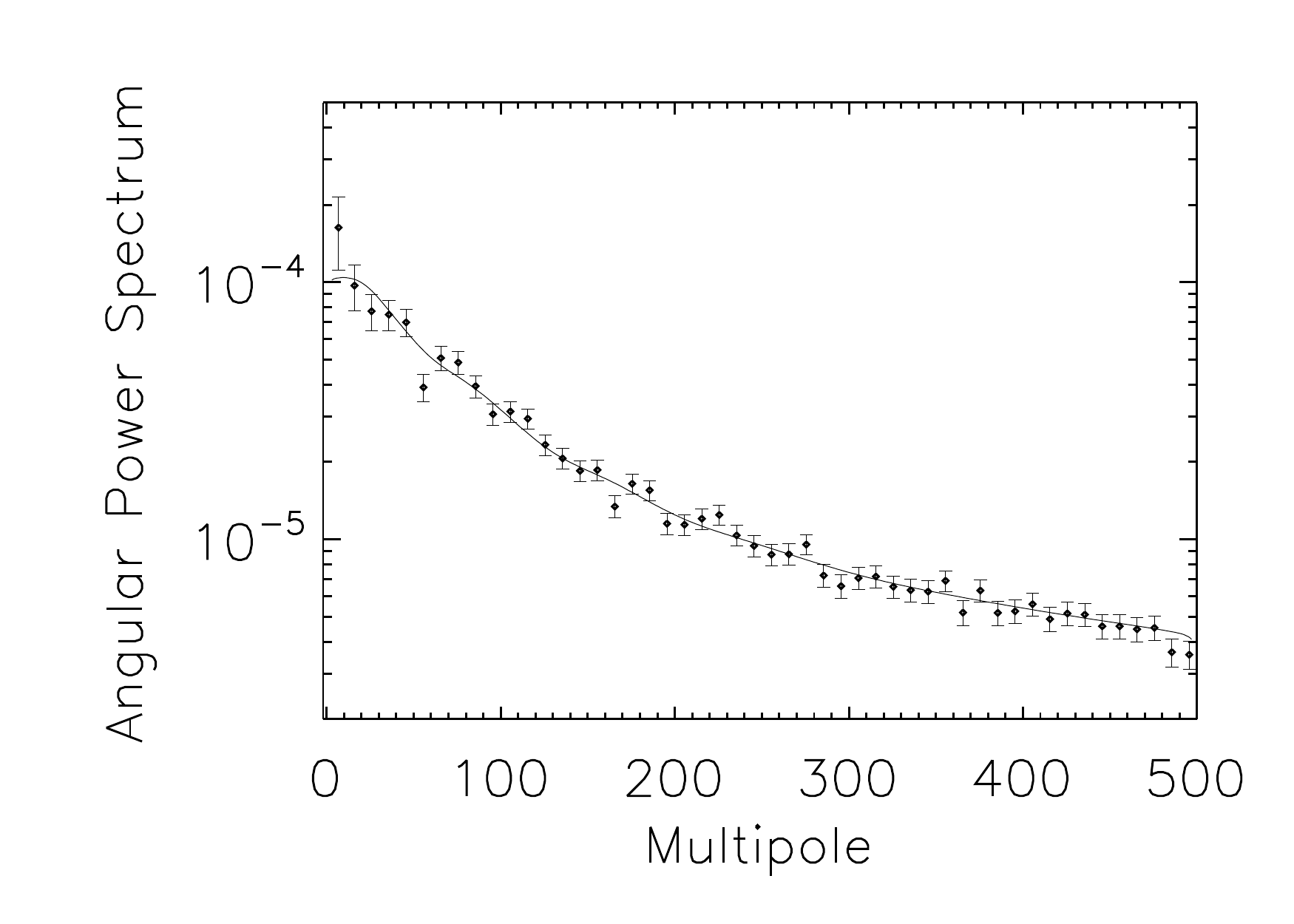} 
    \end{minipage}
    \begin{minipage}[c]{1.00\textwidth}
      \centering
      \includegraphics[width=2.87in,height=2.87in]{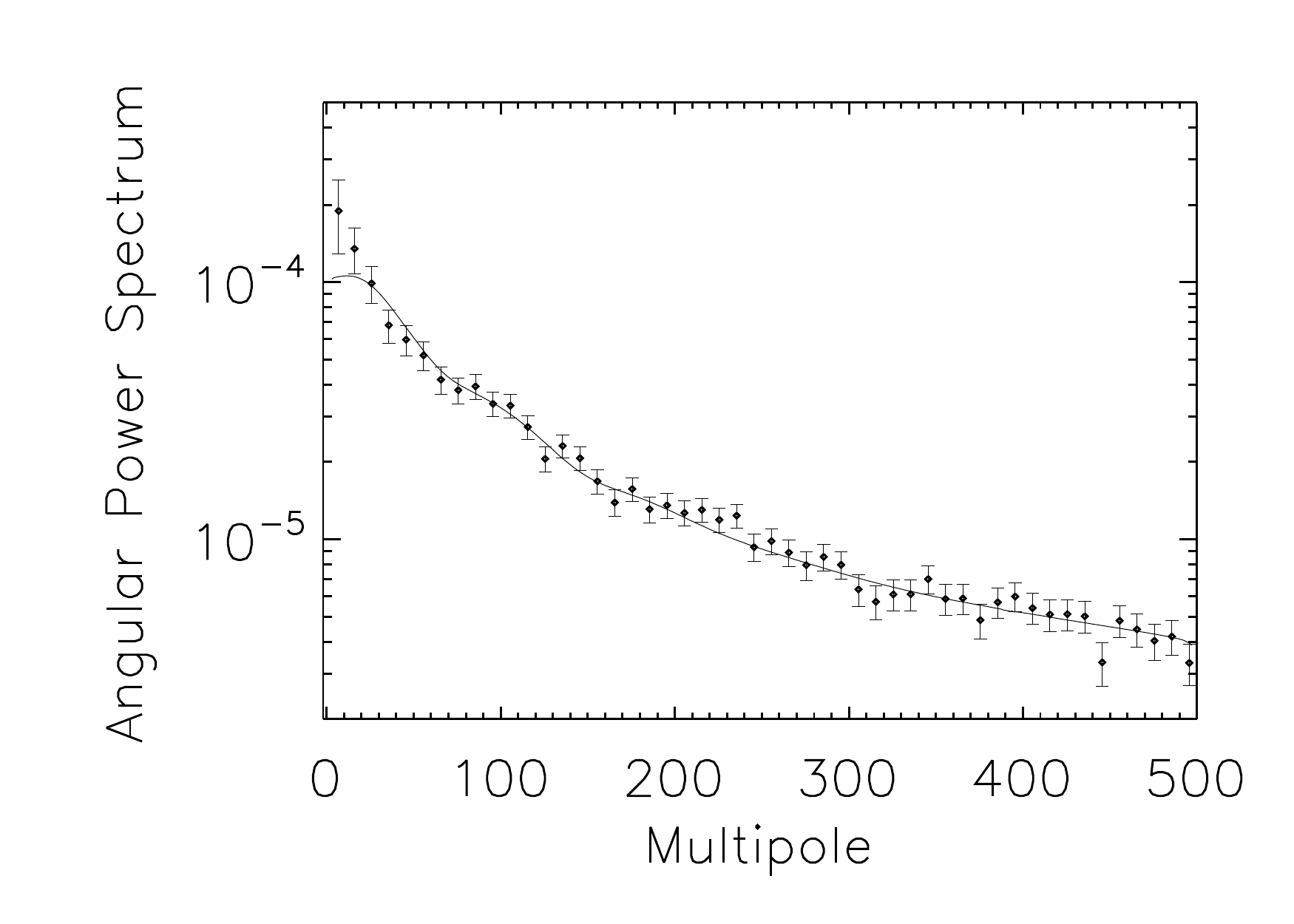} 
      \includegraphics[width=2.87in,height=2.87in]{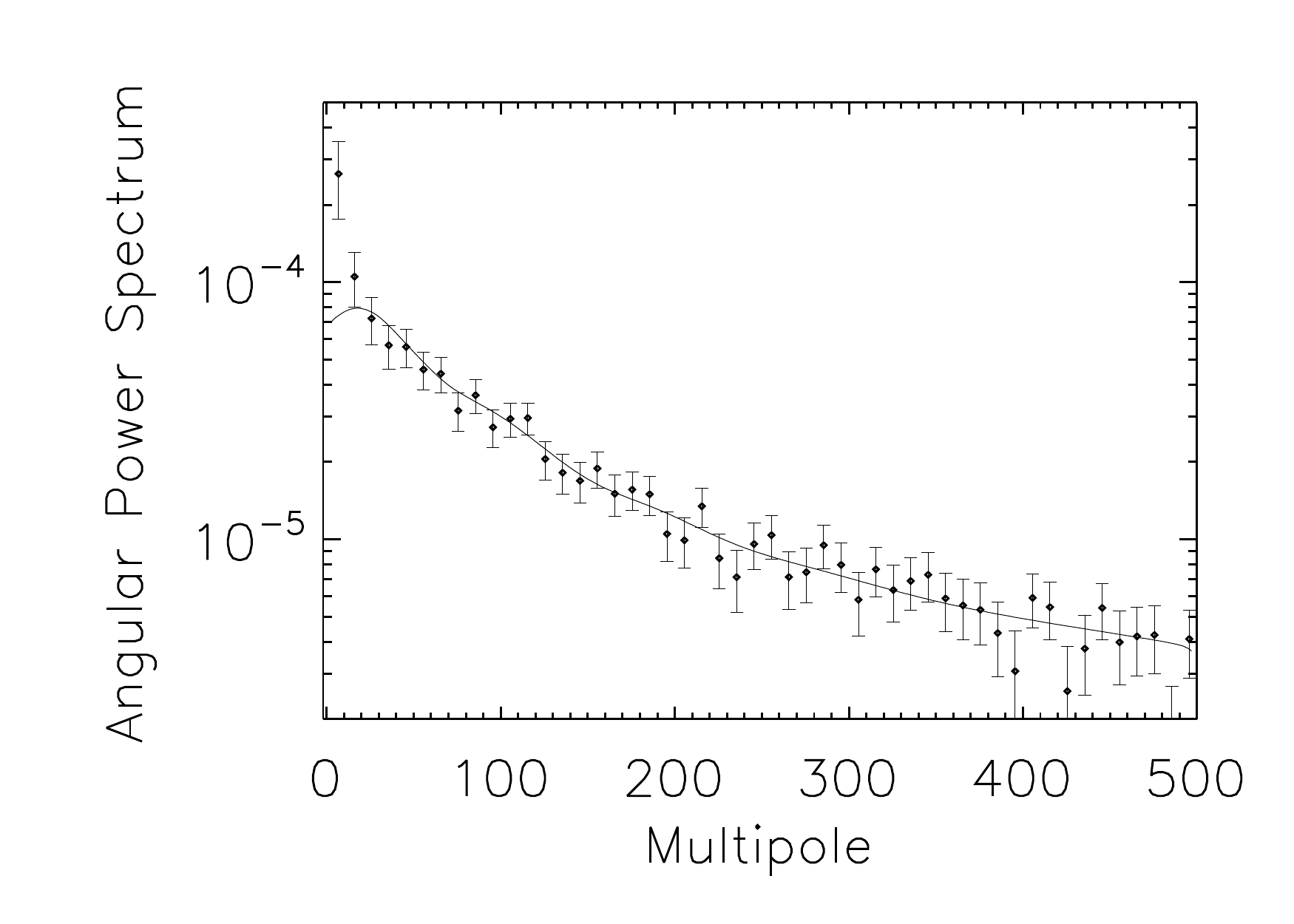} 
    \end{minipage}
    \caption{\small{The measured (auto) Angular Power Spectra ($C_{\ell}$) for the photometric SDSS MegaZ-LRG (DR7) population evaluated using Equation~\ref{eq:weightedCls}. The error bars correspond to those calculated with Equation~\ref{eq:gaussianerror} using the measured power spectrum. These include contributions from cosmic variance and shot noise, while accounting for the fraction of the sky surveyed. The solid line is evaluated for the the best fit parameters found in Section~\ref{sec:result} using the \citealt{Smith03} non-linear prescription. The panels are: Bin 1 (top left), Bin 2 (top right), Bin 3 (bottom left) and Bin 4 (bottom right), containing $259,498$; $237,564$; $155,293$ and $71,201$ galaxies, respectively. In the furthest redshift bin an excess of power is observed over the largest scale.}}
    \label{fig:Cl_redshift_bin_1}  \end{flushleft}
\end{figure*}
\noindent
We have constructed the galaxy clustering angular power spectra $C_{\ell}$ for SDSS MegaZ-LRG (DR7), an extension to the earlier analysis \citep{Blake07} of the original MegaZ-LRG catalogue \citep{Collister07}. Including $723,556$ photometrically determined LRGs and encapsulating $7746$ $\mathrm{deg^{2}}$ the measured values in four redshift bins extending $\Delta z=0.05$ in redshift, from $0.45$ to $0.65$, are included online\footnote{http://zuserver2.star.ucl.ac.uk/$\sim$sat/MegaZ/MegaZDR7.tar.gz}. These values are illustrated in Figure~\ref{fig:Cl_redshift_bin_1}. The full measurement procedure was detailed in Section~\ref{sec:measurement}. In addition, the measured cross power spectra between bins are described in Section~\ref{sec:crosspowerspectrameasured} and are also included online.

With the angular power spectra the statistical errors $\sigma(C_{\ell})$ on each power spectrum measurement are also included as given by Equation~\ref{eq:gaussianerror}, but calculated with the \emph{measured} $C_{\ell}$. They have been further weighted over the $\Delta \ell = 10$ band. This was shown in Figure~\ref{fig:errorratio} to be a good approximation. Note that for the cosmological parameter estimation in Section~\ref{sec:result} we utilise the Gaussian expression but evaluated with \emph{model} $C_{\ell}$s. 

In addition to the simulations described previously we also test the measurement pipeline by reconstructing the observed $C_{\ell}$ as found in the DR4 catalogue. We find these values to be identical to \citealt{Blake07}. As hinted in the DR4 results we find that DR7 also exhibits an excess of power over the largest scale ($\ell \sim 6$ band) in the furthest redshift bin. The effects and potential cause of this will be discussed in a companion paper.

\section{Theoretical Power Spectrum} \label{sec:powerspectrum}

In order to deduce the cosmology to match the measured angular distribution above one must first have a method for connecting the underlying 3D mass distribution to $C_{\ell}$. The outline description below simply follows the approach and notation of \citealt{Huterer01}, \citealt{Tegmark02}, \citealt{Blake07} and, most clearly, \citealt{Padmanabhan07}.  

One starts by noting that before the statistical decomposition of the density field into spherical harmonics in Section~\ref{sec:measurement} the field was projected. The same procedure is initially followed for the theoretical angular power spectra with the 3D mass distribution $\delta$ projected along the line-of-sight $\delta^{2D}$. This gives,

\begin{equation} \label{eq:deltalegendre}
\delta^{2D} = i^{l} \int \frac{\ud^{3}k}{(2\pi)^{3}} \; \delta (\mathbf{k}) W_{l}(k),
\end{equation}
\noindent
where $\delta$ has also undergone a Fourier transformation. The resulting spherical Bessel function $j(kz)$ and the projection's weight $f(z)$ have been absorbed into the window function given by, 

\begin{equation} \label{eq:windowfunction}
W_{l}(k) = \int f(z)j_{l}(kz) \ud z.
\end{equation}
\noindent
The weight naturally depends on the normalised redshift distribution of the objects under consideration $\int n(z) \ud z = 1$ and the linear growth factor $D(z)$,
\begin{equation} \label{eq:weight}
f(z) = n(z) D(z) \Big(\frac{\ud z}{\ud x}\Big)
\end{equation}
\noindent
with the Jacobian relating to the radial comoving coordinate $x$. Using the definition of the power spectrum $P(k)$ for the 3D density field $\delta({\bf k})$,
\begin{equation} \label{eq:powerspectrum}
<\delta(\mathbf{k}) \delta^{*}(\mathbf{k'})> = (2\pi)^{3} \delta^{3}(\mathbf{k} - \mathbf{k'}) P(k)
\end{equation}
\noindent
the angular power spectrum $C_{\ell}$ is found and similarly defined to be,
\begin{equation} \label{eq:angularpowerspectrum}
C_{\ell} \equiv <\delta^{2D} \delta^{* 2D}> = 4\pi \int \Delta^{2}(k)W_{\ell}^{2}(k) \frac{\ud k}{k}.
\end{equation}
\noindent
The spectrum has been recast into the dimensionless power spectrum defined in Equation~\ref{eq:dimensionlesspowerspectrum}. This power spectrum describes the variance of the matter field in logarithmic bands and so the equation for $C_{\ell}$ is subsequently a weighted integral of this quantity over logarithmic intervals ($\ud k/ k = \ud \mathrm{ln} \; k$). 
\begin{equation} \label{eq:dimensionlesspowerspectrum}
\Delta^{2}(k) \equiv \frac{4\pi k^{3} P(k)}{(2\pi)^{3}}
\end{equation}
\noindent
This can be further written in terms of the galaxy power spectrum with the addition of a linear galaxy bias $b$,
\begin{equation} \label{eq:bias}
P_{g}(k) = b^{2} P(k).
\end{equation}
\noindent
For an analysis between redshift bins the above outline can be easily extended. The cross correlation of two distinct projected mass distributions $<\delta_{i}^{2D} \delta_{j}^{* 2D}>$ leads simply to a slight modification in Equation~\ref{eq:angularpowerspectrum}; with the window function for each bin treated separately,
\begin{equation} \label{eq:crosspowerspectrummodel}
C^{ij}_{\ell} = 4\pi \int \Delta^{2}(k) W_{i}(k) W_{j}(k) \frac{\ud k}{k}.
\end{equation}
\noindent
For $\ell \gtrsim 60$ the exact expression (Equation~\ref{eq:angularpowerspectrum}) can be simplified by the small angle approximation (e.g. \citealt{Blake07}),
\begin{equation} \label{eq:appromationcl}
C_{\ell} = b^{2} \int P(k,z) \frac{n(z)^{2}}{x(z)^{2}} \Big( \frac{\ud x}{\ud z}  \Big)^{-1} \ud z.
\end{equation} 
\noindent
On larger scales (smaller $\ell$) this approximation becomes invalid as it seriously underestimates the power in $C_{\ell}$ and is not used in this regime. In fact, even the exact expression does not capture the shape of the true power spectrum below  $\ell \sim 60$. The main reason is because of redshift space distortions, which lead to a significant boost in the angular power spectrum.

\subsection{Redshift Space Distortions} \label{sec:reddist}

The peculiar velocity of a galaxy will cause it to appear shifted along the line-of-sight in redshift coordinates (E.g. \citealt{Sargent77}, \citealt{Peebles80}, \citealt{Kaiser87}, \citealt{Fisher94}, \citealt{Heavens95}, \citealt{Hamilton98} and \citealt{Guzzo08}). This is relative to the same galaxy carried along only by the background Hubble flow. That is, the redshift distance $s$ of a body will be altered from its true distance $r$, by its own peculiar velocity $v \equiv \mathbf{\hat{r}.v}$, radially from the observer,
\begin{equation} \label{redshiftdistance}
s = r + \mathbf{\hat{r}.v} \equiv r + v.
\end{equation}
\noindent
In redshift space this deviation alters the apparent clustering of galaxies and collectively the effect is said to be the result of \emph{redshift space distortions}. 

\begin{figure}
  \centering
      \includegraphics[width=3.25in,height=3.25in]{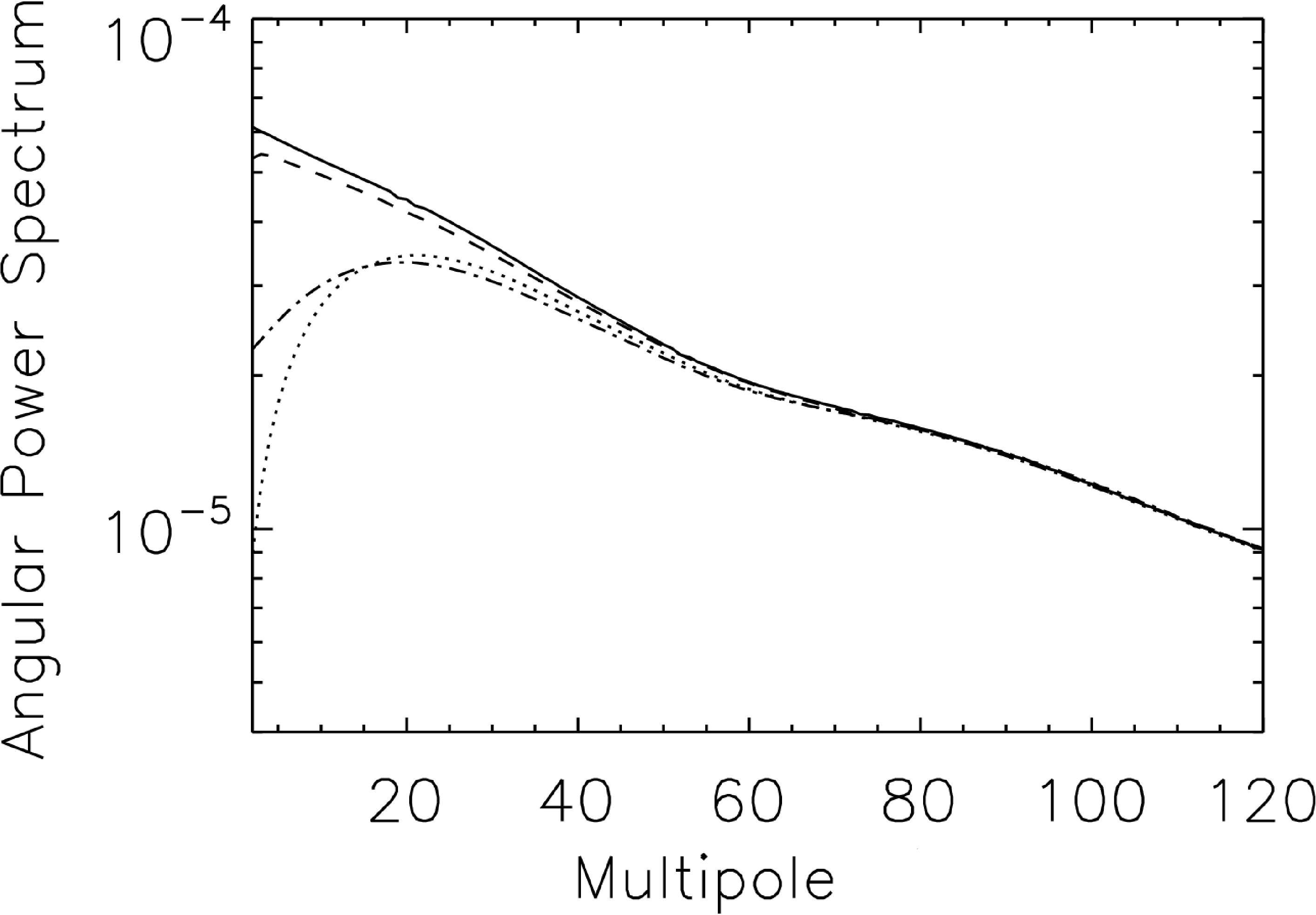}
    \caption{\small{A range of theoretical angular power spectra for the lowest redshift bin used in this survey ($0.45 \le z \le 0.5$). This includes the small angle approximation (Equation~\ref{eq:appromationcl}; dotted line), the exact expression with no redshift space distortions (Equation~\ref{eq:angularpowerspectrum}; dot-dashed line), the exact expression including redshift space distortions (Equation~\ref{eq:recurrancerelation}; solid line) and also with the addition of the partial sky mixing matrix convolution (Section~\ref{sec:the_mixing_matrix}; Equation~\ref{eq:convolvedCl}; dashed line). The input parameters are taken to be: $\Omega_{b} = 0.05$, $\Omega_{m} = 0.3$, $h=0.75$, $\sigma_{8} = 0.8$ and $b = 1$ for all four profiles. The small angle approximation is used for multipole scales $\ell \gtrsim 60$ for faster computation in the cosmological analyses. }}
    \label{fig:theoretical_cl_profile}
\end{figure}

To include redshift space distortions in the angular power spectrum the window function $W_{\ell}(k)$ in Equation~\ref{eq:angularpowerspectrum} is modified such that $W_{\ell}(k) \to W_{\ell}(k) + W^{R}_{\ell}(k)$ (E.g. \citealt{Fisher94} and \citealt{Padmanabhan07}). This is a result of writing the weight properly as a function of redshift distance $f(s)$ and assuming that the magnitude of the peculiar velocities are small. This is because with this assumption one can perform a Taylor expansion of the weight,
\begin{equation} \label{eq:taylorweight}
f(s) \approx f(r) + \frac{\ud f}{\ud r} (\mathbf{v}(r\mathbf{\hat{r}}).\mathbf{\hat{r}}  ).
\end{equation}
\noindent
The subsequent window function (remembering Equation~\ref{eq:windowfunction}) therefore now has the two components, $W_{\ell}(k) + W^{R}_{\ell}(k)$, with the latter currently a function of $\mathbf{v}$ from above. The Fourier transform of $\mathbf{v}$ is in turn related to the density perturbation through the linear continuity equation,
\begin{equation} \label{eq:continuityvelocitytodensity}
\mathbf{v}(\mathbf{k}) = -i\beta \delta_{g}(\mathbf{k}) \frac{\mathbf{k}}{k^{2}}
\end{equation}
\noindent
with the constant of proportionality $\beta$ known as the redshift distortion parameter. This is commonly approximated by $\beta \approx \Omega_{m}^{\gamma}/b$, with $\gamma = 0.55$ in LCDM. Substituting this into the expression for the window function and Legendre transforming (see \citealt{Padmanabhan07} for further details) eventually leaves one with,

\begin{small}
\begin{multline} \label{eq:recurrancerelation}
W_{l}^{R}(k) = \beta \int f(y)\Big[\frac{(2l^{2} + 2l -1)}{(2l+3)(2l-1)}j_{l}(ky) + \\ - 
\frac{l(l-1)}{(2l-1)(2l+1)} j_{l-2}(ky) - \frac{(l+1)(l+2)}{(2l+1)(2l+3)}j_{l+2}(ky)\Big] \ud y.
\end{multline}
\noindent
\end{small}

For large values of $\ell$ the integral within Equation~\ref{eq:recurrancerelation} tends to zero and so the total window function is reduced to the previous form. In this way, even with the inclusion of redshift distortions, the small angle approximation is an efficient and accurate estimate of the angular power spectrum at small scales. The behaviour of this approximation and the effects of the redshift space distortions on the angular power spectra are illustrated further in Figure~\ref{fig:theoretical_cl_profile}. In addition, we later recast our constraints into limits on the distortion parameter $\beta$ as can be seen in Section~\ref{sec:redshiftspacedistortionlimits} and Figure~\ref{fig:redshiftdistortions}.

\subsection{The Mixing Matrix: Partial Sky Convolution} \label{sec:the_mixing_matrix}

\begin{figure}
  \centering
      \includegraphics[width=3.25in,height=3.25in]{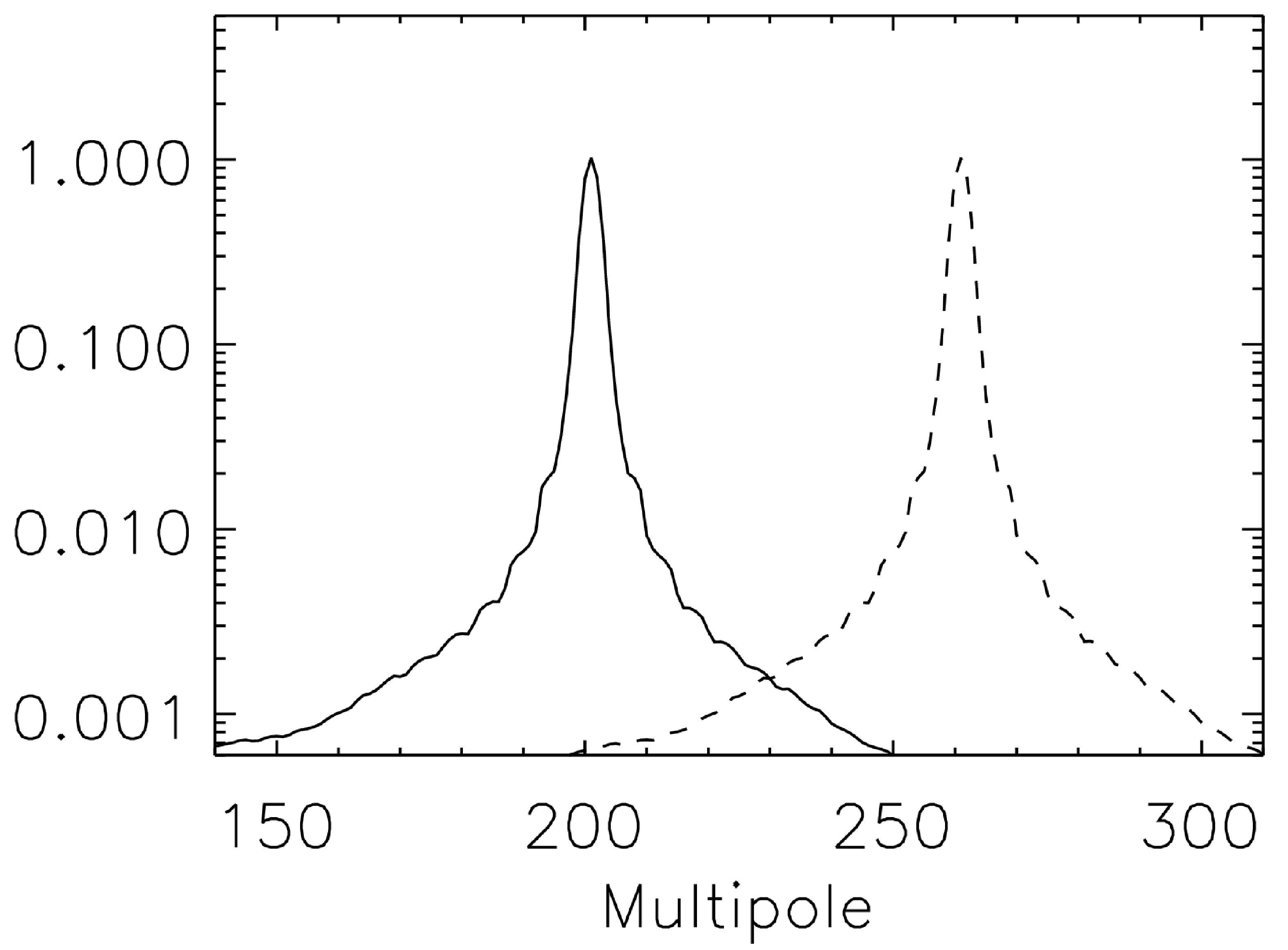}
    \caption{\small{A slice through the mixing matrix $R_{\ell, \ell'}$ is plotted for two fixed multipole values given by $\ell'=200$ (solid curve) and $\ell'=260$ (dashed curve). The amplitude of the matrix peaks at those fixed values and decays rapidly within the size of a $\Delta \ell$ band. This establishes how little correlation is induced by the survey's window function. Furthermore, the behaviour is observed similarly across all angular scales. Note that the matrix profiles have been normalised to unity at their peaks and the vertical axis is in logarithmic space.}}
    \label{fig:mixingmatrixplot}
\end{figure}
\noindent
An additional alteration in the shape of $C_{\ell}$ at large scales is to account for the partial sky coverage of the real survey. As stated in Section~\ref{sec:measurement} this correlates the usually orthonormal spherical harmonic coefficients, effectively creating a dependency on neighbouring scales. The net effect is to slightly suppress the shape of the power spectrum $C_{\ell}$ below $\ell \sim 60$ as seen in Figure~\ref{fig:theoretical_cl_profile}. The effect can be calculated by convolving with the \emph{mixing matrix} $R_{l,l'}$ (\citealt{Hauser73}, \citealt{Hivon02} and \citealt{Blake07}),

\begin{equation} \label{eq:convolvedCl}
C_{l} = \sum_{l'} R_{l,l'} C_{l'}.
\end{equation}
\noindent
The mixing matrix can be pre-calculated and depends purely on the survey geometry. It is described by,
\begin{equation} \label{eq:mixingmatrix}
R_{l,l'} = \frac{2l'+1}{4\pi} \sum_{l''} (2l''+1) W_{l''} \left( \begin{array}{ccc} l & l' & l'' \\ 0 & 0 & 0 \\\end{array} \right)^{2}
\end{equation}
\noindent
with $W_{l}$, the power spectrum of the survey's mask, calculated using Equation~\ref{eq:maskpowerspectrum}. The $2 \times 3$ matrix within $R_{l,l'}$ is a Wigner coefficient. For a full sky survey the convolution should have no effect on the angular power spectrum and accordingly the mixing matrix reduces to the identity matrix $R_{\ell,\ell'} \to \delta_{\ell \ell'}$.
\begin{equation} \label{eq:maskpowerspectrum}
W_{l} = \frac{\sum_{m=-l}^{l} |I_{l,m}|^{2}}{2l+1}
\end{equation}
\noindent
For the DR7 survey geometry the mixing matrix at a given $\ell$ is seen to be heavily peaked as a function of $\ell'$ about that multipole value. The profile rapidly falls within the chosen $\Delta \ell = 10$ bin, implying that only a small correlation between the $\ell$ bands is introduced by the mask. This is illustrated in Figure~\ref{fig:mixingmatrixplot} for two different multipole scales.

\section{The Cosmological Analysis} \label{sec:result}

We calculate $P(k)$ for the angular power spectrum $C_{\ell}$ using {\sc camb} \citep{Lewis00}. The {\sc halofit} fitting function \citep{Smith03} is then used to map the linear power spectrum into the non-linear regime (large $\ell$). To increase the speed of calculation we use the small angle approximation (Equation \ref{eq:appromationcl}) for $\ell \gtrsim 60$ and the full and exact window function, including redshift distortions (Equation \ref{eq:angularpowerspectrum} and Equation \ref{eq:recurrancerelation}), otherwise. This is all convolved with the mixing matrix $R_{\ell, \ell'}$ as described in the previous subsection.

\subsection{The Redshift Distribution} \label{2SLAQredshiftdistribution}

\begin{figure}
  \centering
      \includegraphics[width=3.25in,height=3.25in]{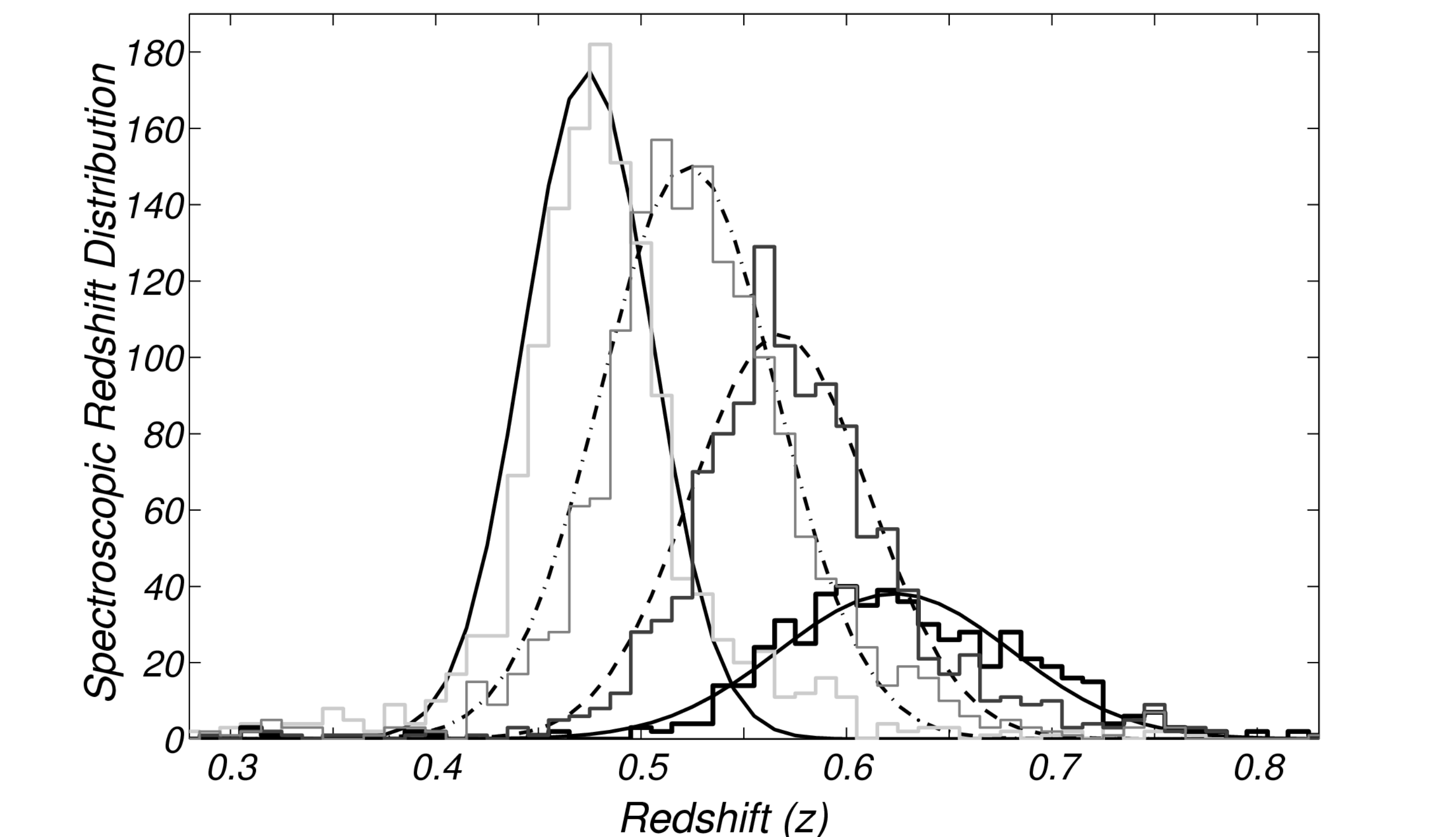}
    \caption{\small{The spectroscopic redshift distribution $n(z)$ for each photometric bin in DR7 is illustrated as a series of histograms. Each redshift distribution is fit by a Gaussian function $\mathrm{exp} [ -(z - \mu)^{2}/2\sigma^{2}]$, where $\mu$ and $\sigma$ are specified in Table~\ref{table:gaussianredshiftfits}. The associated Gaussian fits are represented by the smooth curves. }}
    \label{fig:gaussianredshiftdistribution}
\end{figure}
\noindent
\begin{table}
\centering
\begin{tabular}{cccc} 

\hline
$\mu$ & $\sigma$ & Redshift Bin & Photometric Range \\
\hline
$0.474$ & $0.0312$ & Bin 1 & $0.45<z<0.50$ \\
$0.523$ & $0.0428$ & Bin 2 & $0.50<z<0.55$ \\
$0.568$ & $0.0433$ & Bin 3 & $0.55<z<0.60$ \\
$0.624$ & $0.0568$ & Bin 4 & $0.60<z<0.65$ \\
\hline
\hline
\\
\end{tabular}
\caption{\small{The mean $\mu$ and deviation $\sigma$ of the Gaussian fitting to the spectroscopic redshift distribution $n(z)$ in each photometric bin. This is highlighted in Figure~\ref{fig:gaussianredshiftdistribution}.  }}
\label{table:gaussianredshiftfits}
\end{table}
\noindent
The model redshift distribution $n(z)$ in each redshift slice is taken to be the form of the spectroscopic 2SLAQ evaluation set, with the same LRG selection criteria, in that \emph{photometric} bin. This is possible because the 2SLAQ evaluation objects have both a spectroscopic and photometric redshift. These $n(z)$ were fit with a Gaussian function given by,
\begin{equation} \label{eq:gaussianredshift distribution} 
n(z) \propto \mathrm{exp}\Big[ -\frac{(z - \mu)^{2}}{2\sigma^{2}}  \Big].
\end{equation}
For the cosmological analyses $\mu$ and $\sigma$ are fixed to their best fit values in each bin. We address this assumption as a potential calibration systematic in Section~\ref{sec:crosspowerspectrameasured}. The best fit quantities are summarised in Table~\ref{table:gaussianredshiftfits} for the current (DR7) data release. In addition, the Gaussian fits to the spectroscopic distributions are illustrated in Figure~\ref{fig:gaussianredshiftdistribution}. The vertical axis represents the number of spectroscopic 2SLAQ objects within a small histogram band $(\delta z)$.

\subsection{Parameter Constraints: The Single Redshift Bins}

\begin{figure*}
  \begin{flushleft}
    \centering
    \begin{tabular}{ll}
      \includegraphics[width=3.25in,height=3.25in]{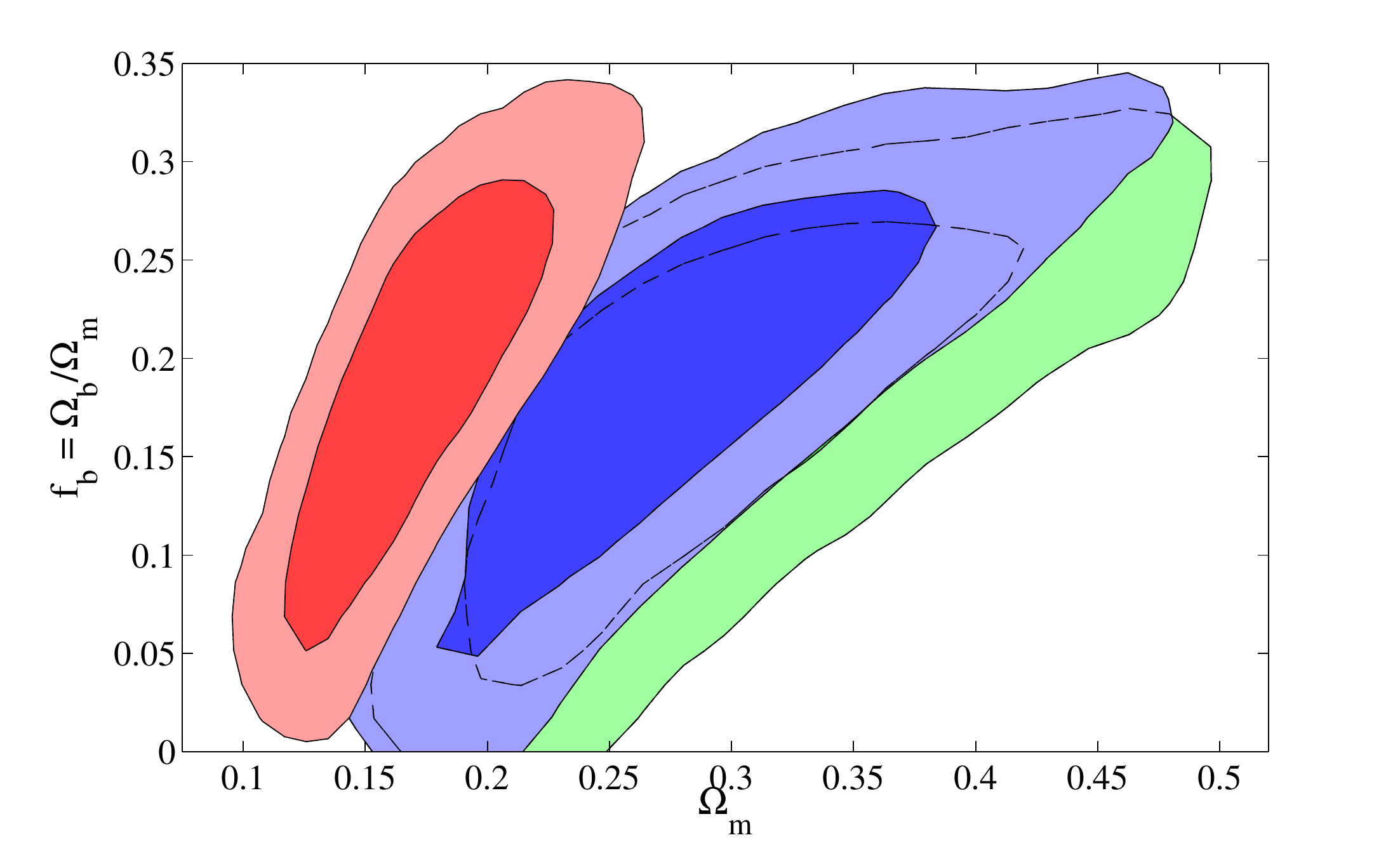} &
      \includegraphics[width=3.25in,height=3.25in]{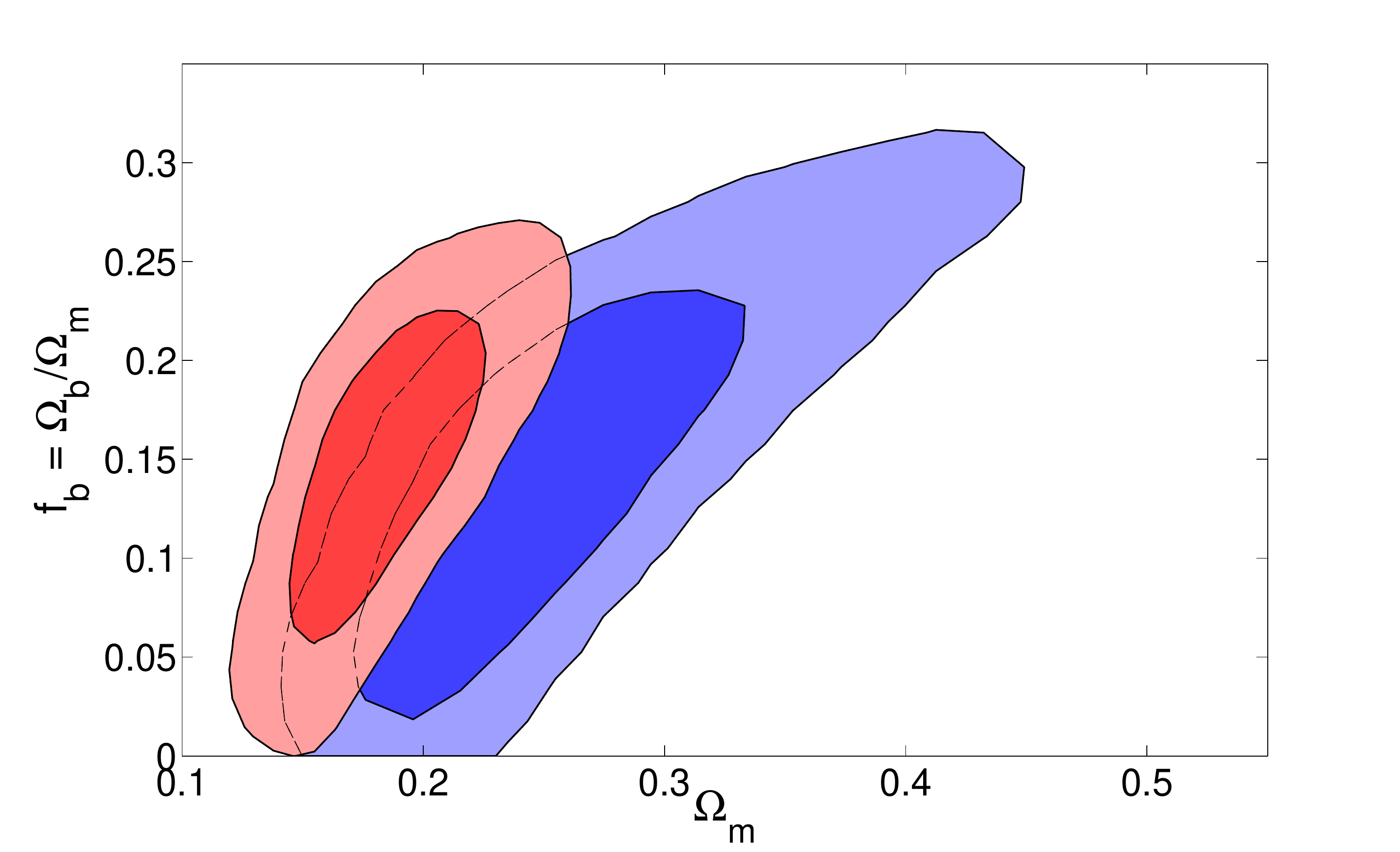}
    \end{tabular}
    \caption{\small{{\it Left Panel:} Constraints on the MegaZ LRG (DR4) highest redshift bin ($0.6 < z \le 0.65$) using model $C_{\ell}$s in the error expression (red/leftmost contour), data $C_{\ell}$s in the error (green/rightmost contour) and model errors with the lowest multipole removed (blue/central contour). The last analysis gives constraints consistent with the previous \citet{Blake07} study. {\it Right Panel:} DR7 constraints on the same bin using model errors (red/left contour) and model errors with the lowest multipole band removed (blue/right contour). Despite a slight decrease in the excess power in DR7 the observed shift in constraints above show the contribution from the anomalous low band to still be significant. The blue contour analysis is consistent with the other three redshift bins (Figures~\ref{fig:bin_contour} and \ref{fig:bin_contour_bias}) and as such this point is removed from all subsequent analyses.}}
    \label{fig:excesspowerDR7}
  \end{flushleft}
\end{figure*}
\noindent
We start by undertaking a preliminary cosmological analysis in each of the four separate redshift bins described previously. A conservative choice of parameters is studied such that we can test for consistency against the previous MegaZ LRG analysis \citep{Blake07}. We therefore vary four quantities: $f_{b} = \Omega_{b}/\Omega_{m}$, $\Omega_{m}$, $\sigma_{8}$ and $b$; the baryon-to-matter density ratio, the matter density, the normalisation of the power spectrum and the galaxy bias, respectively. The bias is assumed to be scale independent. Along with the earlier MegaZ paper the Hubble constant is fixed to $H_{0} = 75$ km $s^{-1}$ Mpc$^{-1}$ and the spectral index to $n_{s}=1$. Both $\sigma_{8}$ and the bias control the amplitude of the power spectrum and are thus degenerate with one another. A flat prior is therefore enforced on the former such that $0.7 \le \sigma_{8} \le 1.1$. The Universe is assumed to be flat throughout with the equation of state fixed to $w = -1$. We use all the multipole values up to $\ell = 300$. This is the scale at which the non-linear corrections become increasingly significant. For the parameter exploration we use the publicly available {\sc CosmoMC} package \citep{Lewis02}.

\subsubsection{Data Release 4} \label{sec:datarelease4results}

To test for consistency we first perform the cosmological analysis on the previous DR4 angular power spectra found in \citealt{Blake07}. We find a remarkably similar agreement to the previous study in the first three redshift bins over a redshift range $0.45 \le z \le 0.6$. However, for the final and furthest redshift bin ($0.6 < z \le 0.65$) a large discrepancy is discovered when all angular scales to $l_{\mathrm{max}} = 300$ are utilised. It is interesting that for this particular redshift bin a large excess of power is observed in the measurement of the $C_{\ell}$ on the largest angular scale ($\ell \sim 6$ band). Even though this is approximately at the turnover scale of the power spectrum, where one might expect the power to start decreasing, the excess was found to be over $1\sigma$ from the best fit $C_{\ell}$ profile. One might not therefore expect this anomalous point to cause any significant alteration in the cosmological analysis. It is important to remember, however, that the error on this data point, assigned in the previous study, was the error given by the Gaussian expression (Equation~\ref{eq:gaussianerror}) using the \emph{data} value for the $C_{\ell}$. As the magnitude of this point is so much larger than the $C_{\ell}$ corresponding to a smooth fit through the other data points, the associated data error bar is made to appear much larger also. In the parameter estimation performed here and in \citealt{Blake07} the error and therefore covariance matrix are evaluated using the \emph{model} errors. This is because in a Bayesian analysis one implicitly assumes the model to be true. Any model spectrum attempting to fit the other data points will assign a theoretical value at the largest angular scales much lower than that measured and subsequently the error bar will be much smaller. Therefore we find the excess power a much poorer fit than was ascribed previously in \citealt{Blake07}. In order to try and replicate the original DR4 constraint for this furthest bin we remove this irregular point. In addition, we also follow an analysis using the data errors in the covariance matrix while \emph{including} the excess power quantity. The resulting contours are shown in the left panel of Figure~\ref{fig:excesspowerDR7}. 

The plot highlights that the excess power at low multipoles is indeed significant, with the inclusion of the lowest point dragging the constraint to much lower values of $\Omega_{m}$ (red contour). Also, the figure reiterates the notion that the data error (green contour) acts to buffer against this anomaly given that the contour is similar to the model analysis that excludes the excess power (blue contour). When fitting with the model errors and no excess power we find the constraints to be identical to those in \citealt{Blake07} and also consistent with the three other redshift bins. 

\subsubsection{Data Release 7} \label{sec:datarelease7results}

\begin{figure*}
  \begin{flushleft}
    \centering
    \begin{minipage}[c]{1.00\textwidth}
      \centering
      \includegraphics[width=2.91in,height=2.91in]{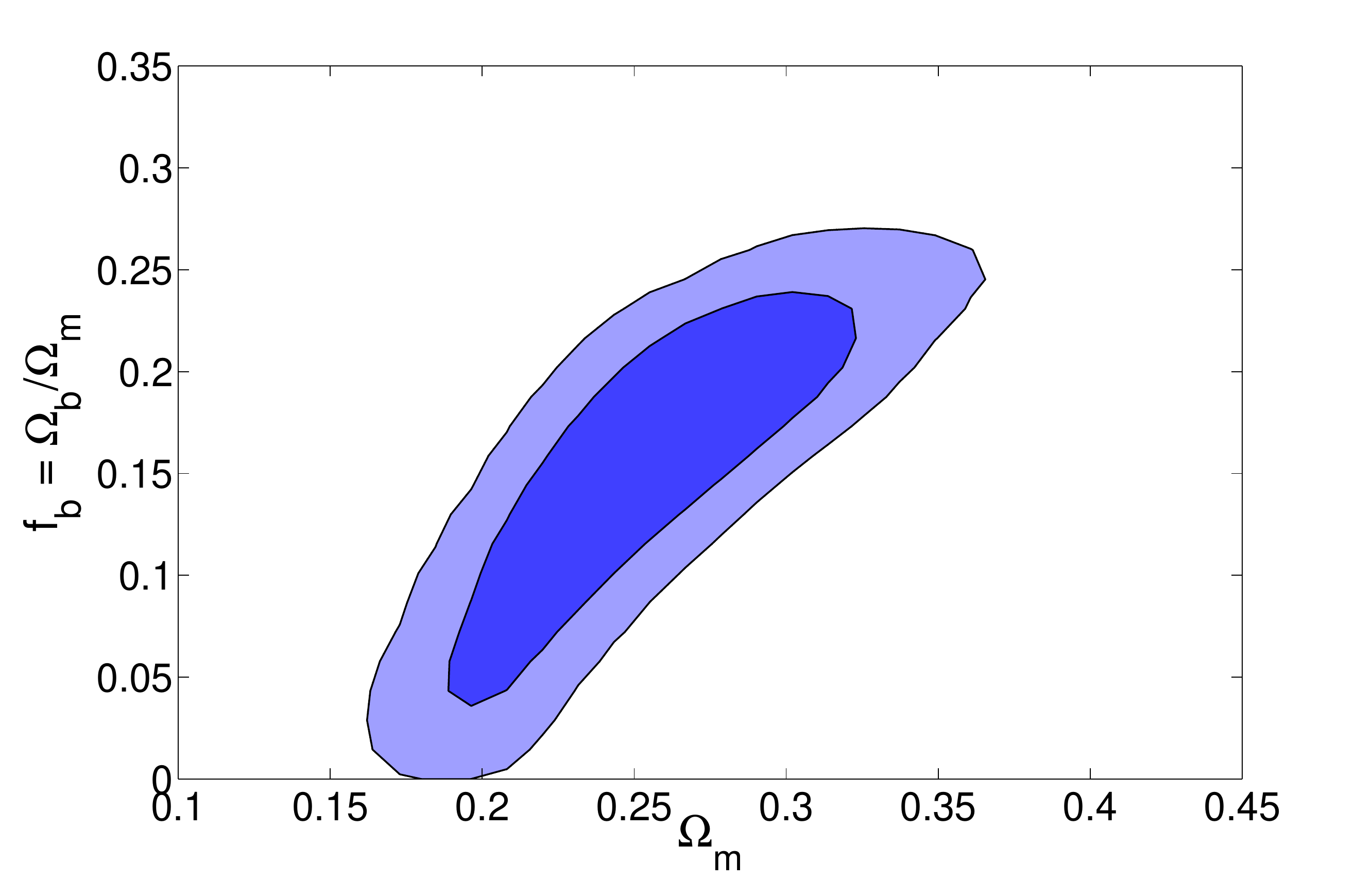} 
      \includegraphics[width=2.91in,height=2.91in]{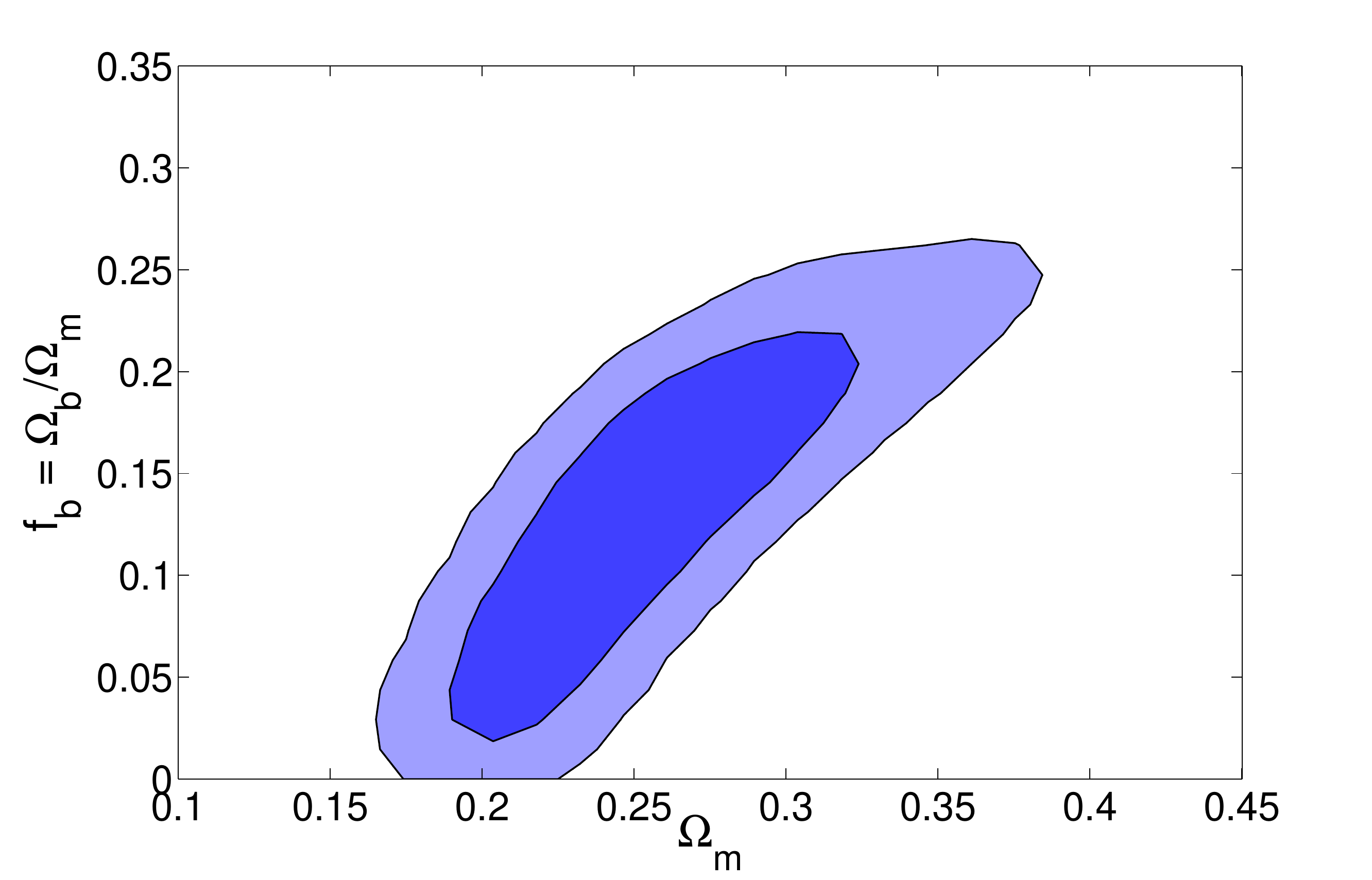} 
    \end{minipage}
    \begin{minipage}[c]{1.00\textwidth}
      \centering
      \includegraphics[width=2.91in,height=2.91in]{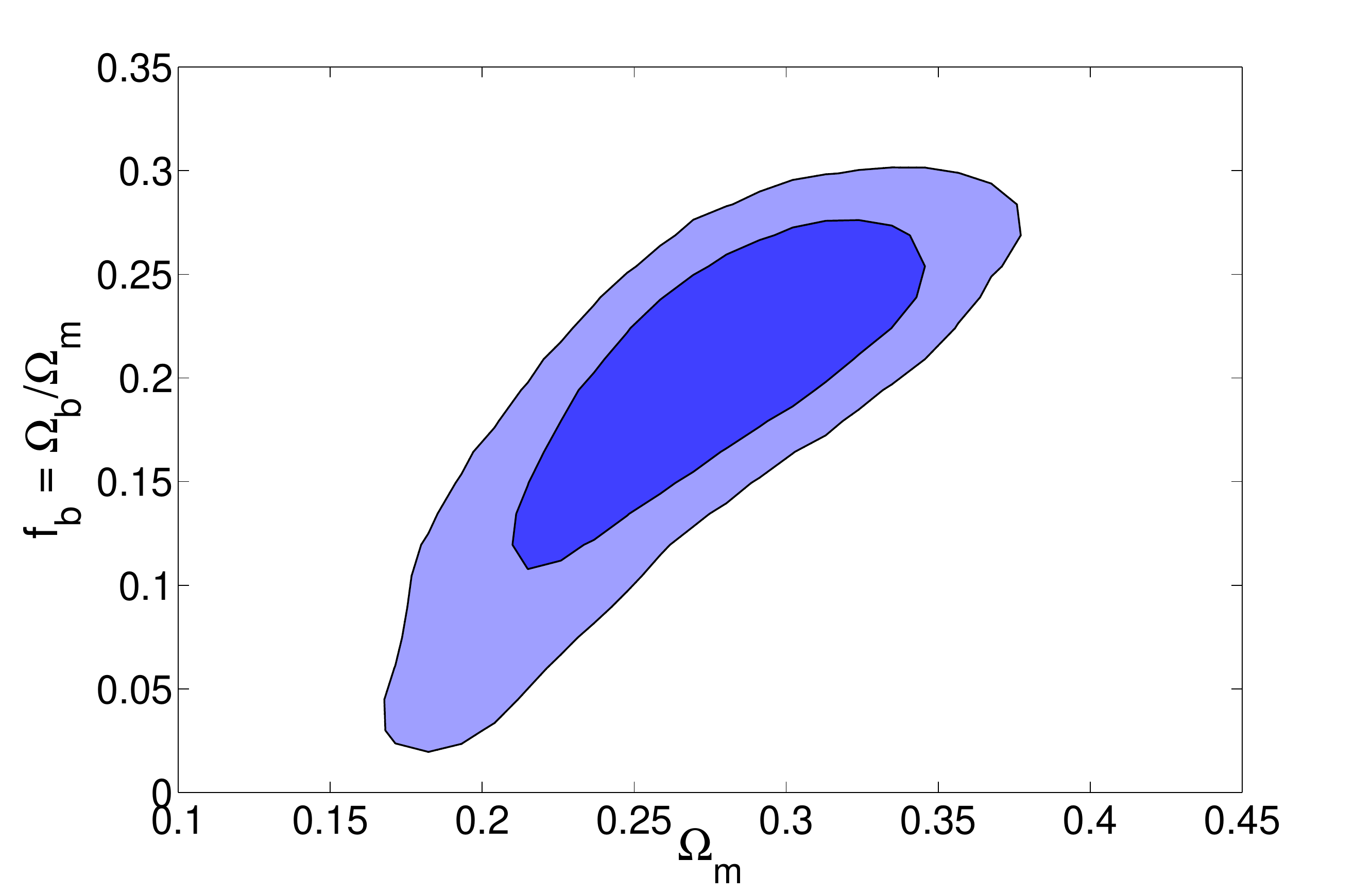} 
      \includegraphics[width=2.91in,height=2.91in]{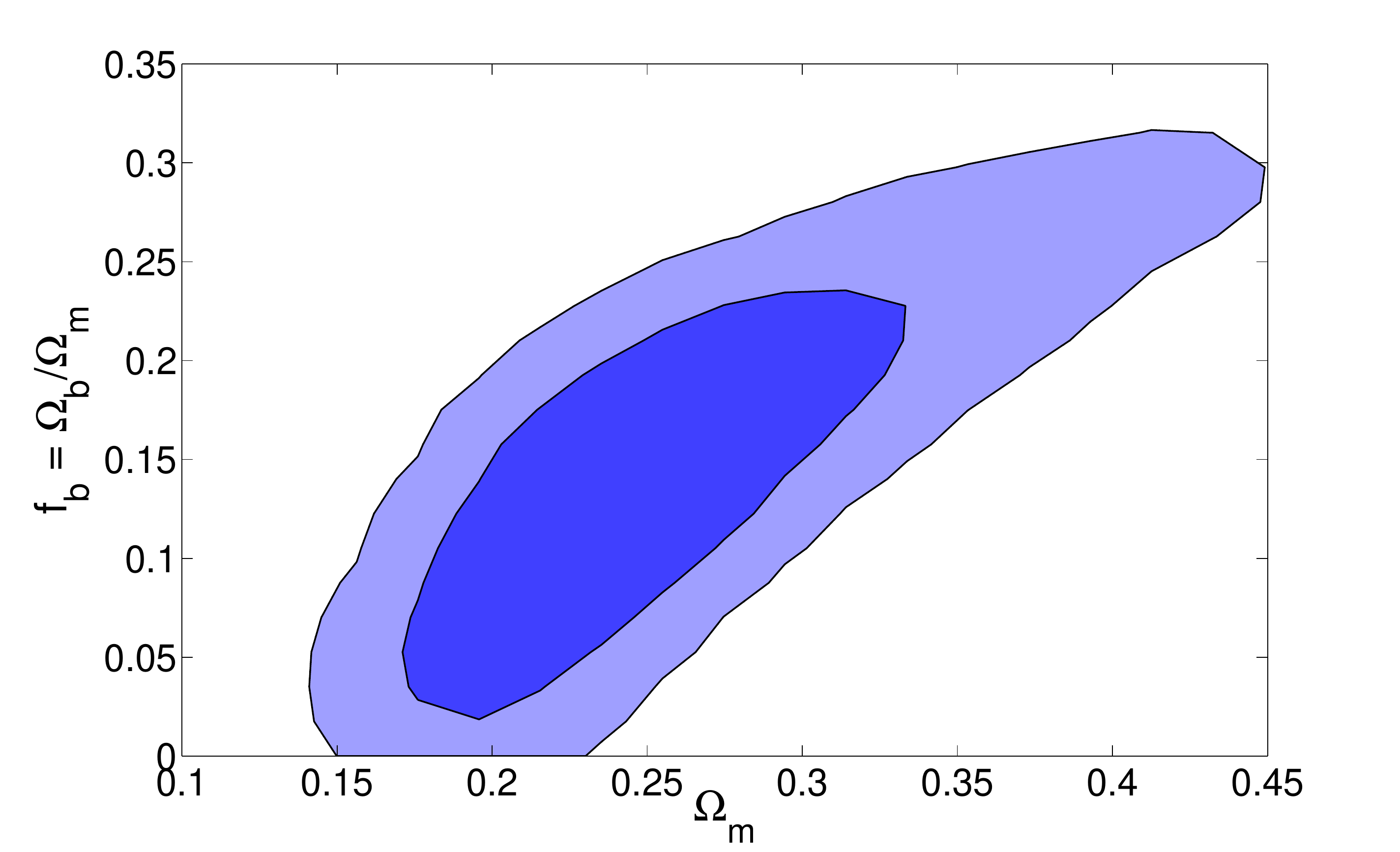} 
    \end{minipage}
    \caption{\small{MegaZ LRG DR7 constraints on $f_{b} = \Omega_{b}/\Omega_{m}$ and $\Omega_{m}$ for four {\it separate} redshift bins. $b$ and $\sigma_{8}$ have been marginalised over and $H_{0}$ and $n_{s}$ are fixed to  $75$ km $s^{-1}$ Mpc$^{-1}$ and $1$, respectively. The panels are: Bin 1 (top left), Bin 2 (top right), Bin 3 (bottom left) and Bin 4 (bottom right). The inner and outer contours are the $68\%$ and $95\%$ confidence levels, respectively.}}
    \label{fig:bin_contour}
  \end{flushleft}
\end{figure*}

\begin{figure*}
  \begin{flushleft}
    \centering
    \begin{minipage}[c]{1.00\textwidth}
      \centering
      \includegraphics[width=2.91in,height=2.91in]{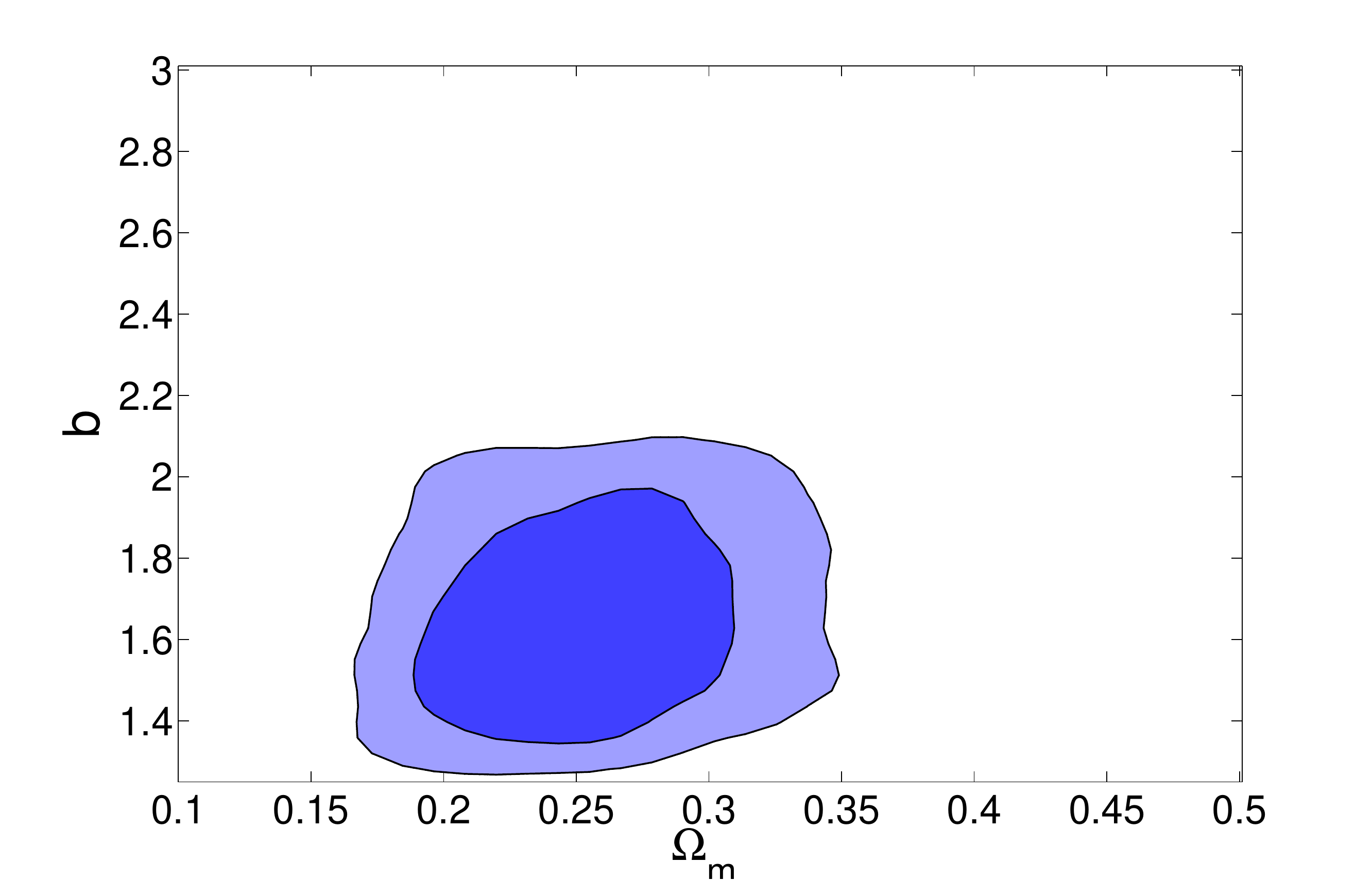} 
      \includegraphics[width=2.91in,height=2.91in]{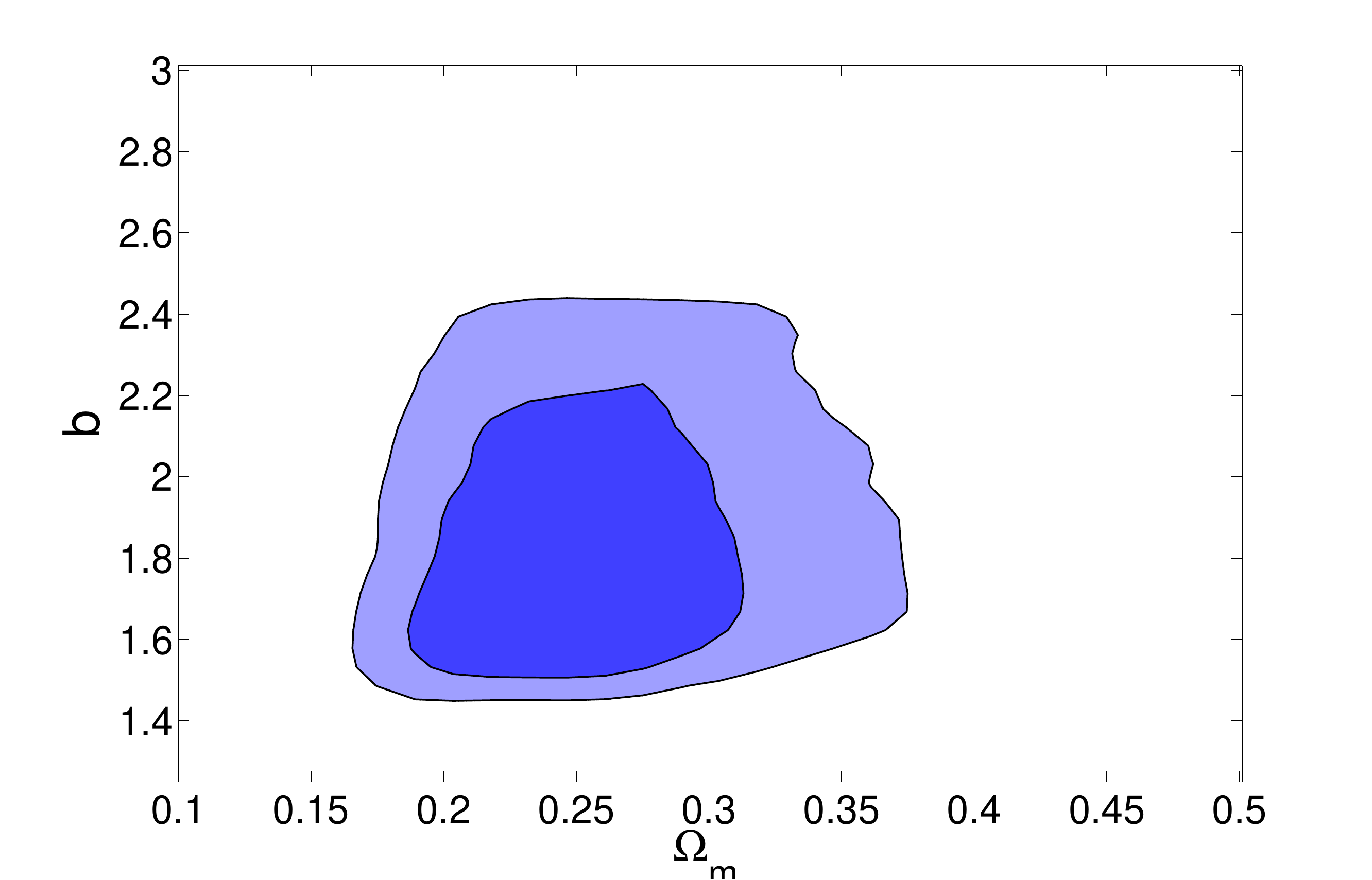} 
    \end{minipage}
    \begin{minipage}[c]{1.00\textwidth}
      \centering
      \includegraphics[width=2.91in,height=2.91in]{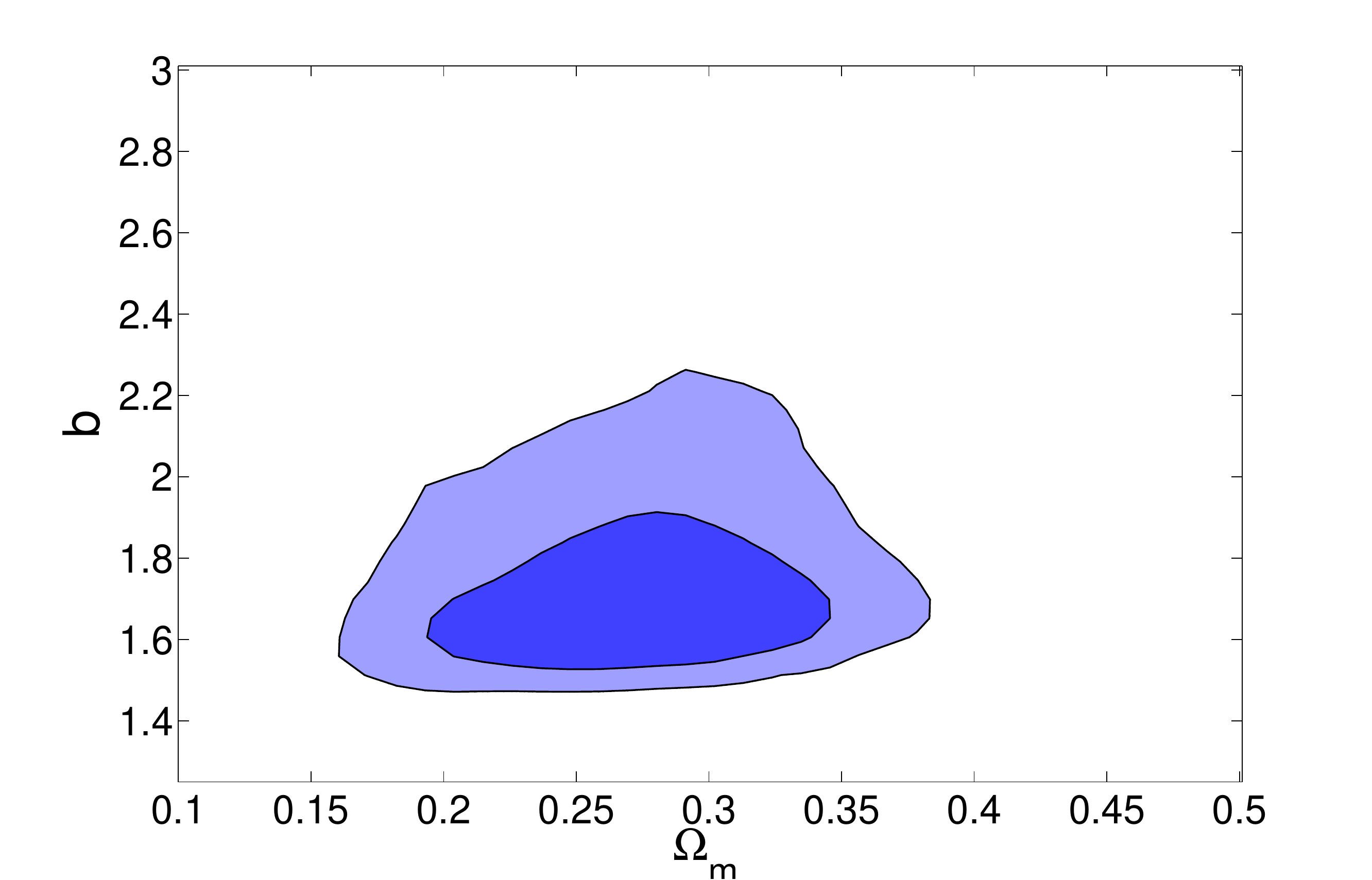} 
      \includegraphics[width=2.91in,height=2.91in]{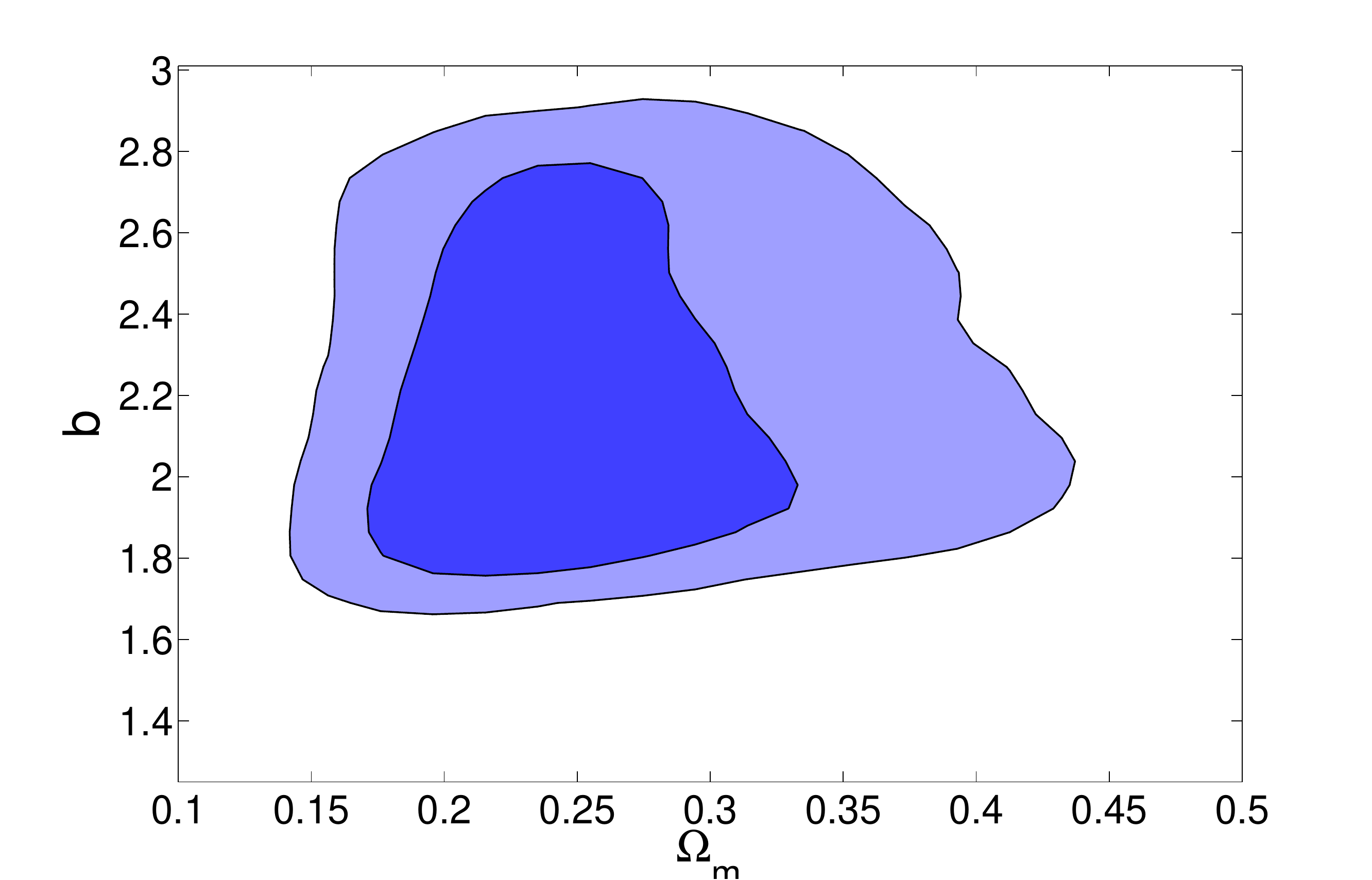} 
    \end{minipage}
    \caption{\small{MegaZ LRG DR7 constraints on $\Omega_{m}$ and the bias $b$ for four {\it separate} redshift bins. $f_{b} = \Omega_{b}/\Omega_{m}$ and $\sigma_{8}$ have been marginalised over and $H_{0}$ and $n_{s}$ are fixed to  $75$ km $s^{-1}$ Mpc$^{-1}$ and $1$, respectively. The panels are: Bin 1 (top left), Bin 2 (top right), Bin 3 (bottom left) and Bin 4 (bottom right). The inner and outer contours are the $68\%$ and $95\%$ confidence levels, respectively.}}
    \label{fig:bin_contour_bias}
  \end{flushleft}
\end{figure*}

The angular power spectrum for the last redshift bin was measured for DR7 in Section~\ref{sec:observedspectra} and shown in Figure~\ref{fig:Cl_redshift_bin_1}. Once again an excess of power is detected at this high redshift. However, there seems to be a slight hint of an ease in tension as the magnitude of the DR4 point is found to be $40\%$ higher than the newly measured DR7 value. We therefore undertake a cosmological run for this bin using the excess power point and also with it removed to test the effects. We find that despite the more recent decrement in the $C_{\ell}$ on these large scales the inclusion of the quantity still significantly affects the parameter constraints found with the bin. This is illustrated clearly in the right panel of Figure~\ref{fig:excesspowerDR7}. Again, with this point excluded the fourth redshift bin is found to be consistent with the other three slices. For the new DR7 release the associated constraints for every redshift bin are displayed in Figure~\ref{fig:bin_contour} and Figure~\ref{fig:bin_contour_bias}.

We therefore choose to continue the galaxy clustering study by excluding the anomalous excess power in the $\ell \sim 6$ band for the furthest redshift bin. It is intriguing that slight hints of excess power have also been seen in \citealt{Padmanabhan07} and in the study of the maxBCG cluster power spectrum by \citealt{Huetsi09}. We discuss this signal and potentially related systematics or causes further in a companion paper. 

When obtaining $f_{b}$, $\Omega_{m}$ or $b$ all the other parameters are marginalised over. The bias is subsequently seen to enlarge with an increase in redshift. This is partially due to the observed galaxies in the furthest redshift bin necessarily being more luminous, resulting from the pseudo-magnitude limit in the survey. They are therefore observed to be more highly clustered \citep{Blake07}. All the inferred constraints are summarised in Table~\ref{table:binconstraints}.

\subsection{Parameter Constraints: The Combined Redshift Bins} \label{sec:combinedbinsparameters}

\begin{figure*}
  \begin{flushleft}
    \centering
    \begin{minipage}[c]{1.00\textwidth}
      \centering
      \includegraphics[width=2.88in,height=2.87in]{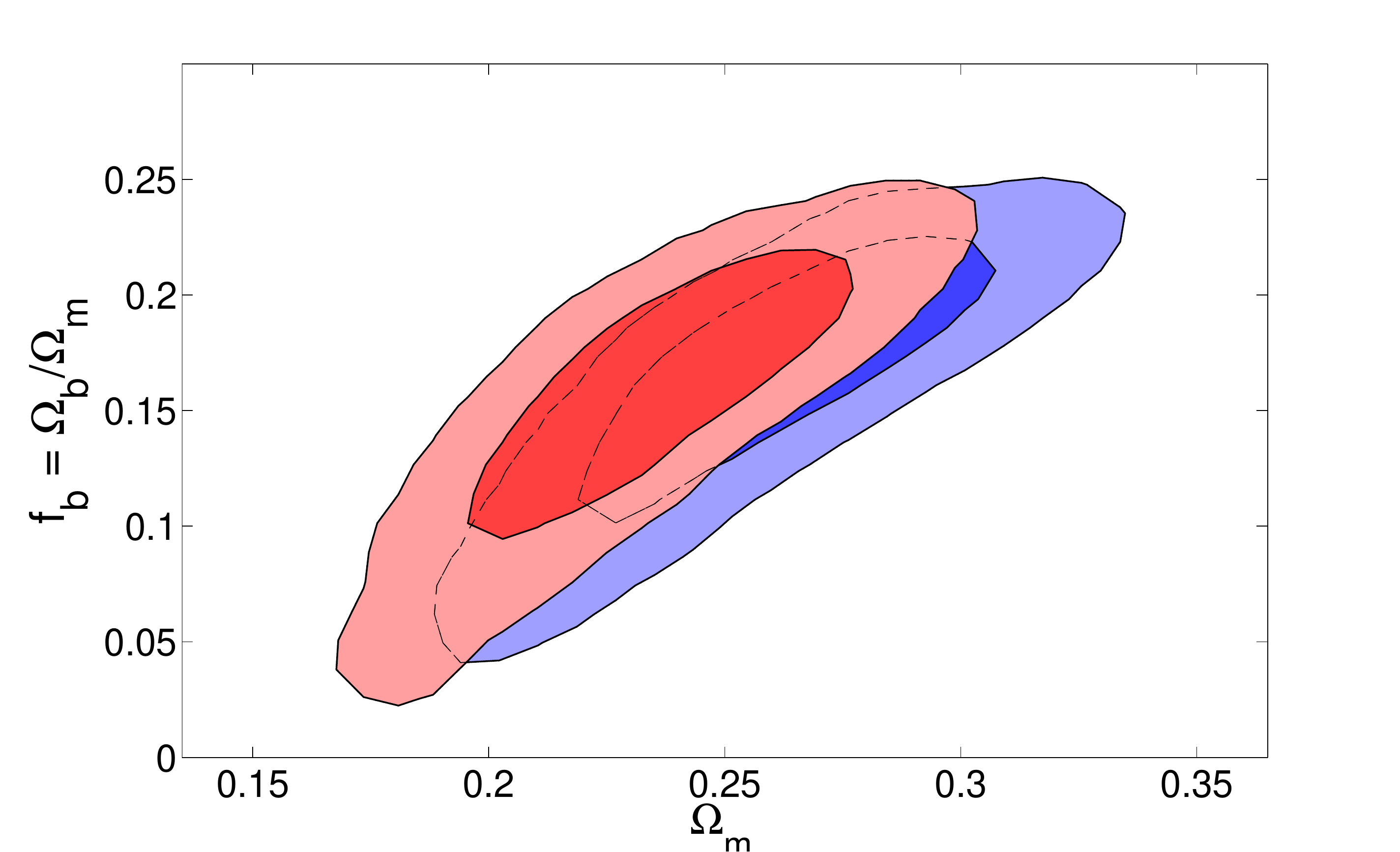} 
    \end{minipage}
    \begin{minipage}[c]{1.00\textwidth}
      \centering
      \includegraphics[width=2.88in,height=2.87in]{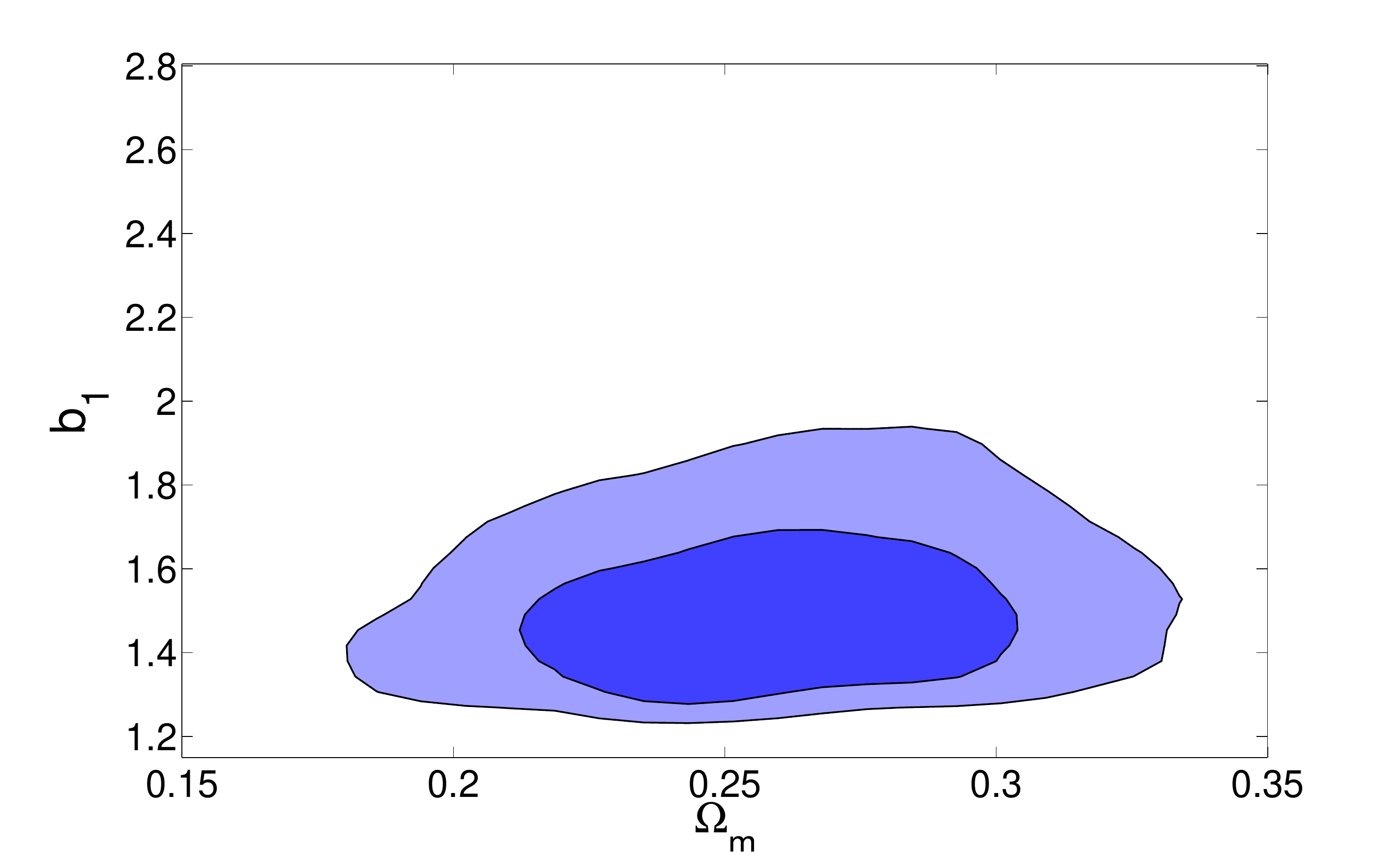} 
      \includegraphics[width=2.88in,height=2.87in]{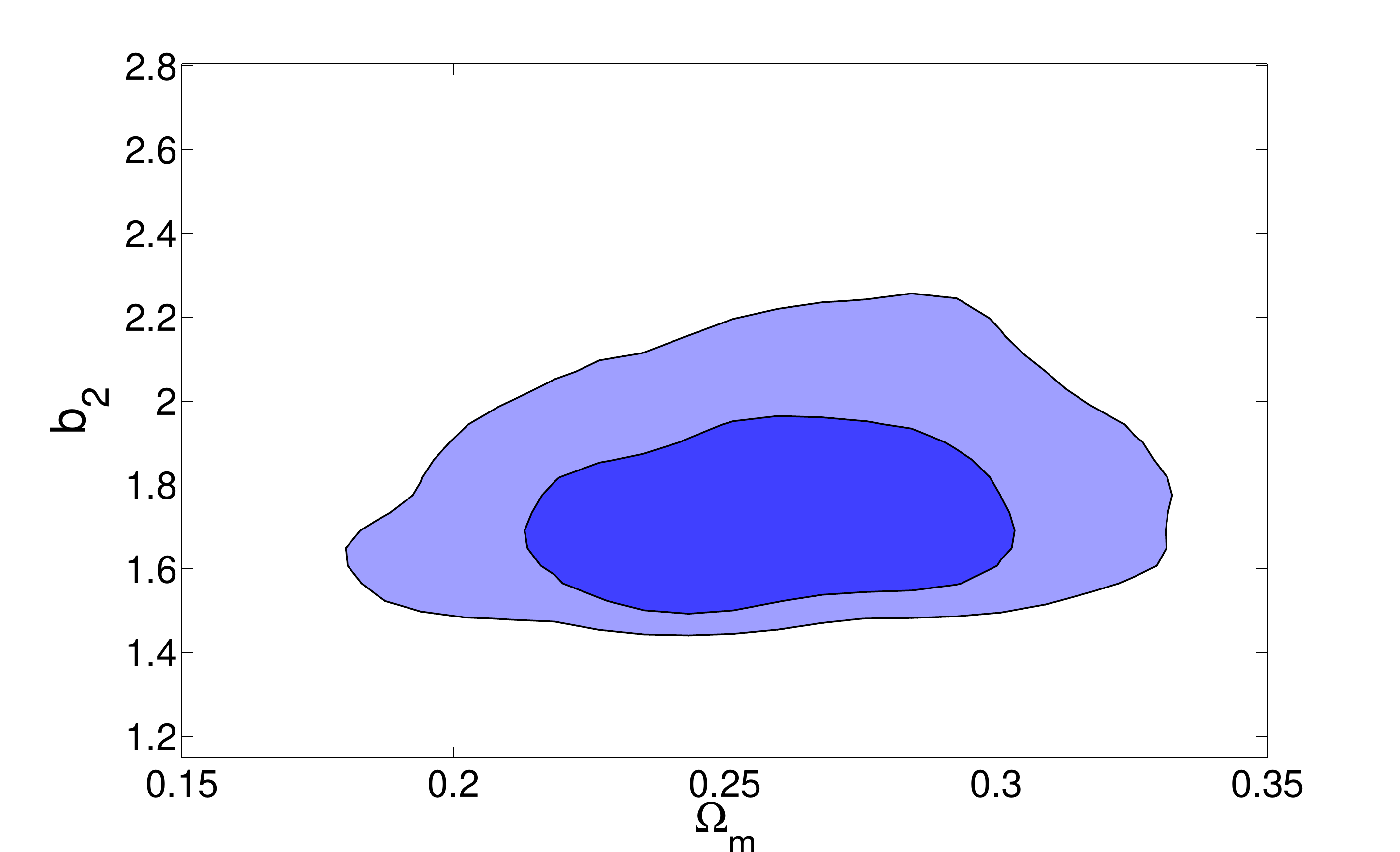} 
    \end{minipage}
    \begin{minipage}[c]{1.00\textwidth}
      \centering
      \includegraphics[width=2.88in,height=2.87in]{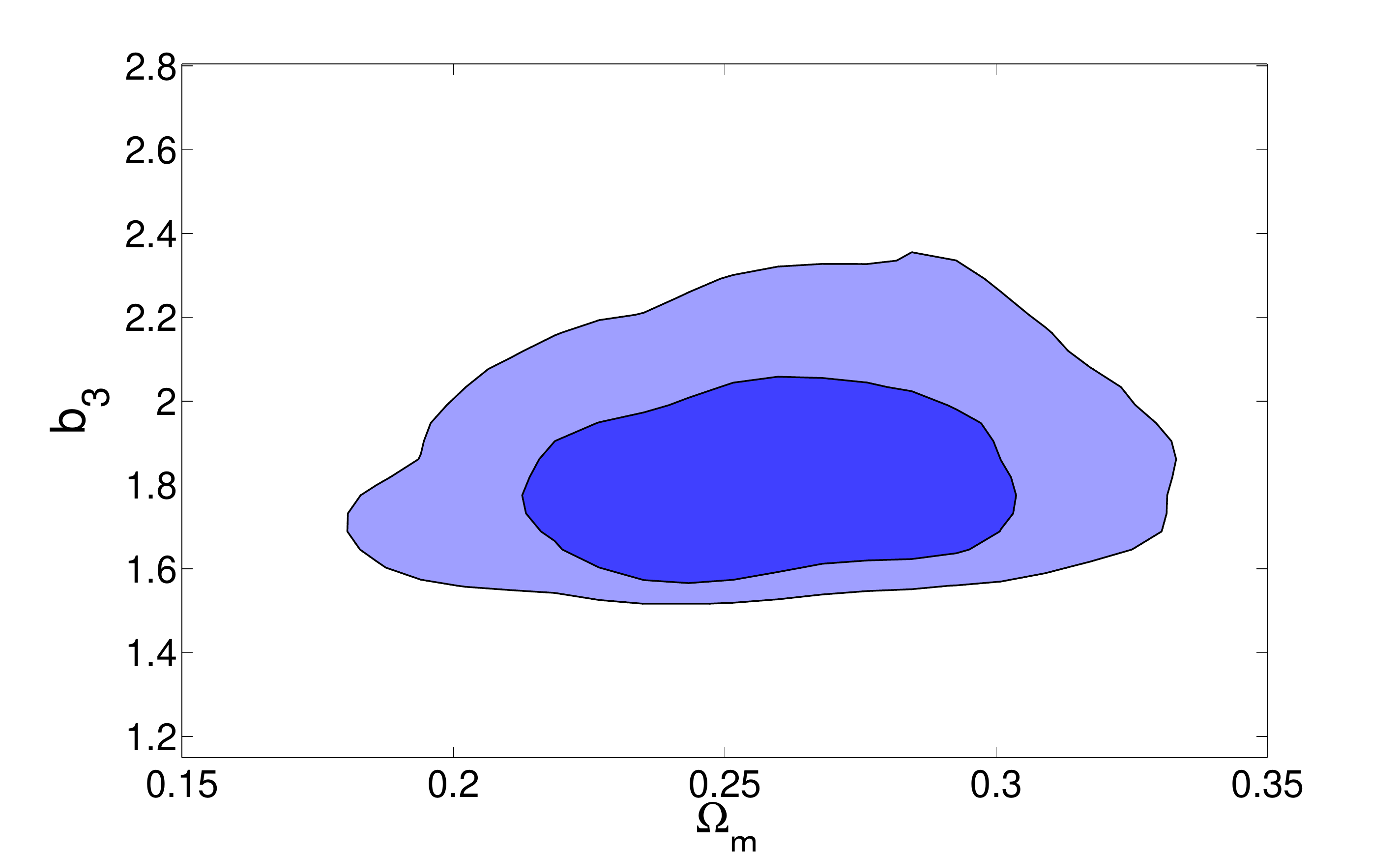} 
      \includegraphics[width=2.88in,height=2.87in]{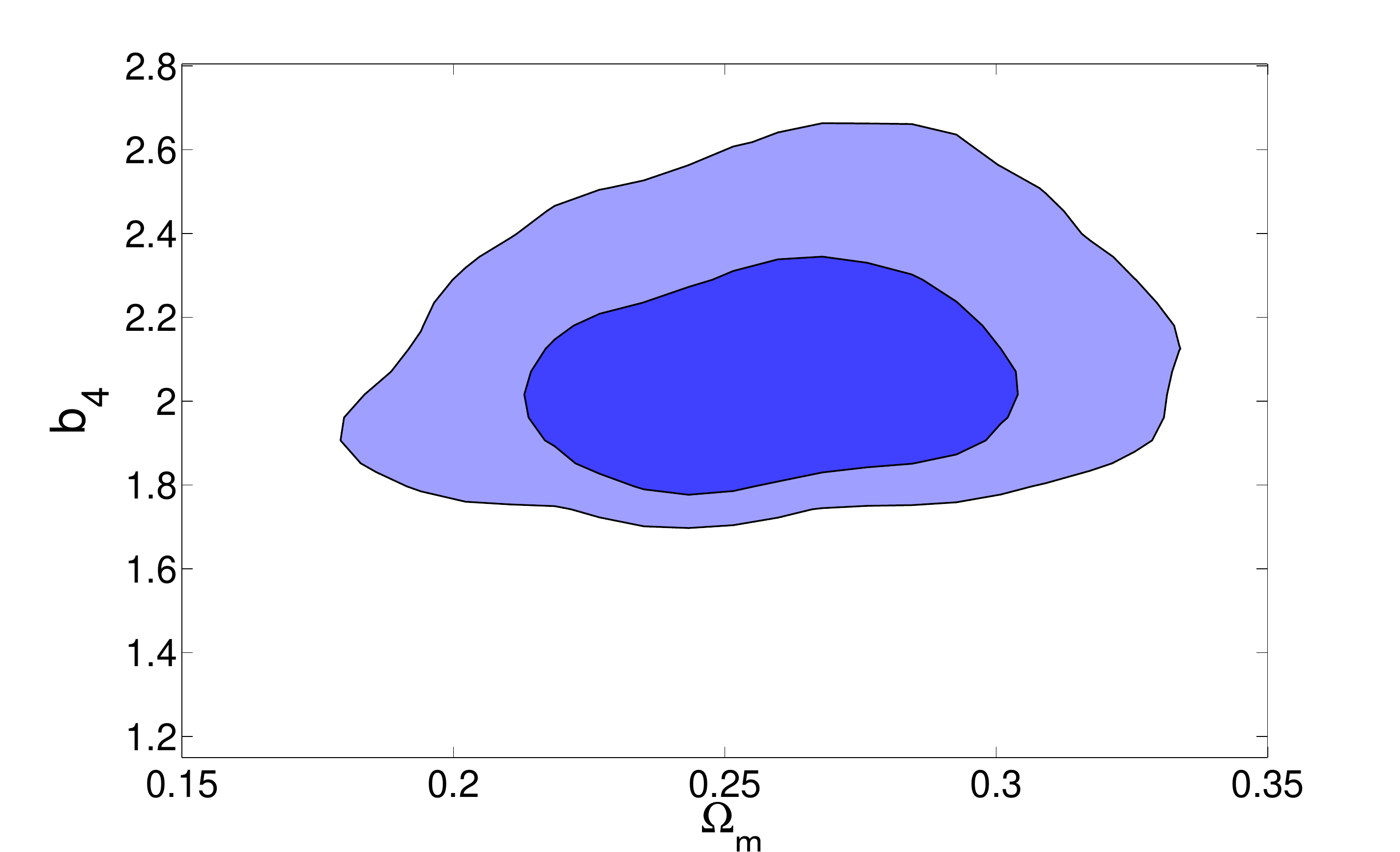} 
    \end{minipage}
    \caption{\small{{MegaZ LRG DR7 cosmological constraints given by the {\it combination} of four redshift bins between $0.45 < z < 0.65$. The red/left 2D distribution in the top panel shows the systematic shift induced by including the excess power measured over large scales in the highest redshift bin, whereas the normal blue contours have this anomalous point removed. The bottom four panels show the determination of the bias where $b_{1}$, $b_{2}$, $b_{3}$ and $b_{4}$ are the quantities in sequentially higher redshift bins. These correspond to marginalised limits of $b_{1} = 1.47 \pm 0.15$, $b_{2} = 1.71 \pm 0.17$, $b_{3} = 1.80 \pm 0.18$ and $b_{4} = 2.05 \pm 0.21$.}}}
    \label{fig:combinedredshiftbins}
  \end{flushleft}
\end{figure*}
\noindent

\begin{table}
\centering
\begin{tabular}{ccc} 

\hline
$f_{b}$ & $\Omega_{m}$ & Redshift Slice \\
\hline
\hline
$0.166 \pm 0.066$ & $0.253 \pm 0.049$ &  Bin 1 \\
$0.136 \pm 0.069$ & $0.251 \pm 0.051$ &  Bin 2 \\
$0.206 \pm 0.062$ & $0.274 \pm 0.052$ &  Bin 3 \\
$0.146 \pm 0.076$ & $0.248 \pm 0.067$ &  Bin 4 \\
$0.173 \pm 0.0462$ & $0.260 \pm 0.0351$ &  All bins \\
\hline
$0.163 \pm 0.0480$ & $0.234 \pm 0.0309$ &  All bins\\
\hline
\hline
\\
\end{tabular}
\caption{The marginalised mean values obtained from the analyses of the galaxy clustering angular power spectra $C_{\ell}$. $f_{b} = \Omega_{b}/\Omega_{m}$, $\Omega_{m}$, $\sigma_{8}$ and $b$ are varied for each single bin run. In the `all bins' analyses all the bins were combined together using the full covariance matrix and a bias parameter for each bin ($b_{1}$, $b_{2}$, $b_{3}$, $b_{4}$). In the last analysis the lowest multipole band in the highest redshift slice is included. }
\label{table:binconstraints}
\end{table} 
\noindent
We now combine the data from each of the four redshift bins. These bins are not independent, however, as photometric redshift errors act to disperse galaxies throughout the bins. Another way of noting this is to observe that the Gaussian redshift distributions, as seen in Figure~\ref{fig:gaussianredshiftdistribution}, overlap for each bin. We therefore use the full covariance matrix in the analysis. The variance element corresponding to the same redshift bin (e.g. between $C_{\ell}^{i}$ and $C_{\ell}^{i}$) is given by the square of Equation~\ref{eq:gaussianerror} using the \emph{theoretical} expression for $C_{\ell}$ as before. The covariance elements between different bins are described by,
\begin{equation} \label{eq:different_bin_covariance} 
\mathrm{Cov}(C_{\ell}^{i},C_{\ell}^{j}) = \frac{2}{f_{\mathrm{sky}} (2\ell+1)} \Big(C_{\ell}^{i,j}\Big)^{2}.
\end{equation}
\noindent
In this way the whole matrix allows for the covariance between all bin combinations but not multipole bands. This is a good approximation given our earlier discussion of the highly peaked mixing matrix $R_{l,l'}$ (Figure~\ref{fig:mixingmatrixplot}). 

We include a redshift dependence in the galaxy bias, to the extent that each redshift bin is assigned a separate bias parameter ($b_{1}$, $b_{2}$, $b_{3}$ and $b_{4}$) in the cosmological run. Potentially there could also be added complexity in the bias (E.g. \citealt{Swanson08} and \citealt{Cresswell09}) but this is beyond the scope of this current work. 

The marginalised best fit parameters are again listed in Table~\ref{table:binconstraints}, with the corresponding contours displayed in Figure~\ref{fig:combinedredshiftbins}. As found with the individual bins the contours can be seen to visibly rise along the bias axis with an increase in redshift. Moreover, the four bias quantities are seen to be high implying that the LRGs strongly trace the underlying mass distribution. In particular we find: $b_{1} = 1.47 \pm 0.15$, $b_{2} = 1.71 \pm 0.17$, $b_{3} = 1.80 \pm 0.18$ and $b_{4} = 2.05 \pm 0.21$.

Finally, in the top panel we include a calculation of the combined bins with (red/left contour) and without (all blue contours) the lowest multipole band measured in redshift bin 4. As with the individual bin the excess power is seen to systematically displace the marginalised distribution and once again is removed from all other constraints.

\subsection{Measuring Redshift Distortions} \label{sec:redshiftspacedistortionlimits}

\begin{figure*}
  \begin{flushleft}
    \centering
    \begin{tabular}{ll}
      \includegraphics[width=2.28in,height=2.3in]{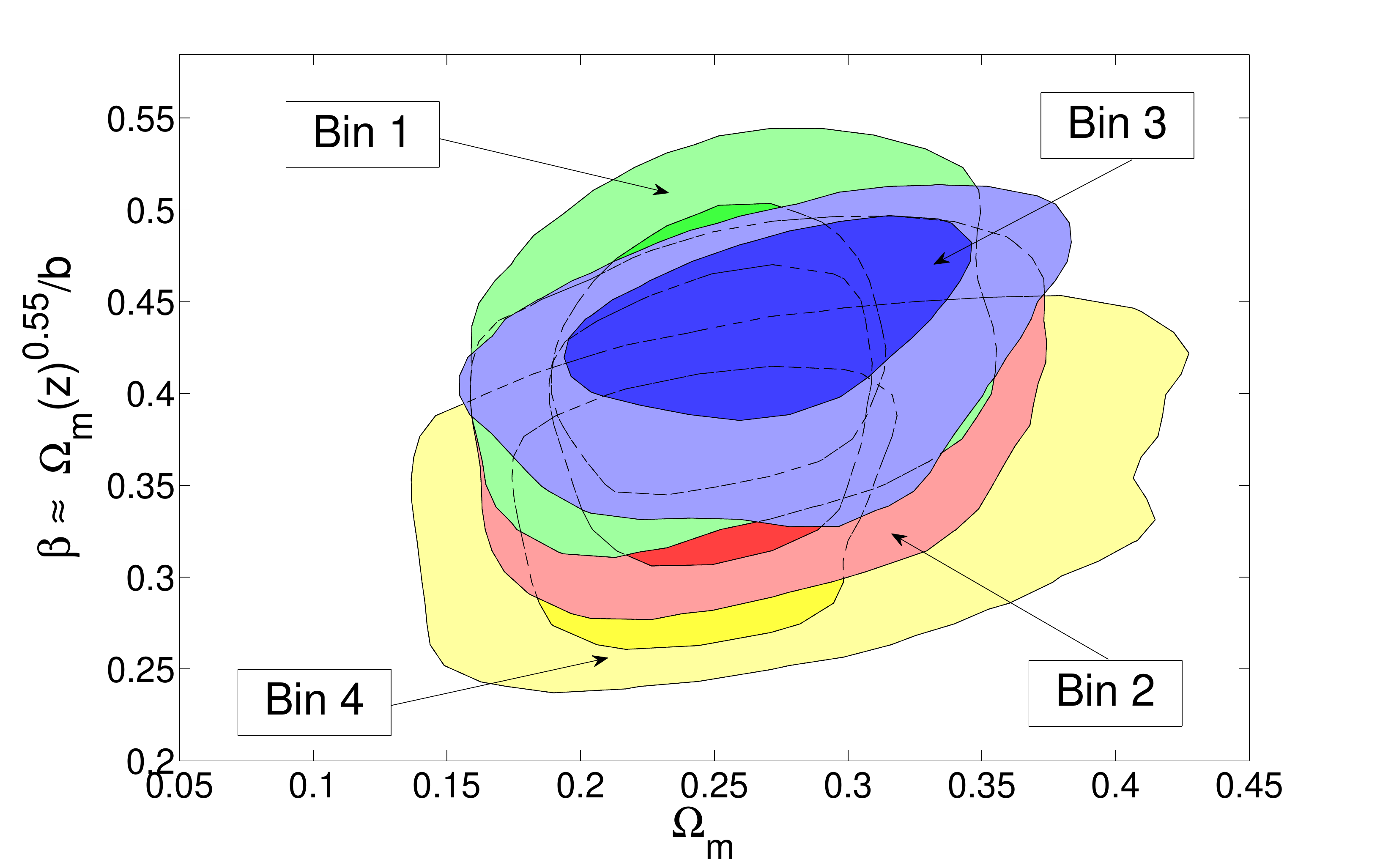} &
      \includegraphics[width=2.28in,height=2.3in]{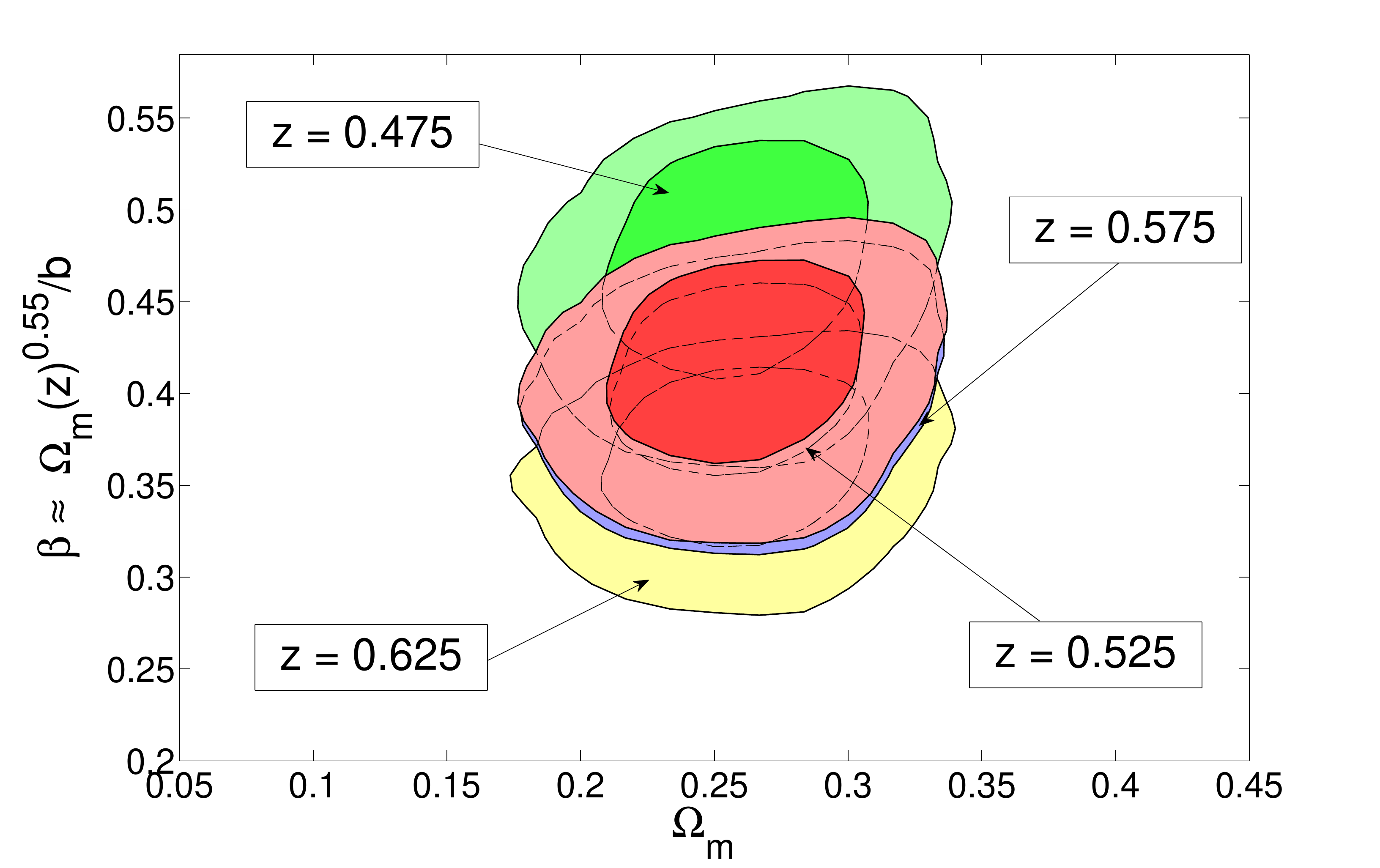}
      \includegraphics[width=2.28in,height=2.3in]{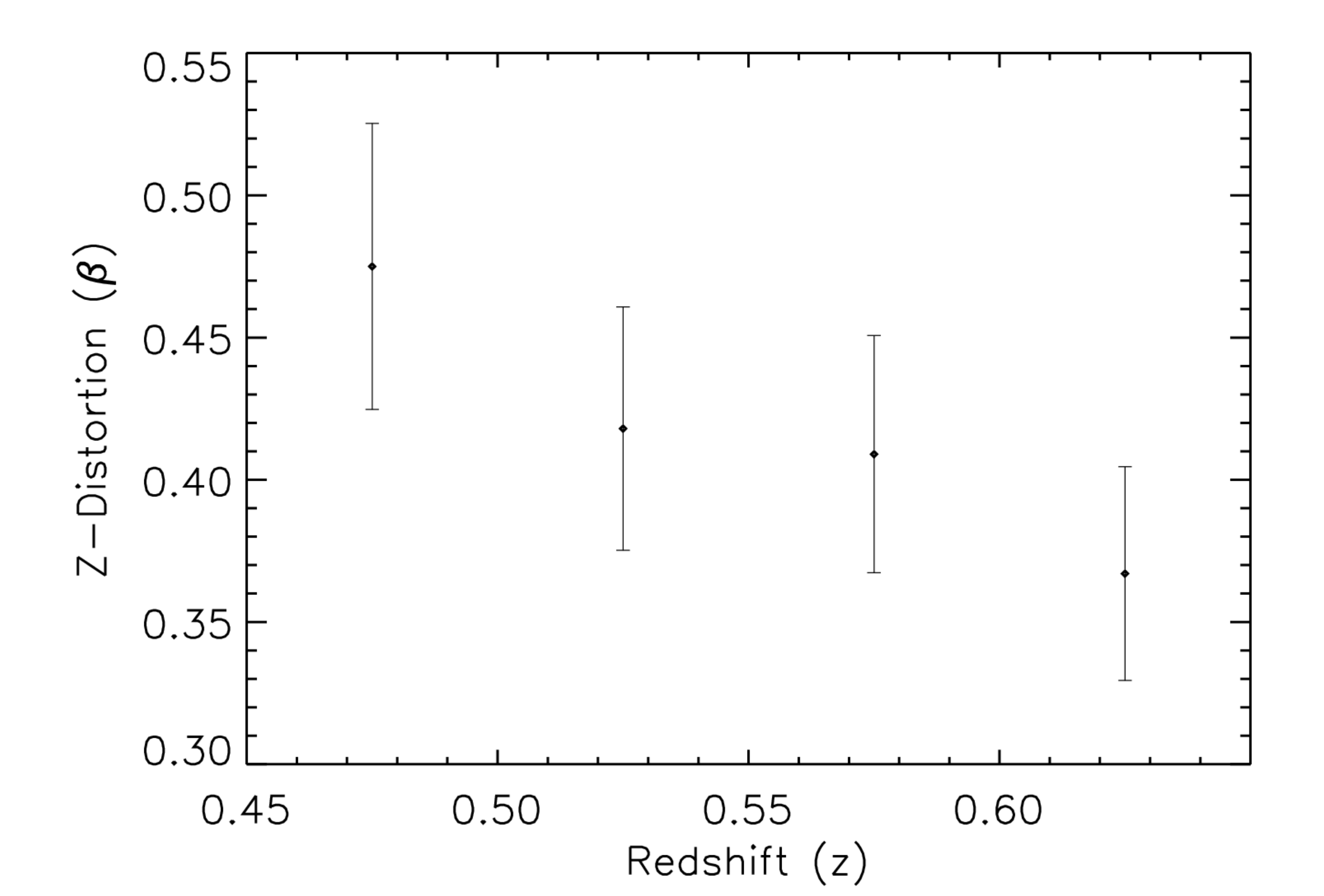}
    \end{tabular}
    \caption{\small{Limits on the redshift distortion parameter $\beta \approx \Omega_{m}(z)^{0.55}/b$ and $\Omega_{m}$ for an analysis of each {\it separate} redshift bin (left panel); for $\beta_{1}$, $\beta_{2}$, $\beta_{3}$, $\beta_{4}$ and $\Omega_{m}$ for the four redshift slices {\it combined} (middle panel); and the corresponding marginalised values and error derived from this {\it combined} bins analysis (right panel). Again, $f_{b}$ and $\sigma_{8}$ have been marginalised over; there is a $0.7 \le \sigma_{8} \le 1.1$ prior on $\sigma_{8}$, and $H_{0}$ and $n_{s}$ are fixed to $75$ km s$^{-1}$ Mpc$^{-1}$ and 1, respectively.  }}
    \label{fig:redshiftdistortions}
  \end{flushleft}
\end{figure*}

We noted previously that the peculiar velocity of a galaxy will cause it to appear shifted along the line-of-sight. The modification to the window function was described and detailed in Section~\ref{sec:reddist}. For angular power spectra $C_{\ell}$ this can cause a significant effect over large scales as seen in Figure~\ref{fig:theoretical_cl_profile}. Using this sensitivity we recast the bias parameter(s) in our MCMC chains into the redshift distortion parameter $\beta(z) \approx \Omega_{m}(z)^{0.55}/b$, where we take $z$ to be the mid-point of the bin in question and also $b$ is the bias of that bin. For the combined bin analysis this renders four distortion parameters $\beta_{1} = \beta(0.475)$, $\beta_{2} = \beta(0.525)$, $\beta_{3} = \beta(0.575)$ and $\beta_{4} = \beta(0.625)$. The resulting limits from each separate bin and from the combined analysis are listed in Table~\ref{table:distortionmeasure} and are illustrated in Figure~\ref{fig:redshiftdistortions}. It is interesting to note that one could use a similar methodology to test for deviations to gravity or dark energy clustering by allowing the exponent in the distortion parameter to vary as a free parameter, e.g. $\beta \approx \Omega_{m}(z)^{\gamma}/b$ (\citealt{Linder2007}, \citealt{Guzzo08} and \citealt{Thomas08}). Further, with this data set and method one could also fit $\beta = (1/b) \ud \, \mathrm{ln} \, D /\ud \, \mathrm{ln} \, a$ and/or $\gamma$ independently.

\begin{table}
\centering
\begin{tabular}{ccc} 

\hline
Separate Bins & Combined Bins & Redshift \\
\hline
\hline
$0.423 \pm 0.058$ & $0.475 \pm 0.050$ &  $\beta(0.475)$ \\
$0.392 \pm 0.057$ & $0.418 \pm 0.043$ &  $\beta(0.525)$ \\
$0.440 \pm 0.045$ & $0.409 \pm 0.042$ &  $\beta(0.575)$ \\
$0.345 \pm 0.055$ & $0.367 \pm 0.038$ &  $\beta(0.625)$ \\
\hline
\hline
\\
\end{tabular}
\caption{The marginalised mean redshift distortion parameters measured from the analyses of the galaxy clustering angular power spectra $C_{\ell}$. $f_{b} = \Omega_{b}/\Omega_{m}$, $\Omega_{m}$, $\sigma_{8}$ have been marginalised over with a prior of $0.7 \le \sigma_{8} \le 1.1$ on $\sigma_{8}$. $H_{0}$ and $n_{s}$ are fixed to $75$ km s$^{-1}$ Mpc$^{-1}$ and 1, respectively.}
\label{table:distortionmeasure}
\end{table} 

\noindent

\subsection{Other Studies} \label{sec:otherstudies}

In as much as other analyses can be compared, with varying parameter choices and assumptions, these results are concordant but competitive with respect to recent studies of SDSS galaxy clustering. These often include alternate or earlier data sets and at different redshifts. This includes \citealt{Padmanabhan07}, an analogous \emph{photometric} study to \citealt{Blake07}, that instead reconstructs the 3D real space power spectrum. Apart from these two works, studies have tended to focus on the spectroscopic samples, such as \citealt{Tegmark04}, \citealt{Tegmark06}, \citealt{Cabre09} and \citealt{Sanchez09} and most recently \citealt{Reid09} (DR7). 

Furthermore, the SDSS galaxies have permitted measurements of the Baryon Acoustic Oscillations with \citealt{Percival07}, \citealt{Gaztanaga2008b} and \citealt{Percival09}.

\section{Systematics and Further Tests} \label{sec:systematics1}

The earlier MegaZ release \citep{Blake07} performed a series of systematic tests based naturally on examining variations across the sky. This included astronomical seeing, overlapping survey stripes, regions of low Galactic latitude, varying completeness and variations in star-galaxy separation. All potential effects were found to have little or no influence on the estimated power spectra. The aforementioned paper also highlighted the impact of photometric errors for LRGs given their location on the galaxy luminosity function. This function $\phi(M)$ describes the \emph{number} of galaxies that have absolute magnitudes $M$ within an interval $M+\ud M$. The position of the galaxy sample under consideration is one where the gradient of this function is high. Therefore, any slight systematic shift in $M$ will impart a large systematic shift in the number of galaxies. If this systematic shift were some function of sky position, for example, it could contribute significantly (and artificially) to the galaxy clustering signal at that scale. With these considerations in mind we therefore choose to examine the redshift distribution, the role of extinction and the process of photometric redshift estimation.

\subsection{Redshift Bin Cross Correlations} \label{sec:crosspowerspectrameasured}

A useful test of any known or unknown systematic present in the study is the cross angular power spectra (Equation~\ref{eq:crosspowerspectrumdata} and Equation~\ref{eq:crosspowerspectrummodel}). A signal in these quantities should be the result of photometric errors scattering galaxies between bins as \emph{predicted} by the spectroscopic redshift distribution defined earlier and the best fit auto-power spectra. Any significant alteration in the measurement relative to the theoretical $C_{\ell}^{i,j}$ could indicate an additional systematic in the photometry, extinction correction or an ill-calibrated redshift distribution, for example. 

\begin{figure*}
  \begin{flushleft}
    \centering
    \begin{minipage}[c]{1.00\textwidth}
      \centering
      \includegraphics[width=2.87in,height=2.87in]{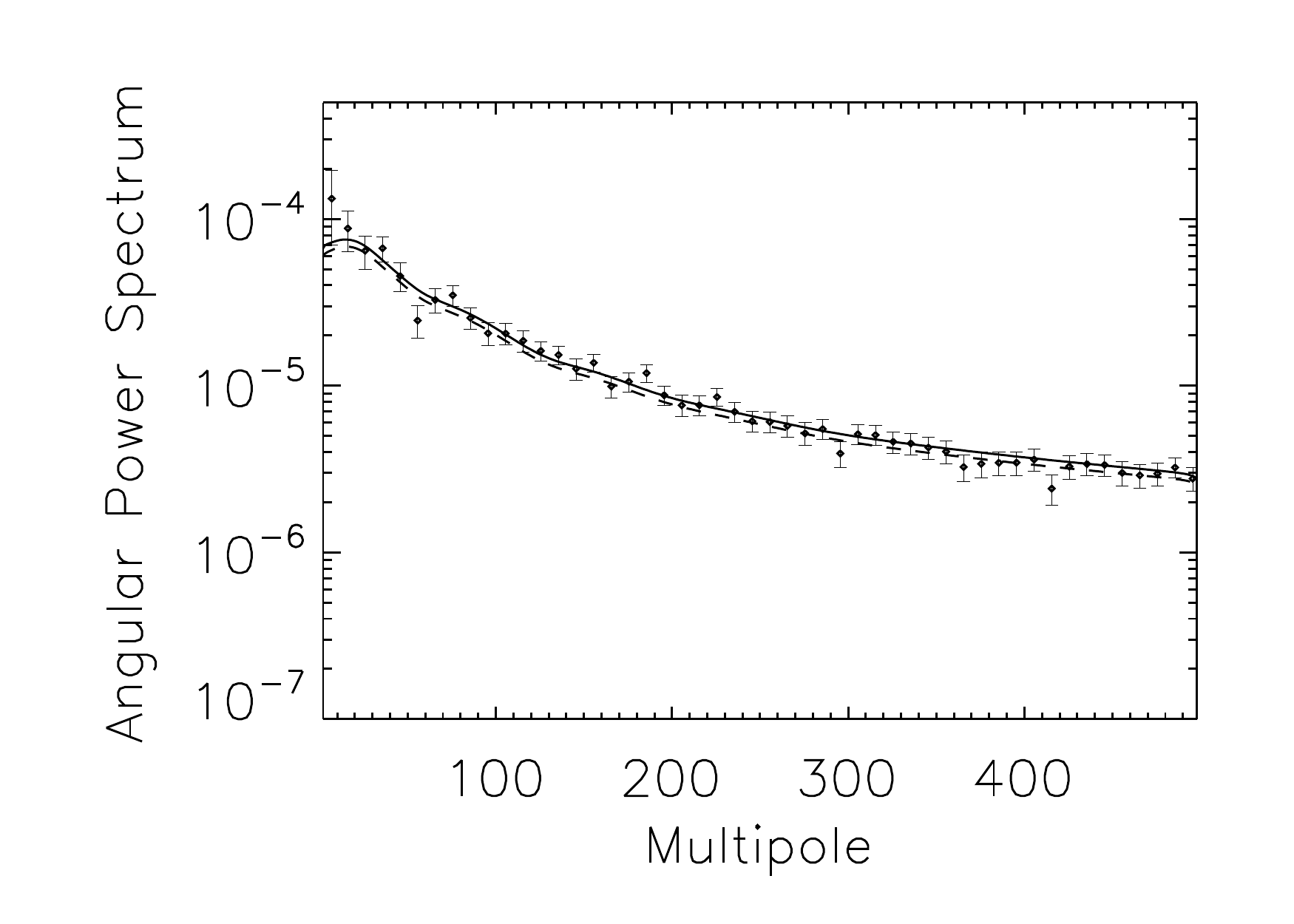} 
      \includegraphics[width=2.87in,height=2.87in]{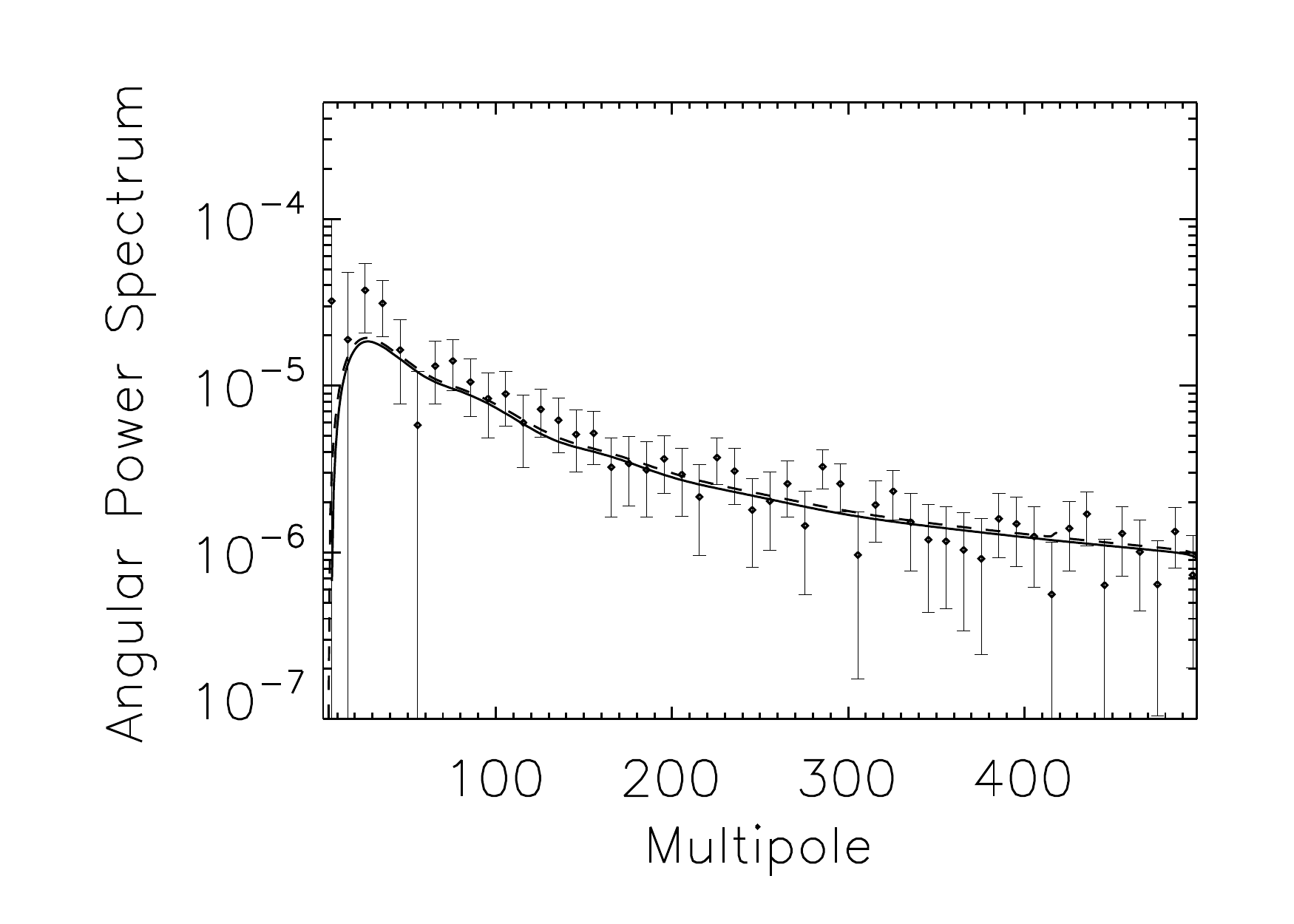} 
    \end{minipage}
    \begin{minipage}[c]{1.00\textwidth}
      \centering
      \includegraphics[width=2.87in,height=2.87in]{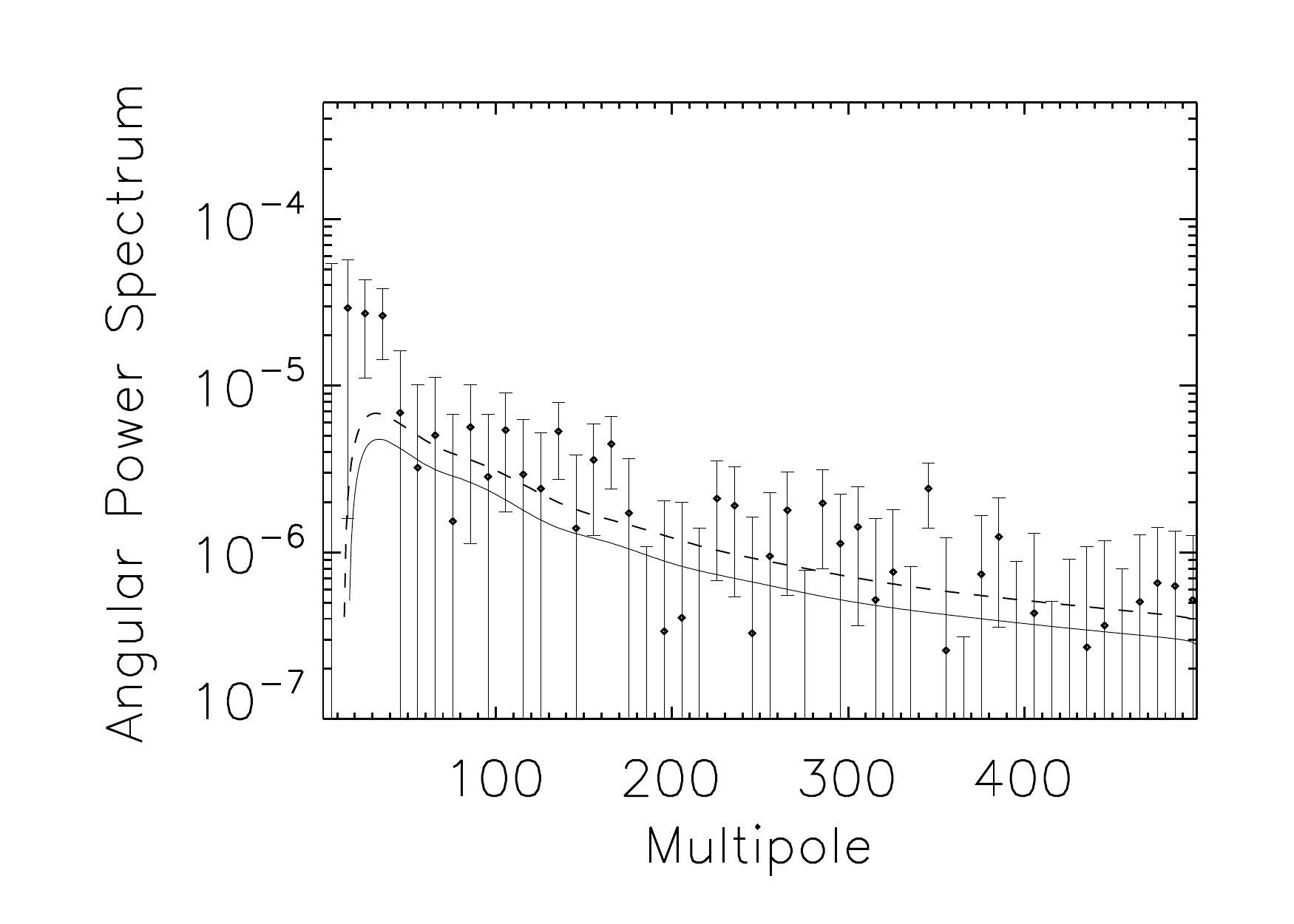} 
      \includegraphics[width=2.87in,height=2.87in]{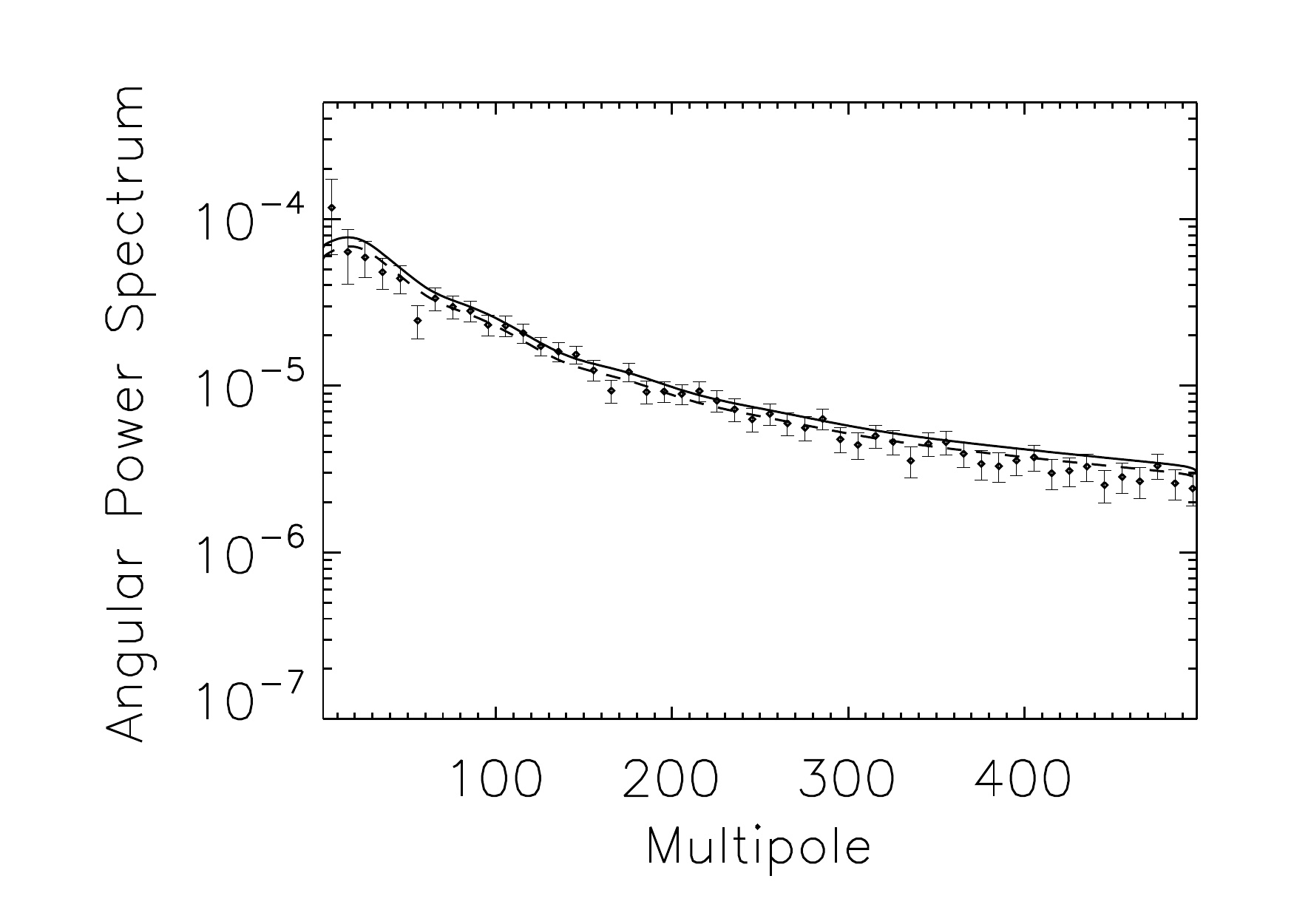} 
    \end{minipage}
    \centering
    \begin{minipage}[c]{1.00\textwidth}
      \centering
      \includegraphics[width=2.87in,height=2.87in]{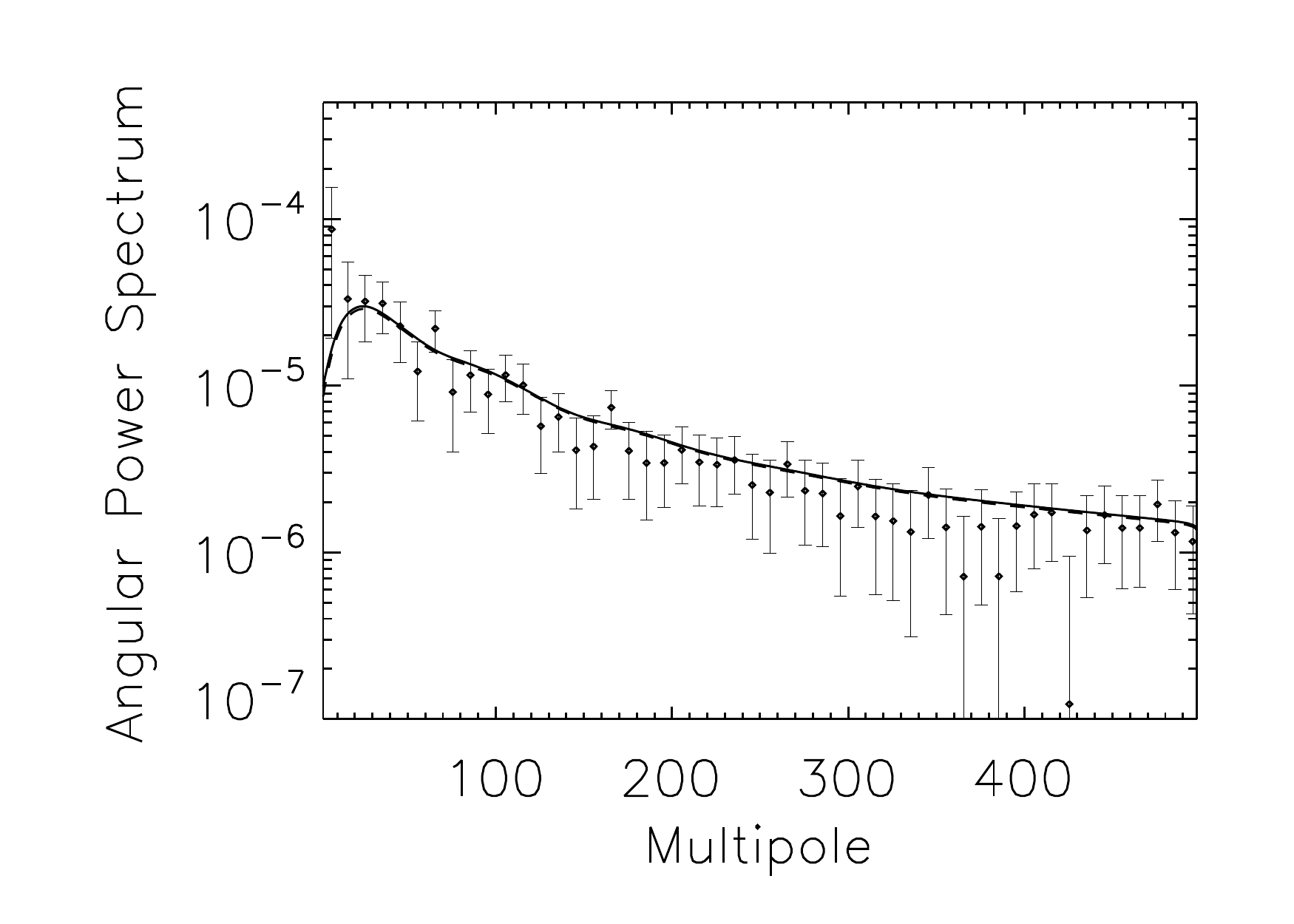} 
      \includegraphics[width=2.87in,height=2.87in]{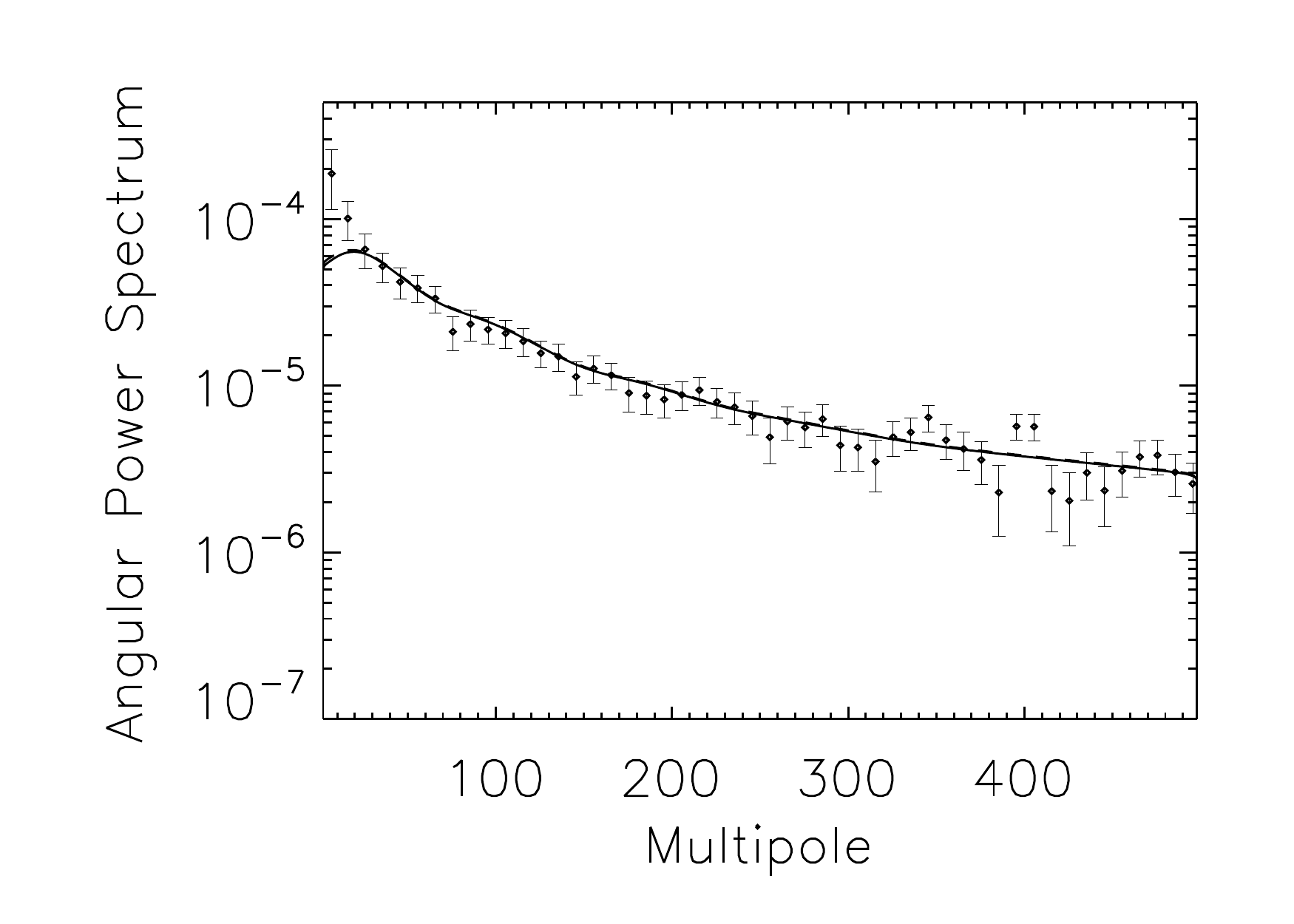} 
    \end{minipage}
    \caption{\small{The measured cross Angular Power Spectra ($C_{\ell}^{i,j}$) for the photometric SDSS MegaZ-LRG (DR7) population evaluated using Equation~\ref{eq:crosspowerspectrumdata}. The error bars correspond to those calculated with Equation~\ref{eq:crossgaussianerror} using the measured power spectrum. The solid lines are evaluated for the the best fit parameters found in Section~\ref{sec:combinedbinsparameters} using the the Gaussian redshift distributions. The dashed lines are the theoretical power spectra using a spline interpolation of the spectroscopic distribution. The panels are: Bin 1,2 (top left), Bin 1,3 (top right), Bin 1,4 (middle left), Bin 2,3 (middle right), Bin 2,4 (bottom left) and Bin 3,4 (bottom right). }}
    \label{fig:estimatedcrosspowerspectra}
  \end{flushleft}
\end{figure*}

We measure the cross power spectra in each of the six cross-bin combinations (note that $C_{\ell}^{i,j} = C_{\ell}^{j,i}$) in multipole bands of $\Delta \ell = 10$ up until $l_{\mathrm{max}} = 500$ as performed previously for the auto power spectra. The observed values are included online and are plotted in Figure~\ref{fig:estimatedcrosspowerspectra} along with their associated error bars. The solid lines in these plots show the predicted theoretical spectra using the best fit values from the combined bins analysis and the corresponding Gaussian redshift distributions. For nearby bins there is excellent agreement in the values. However, the anticipated cross spectrum between bin 1 and bin 4 (middle left panel) suffers from a lack of amplitude and consequently does not fit the mean profile of the data well. This is most likely the result of the Gaussian redshift distributions being weak fits to the spectroscopic profiles far from the mean of the distribution. As can be seen in Figure~\ref{fig:gaussianredshiftdistribution} the Gaussian underestimates the number of galaxies far from the bin centre. This will lead to an under prediction in the cross term. Less dramatic is the apparent marginal overestimation of the cross spectrum between bins 2 and 3 (middle right panel). This might be the result of the Gaussian smoothing adding slightly more galaxies in the overlap region between the redshift slices.

To test this hypothesis we interpolated the spectroscopic distribution with a spline through the $n(z)$ histogram. Then using this more `realistic' profile we re-evaluated the theoretical cross power spectra in the bin. These are shown as the dashed lines in the cross spectrum panels. For the most physically separated bins (1 and 4) this is seen to give, as predicted, a boost in amplitude and a better fit to the data points. This could hint that the use of a fixed $\mu$ and $\sigma$ in the Gaussian is not completely optimal. Potentially, one could use a splined distribution in the cosmological parameter estimation, however it is important to note that this could introduce errors of its own. For example, it might propagate inherent fluctuations in the profile that are particular to that bin and patch of the sky into the analysis. In any case the main effect seen from these figures will be to slightly alter the furthest (and least contributing) corners of the covariance matrix, whereas the majority are good fits. Finally, with the redshift function now more fully tested in the cross correlation measurement it is interesting to see that in several of the bins correlated with bin 4 there exists a slight excess of power. This could point towards a residual systematic in the catalogue.

\noindent
\subsection{Extinction} \label{sec:systematicextinction}

Light from galaxies is potentially absorbed, scattered or re-emitted by the dust and gas within our own Galaxy. This \emph{Galactic extinction} has the capacity to be one of the dominant systematics in a galaxy survey such as this. For example, extinction can preferentially absorb light at the blue end of a galaxy's spectral energy distribution thus making it appear redder and more LRG-like. Alternatively, it can have the effect of scattering faint galaxies from the sample. As the contribution from our own Galaxy changes as a function of position (Figure~\ref{fig:SDSSmask}) this is a cause for concern given that we are interested in inferring cosmological quantities through statistical variations across the sky. Worse still, it could act to further systematically bias our redshift estimates given that the ANNz derived galaxy catalogue is a spatial extrapolation of the 2SLAQ \emph{training set}, which confined to a stripe at $\delta \approx 0\,^{\circ}$, covers a limited region of Galactic extinction.

\begin{figure}
  \centering
      \includegraphics[width=3.25in,height=3.25in]{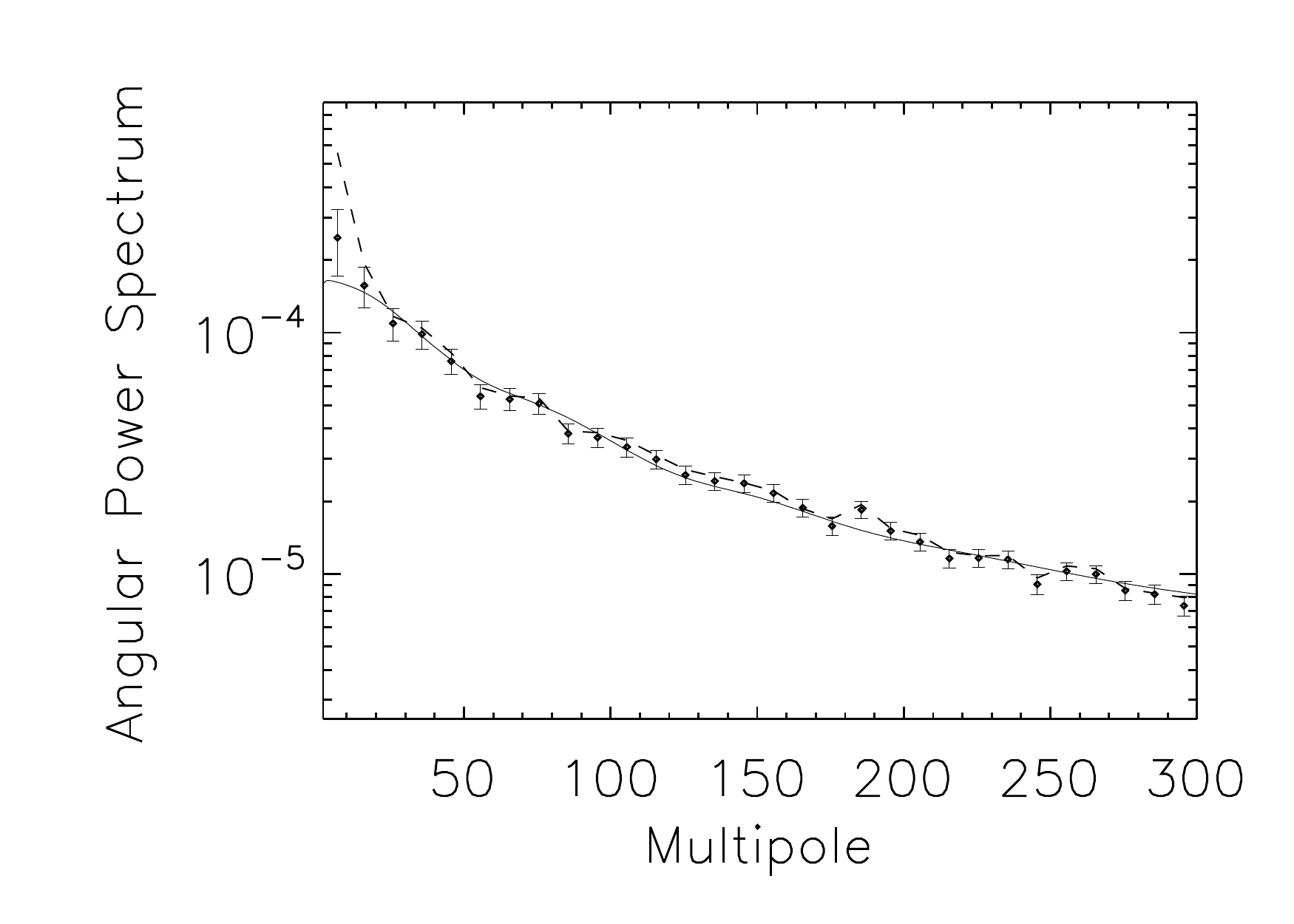}
    \caption{\small{The exaggerated effect caused by neglecting the extinction correction for the $i_{\mathrm{dev}}$ magnitude cut in bin 1 (dashed line). Although this is not used in the study it highlights the range of scales that \emph{could} be affected by any such systematic error in the correction. The extinction corrected spectrum is shown by the solid points with associated error bars. The solid line is a best fit profile for comparison. }}
    \label{fig:idevnotcorrected}
\end{figure}

Fortunately detailed maps of Galactic extinction are available (\citealt{Schlegel98}; Figure~\ref{fig:SDSSmask}) and subsequently the $u$, $g$, $r$, $i$ and $z$ bands used are dereddened model magnitudes, i.e. they are extinction corrected. Figure~\ref{fig:idevnotcorrected} shows the exaggerated effect that is the result of not adjusting properly for the presence of dust. In this plot the angular power spectrum is evaluated for the catalogue when the $i_{\mathrm{dev}}$ magnitude cut is not extinction corrected. This causes extra galaxies to be scattered from the sample in different regions of the survey area and a large boost of power is observed. Although the values used for our galaxy clustering statistics \emph{are} corrected for extinction it could be that there are errors in the correction map. If these errors were related to the magnitude of extinction or again varied with position, then they too would propagate into the LRG sample. 

\begin{figure*}
  \begin{flushleft}
    \centering
    \begin{minipage}[c]{1.00\textwidth}
      \centering
      \includegraphics[width=2.87in,height=2.87in]{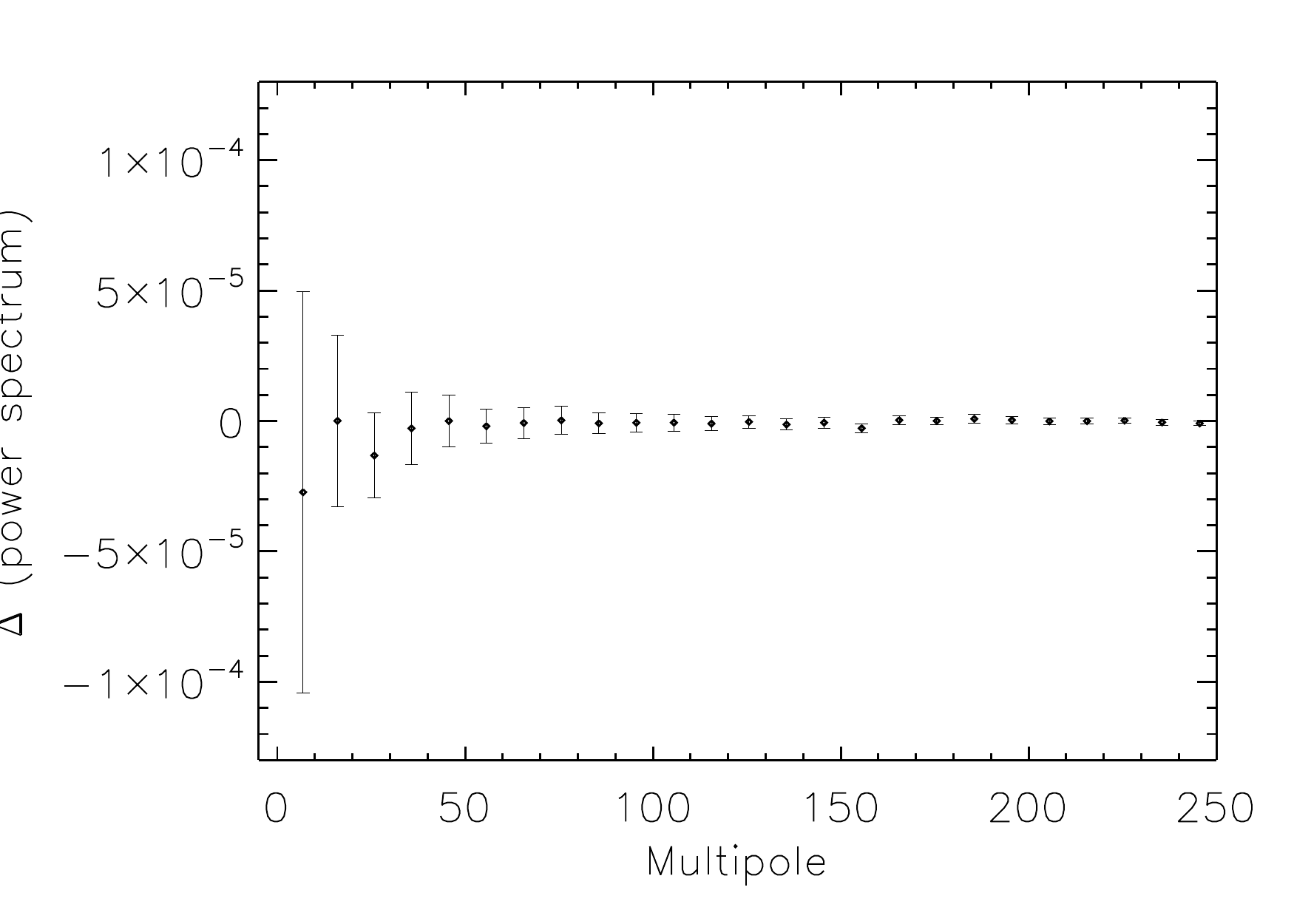} 
      \includegraphics[width=2.87in,height=2.87in]{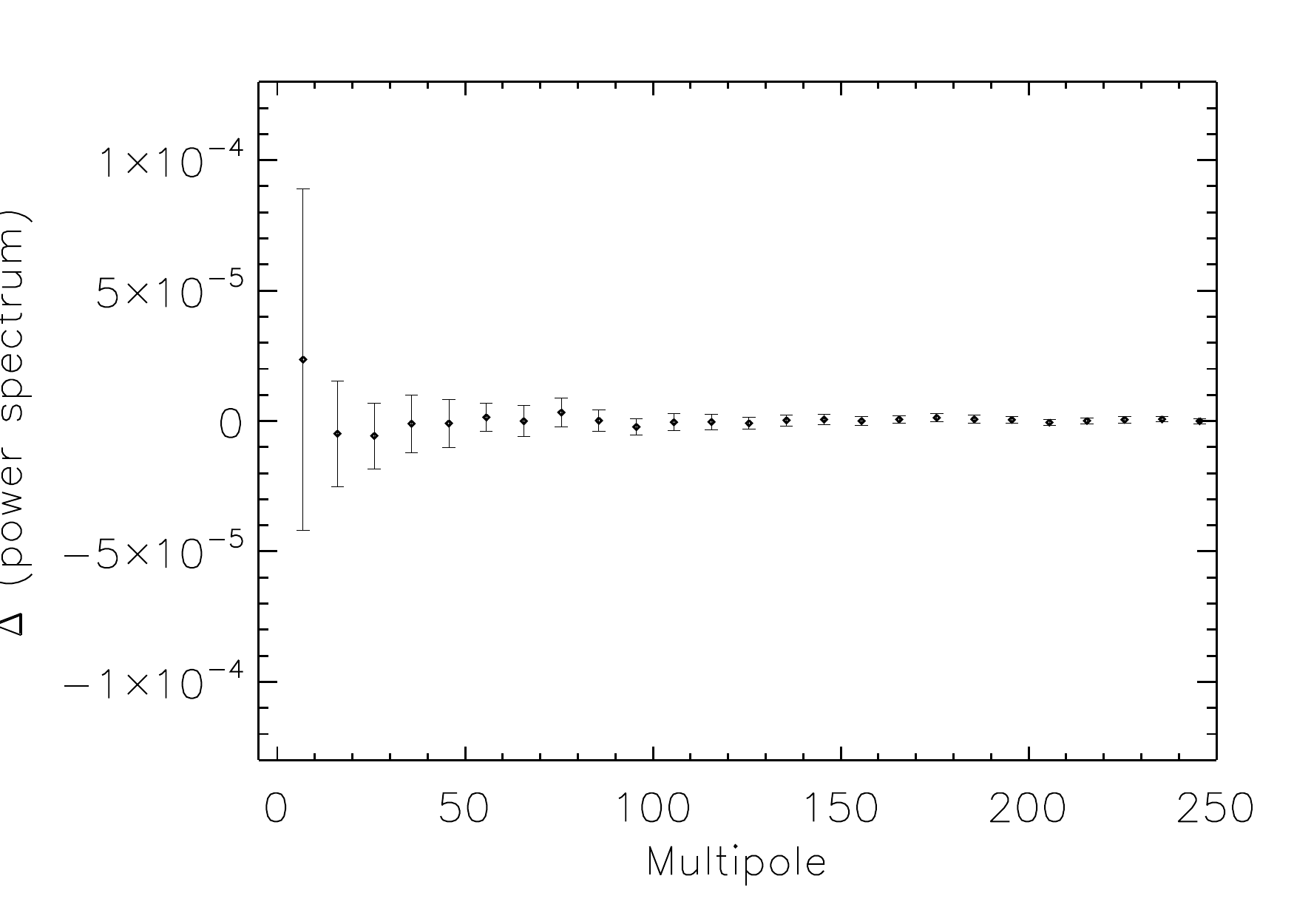} 
    \end{minipage}
    \begin{minipage}[c]{1.00\textwidth}
      \centering
      \includegraphics[width=2.87in,height=2.87in]{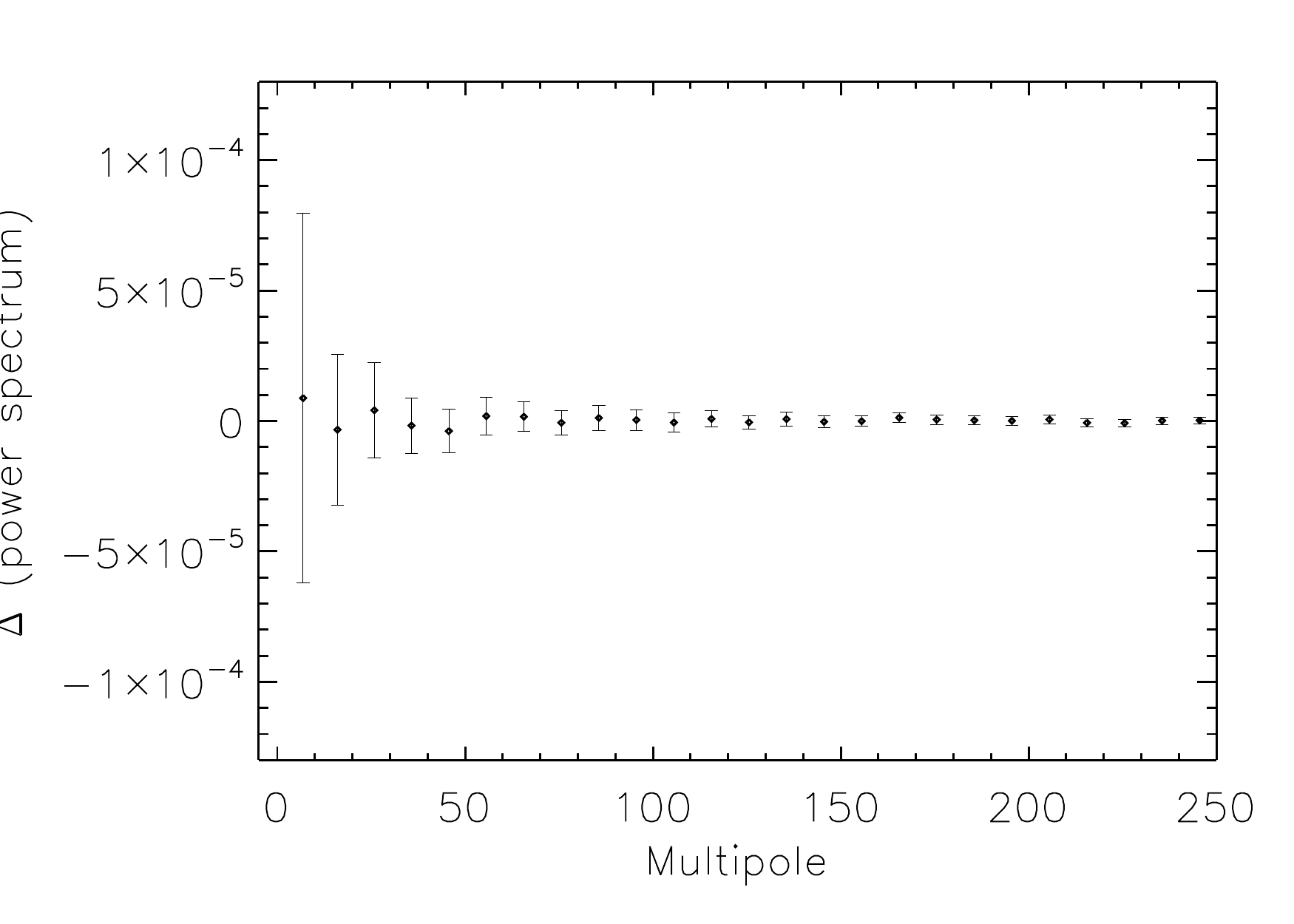} 
      \includegraphics[width=2.87in,height=2.87in]{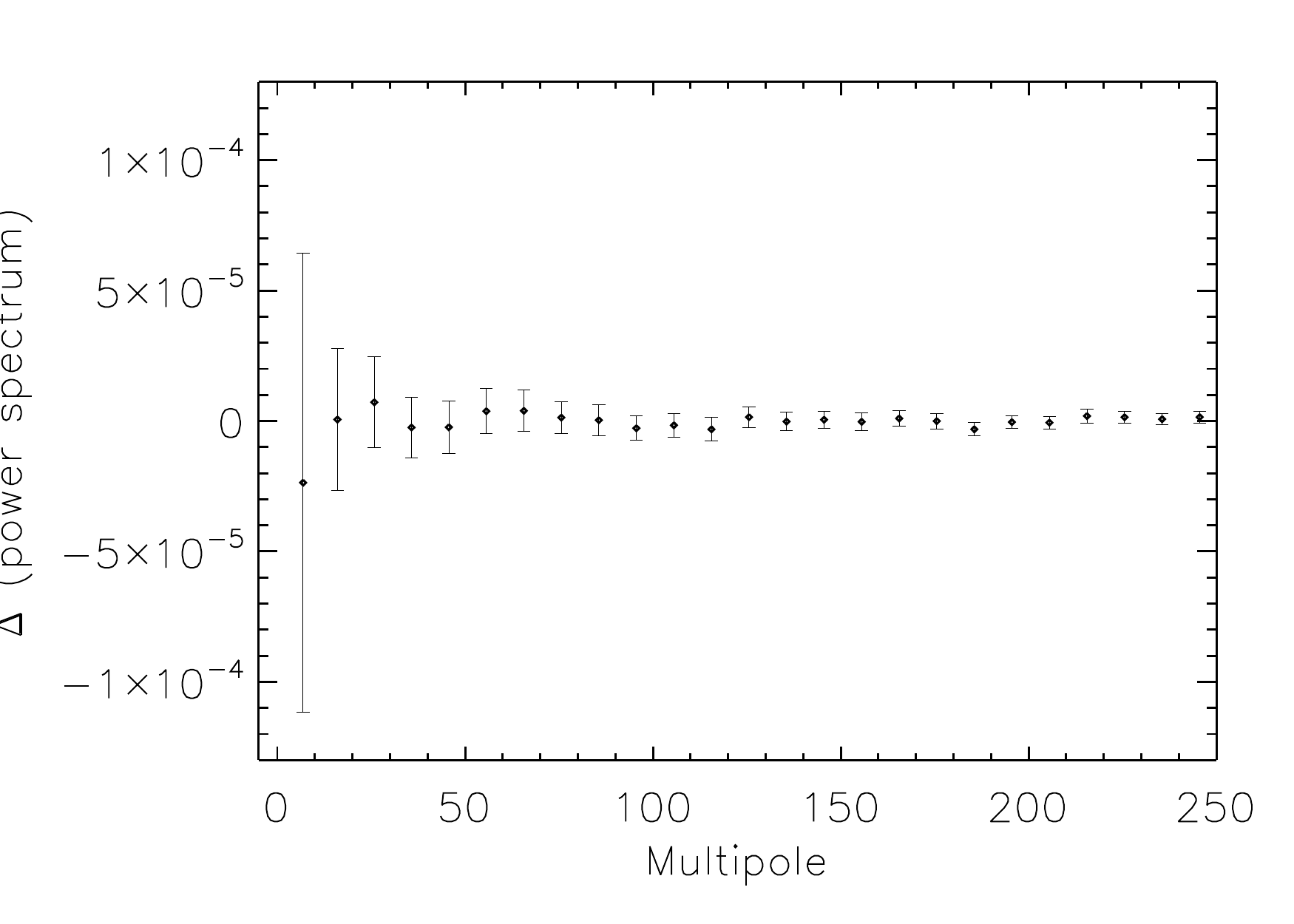} 
    \end{minipage}
    \caption{\small{The difference between the Angular Power Spectra with regions of high Galactic extinction removed ($> 0.1$ mag) and with the whole survey geometry included as before (i.e., $C_{\ell}^{\mathrm{partial}} - C_{\ell}^{\mathrm{all}}$). This is to test for possible extinction contamination in the analysis. The panels are: Bin 1 (top left), Bin 2 (top right), Bin 3 (bottom left) and Bin 4 (bottom right). There is no observable discrepancy between the two calculations.}}
    \label{fig:extinctioncorrecedcls}
  \end{flushleft}
\end{figure*}

To test for extinction correction errors we repeat the whole measurement of the angular power spectra with regions of high extinction removed ($> 0.1$ mag). This constitutes a $15 \%$ removal of the survey area. The changes in the power spectra are plotted in Figure~\ref{fig:extinctioncorrecedcls} with error bars as derived before but for the `cut' spectra. It is clear that the general profiles are not significantly affected. This result is consistent with the preliminary examination in \citealt{Blake07} and \citealt{Abdalla08}.

\subsection{Photometric Redshift Codes} \label{sec:systematiccodes}

\begin{figure*}
  \begin{flushleft}
    \centering
    \begin{tabular}{ll}
      \includegraphics[width=3.25in,height=3.25in]{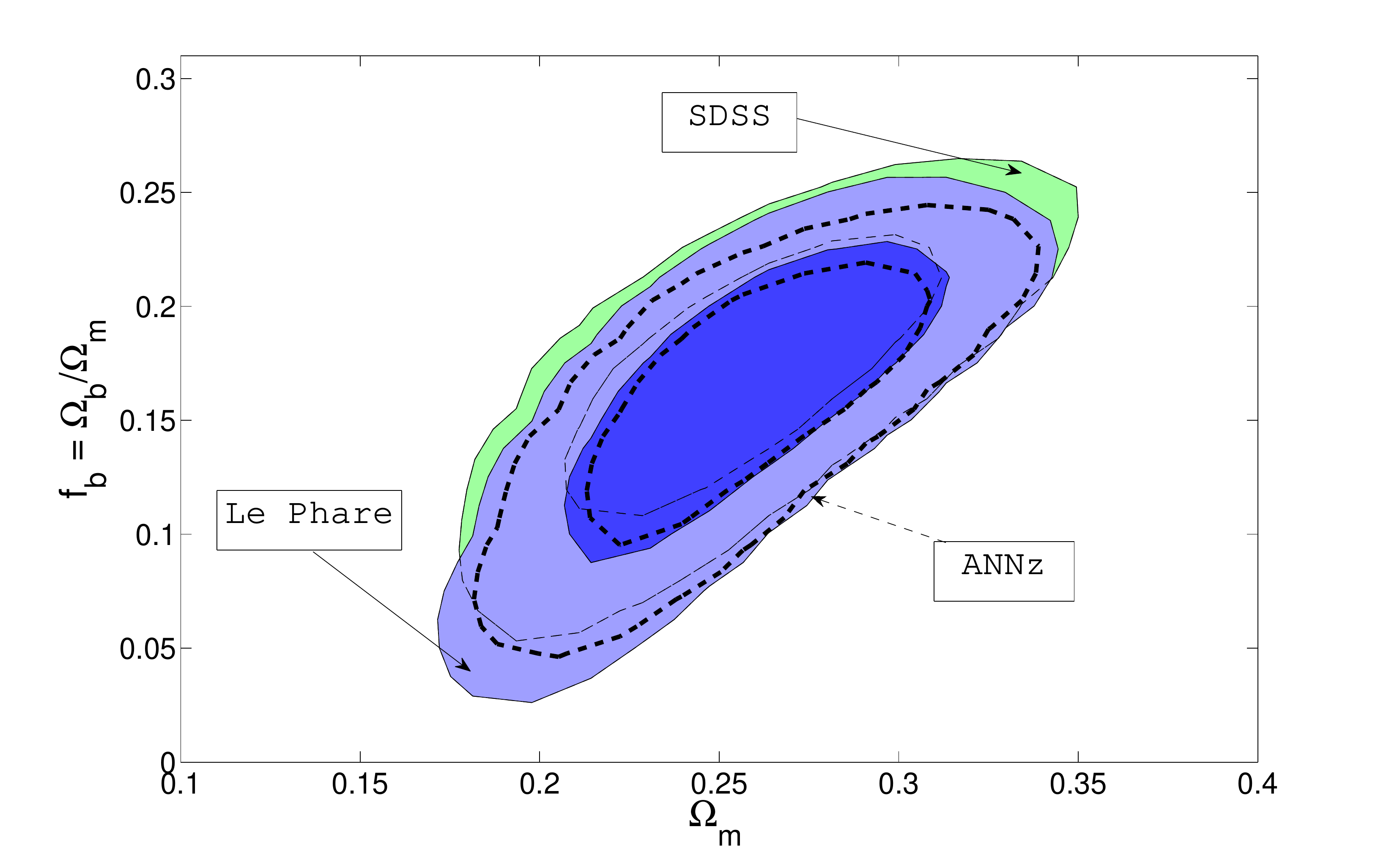} &
      \includegraphics[width=3.25in,height=3.25in]{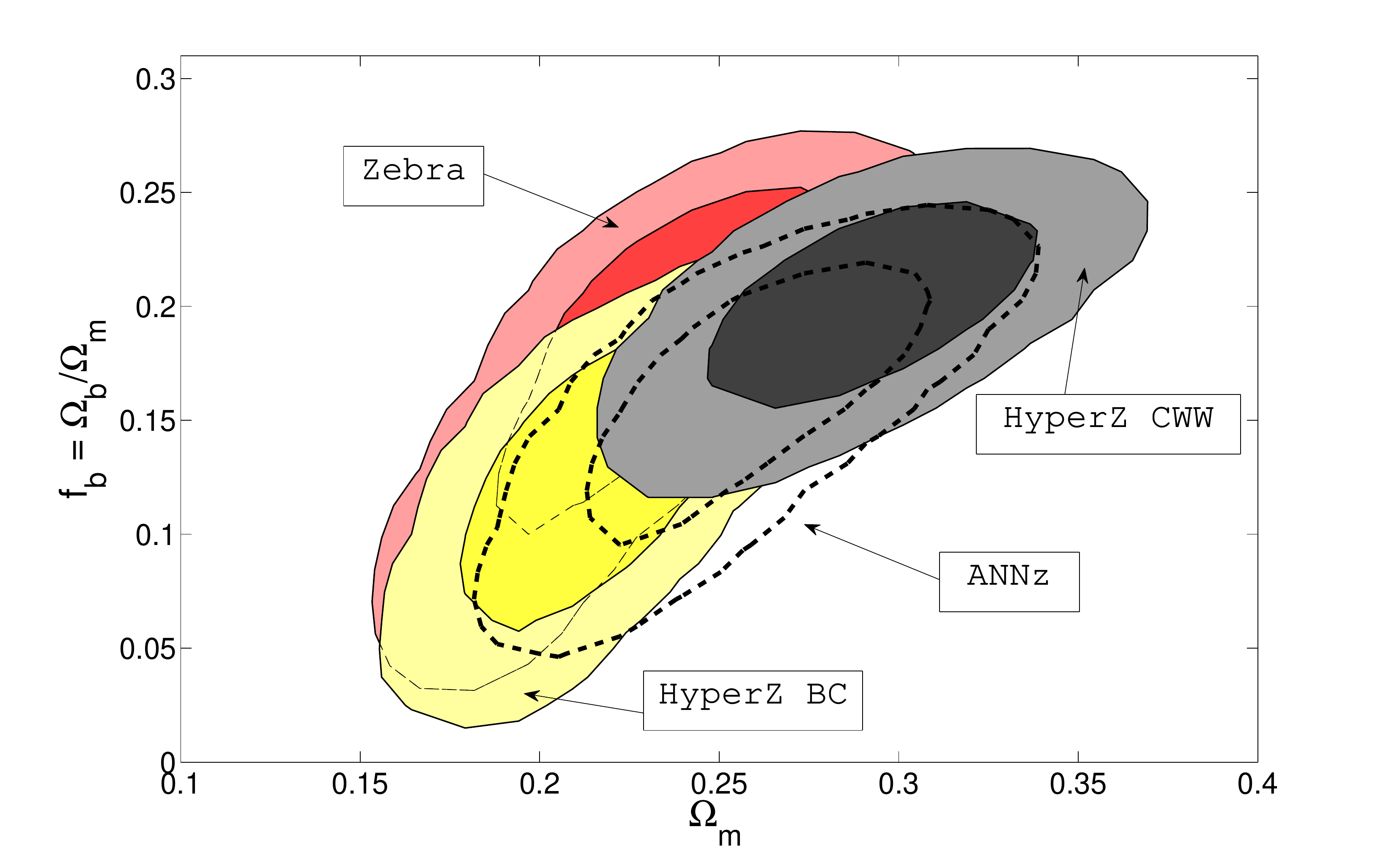}
    \end{tabular}
    \caption{\small{Cosmological constraints on the four combined bins using various photometric codes. This represents a comparison between the codes and templates with current data. {\it Left Panel:} The training set method with ANNz (thick dashed line) is extremely consistent with Le Phare (blue/darker) and SDSS (green/lighter) template procedures and so vindicates the spatial limitation of the 2SLAQ training set, including with extinction, for example. {\it Right Panel:} The only difference is between a subset of \emph{template} procedures highlighted by HyperZ BC (yellow/lightest), Zebra (red/darker) and HyperZ CWW (grey/darkest). However, all contours still overlap within $1\sigma$. Again, ANNz is illustrated by the dashed contour for reference.}}
    \label{fig:photometricestimationplot}
  \end{flushleft}
\end{figure*}
\noindent
We chose to derive the redshift estimates for our catalogue using ANNz given that it was shown in \citealt{Abdalla08} to have the best performance for this survey. This is partially due to the presence of the specific 2SLAQ training set. We noted that this training set was limited in sky position, and therefore extinction, and the application to the wider survey was an extrapolation. We test this limitation now by comparing the results from different redshift codes. Note that this also represents a cosmological comparison of the codes with current data.

\citealt{Abdalla08} evaluated the SDSS LRG DR6 catalogue with six photometric codes: ANNz, HyperZ \citep{Bolzonella:HyperZ}, SDSS \citep{Padmanabhan05}, Le PHARE \citep{Ilbert:LePhare06}, BPZ \citep{Benitez:BPZ} and ZEBRA \citep{Feldmann:ZEBRA}. These other codes do not require representative training set data but instead use a variety of methods including various LRG templates to obtain the redshift\footnote{Please see \citealt{Abdalla08} for an overview of these codes and templates or the specific references themselves for more detail.}. As the template based procedures do not utilise a spatially confined training set one can argue that they are effectively blind to this potential redshift calibration systematic. HyperZ was also analysed using two different sets of templates. This includes observed templates provided by \citealt{Coleman80} (CWW) and those synthetically produced by \citealt{Bruzual03} (BC).

We can use the catalogues presented in \citealt{Abdalla08} as a test of our work because we compare between codes and the DR6 survey area comprises the same contiguous region but with only a $1\%$ smaller survey area than in DR7 ($7670.9$ $\mathrm{deg^{2}}$). To make a fair like-for-like comparison we perform a star galaxy separation based on the ANNz output as the other codes do not have this option.

We plot the overall cosmological constraints for the four combined redshift bins from these different catalogues in Figure~\ref{fig:photometricestimationplot}. It is reassuring that there is excellent agreement between ANNz and several of the template based codes, including SDSS and Le PHARE. Moreover, it is interesting that where there is some difference it is between template based procedures (ZEBRA, HyperZ BC and HyperZ CWW) and not with the independent training set method. This clearly indicates consistency for our chosen procedure and furthermore shows no degrading effects from a limited training set. This is consistent with \citealt{Abdalla08} and all of the contours are consistent within $1\sigma$. However, for more statistically discriminating surveys in the future the template differences may become important. 

Note that we have accounted for the different redshift distributions ($\mu$, $\sigma$) that each code predicts given the varying spectroscopic-photometric relation. These distributions are given and are illustrated in Table~\ref{table:gaussianredshiftfitsDR6} and Figure~\ref{fig:DR6n_z_distributions} in the Appendix, respectively. 

\section{Conclusion} \label{sec:conclusion}

We have measured and constructed the galaxy catalogue and angular power spectra for $723,556$ Luminous Red Galaxies in the Sloan Digital Sky Survey Data Release 7 - called MegaZ DR7. This photometric extension to the previous release \citep{Blake07} represents the largest galaxy survey to date. Covering $7746$ $\mathrm{deg^{2}}$ over the redshift range $0.45 < z < 0.65$ we find constraints of $f_{b} \equiv \Omega_{b}/\Omega_{m} = 0.173 \pm 0.046$ and $\Omega_{m} = 0.260 \pm 0.035$. We also use our constraints to place limits on the redshift distortion parameter $\beta$ given that the $C_{\ell}$s are sensitive to distortions over low multipoles. We find $\beta_{(z = 0.475)} = 0.475 \pm 0.050$, $\beta_{(z = 0.525)} = 0.418 \pm 0.043$, $\beta_{(z = 0.575)} = 0.409 \pm 0.042$ and $\beta_{(z = 0.625)} = 0.367 \pm 0.038$.

It is reassuring that the galaxy clustering results on $f_{b}$ and $\Omega_{m}$ are consistent with the most recent WMAP analysis \citep{Komatsu10}. This is a crucial consistency check as the two surveys probe vastly contrasting cosmic epochs and are subject to different systematics. In addition, it seems the photometric approach to modern cosmological surveys is justifiable with competitive and concordant results compared to the spectroscopic SDSS survey \citep{Reid09}. This high redshift photometric survey could be combined \emph{simultaneously} with spectroscopic BAO measurements, such as \citealt{Percival07}, with no complex cross-covariance (i.e. they are independent). This can be highly complementary for the parameter space, particularly in neutrino mass measurements as shown in \citealt{Thomas09}. The tight constraints on the matter densities show there is now overwhelming and precision evidence for some dark energy-like component to the cosmos when including photometric data from the late-time Universe. A more complete and extended set of {\it cosmological implications} for this data set and in combination with other data sets are to be included in a companion paper, along with the likelihood.

We examined the possibility of residual systematics in the catalogue from photometric calibration, extinction, sky extrapolation and photometric estimation procedure. We found no significant alteration in the power spectra or combined constraints from these tests. In addition, the latter analysis included a comparison of different photometric codes and templates. We found that ANNz is highly consistent with other template based redshift estimation procedures, such as SDSS and Le PHARE. Small differences were seen between some of the template methods but all codes tested produced results that were overlapping within $1\sigma$. 

\section{Acknowledgements} 

It is a pleasure to thank Manda Banerji, Hume Feldman and the referee for useful discussions and suggestions. ST acknowledges a STFC studentship and UCL's Institute of Origins for a Post-doctoral Fellowship. FBA and OL acknowledge the support of the Royal Society via a Royal Society URF and a Royal Society Wolfson Research Merit Award, respectively. OL and FBA acknowledge the Weizmann Institute of Science for an Erna and Jakob visiting professorship and support for a research visit, respectively. We acknowledge use of the Healpix package \citep{Gorski05} and CAMB \citep{Lewis00}.



\appendix

\section{}

\begin{figure*}
  \begin{flushleft}
    \centering
    \begin{minipage}[c]{1.00\textwidth}
      \centering
      \includegraphics[width=4.36cm,height=4.36cm]{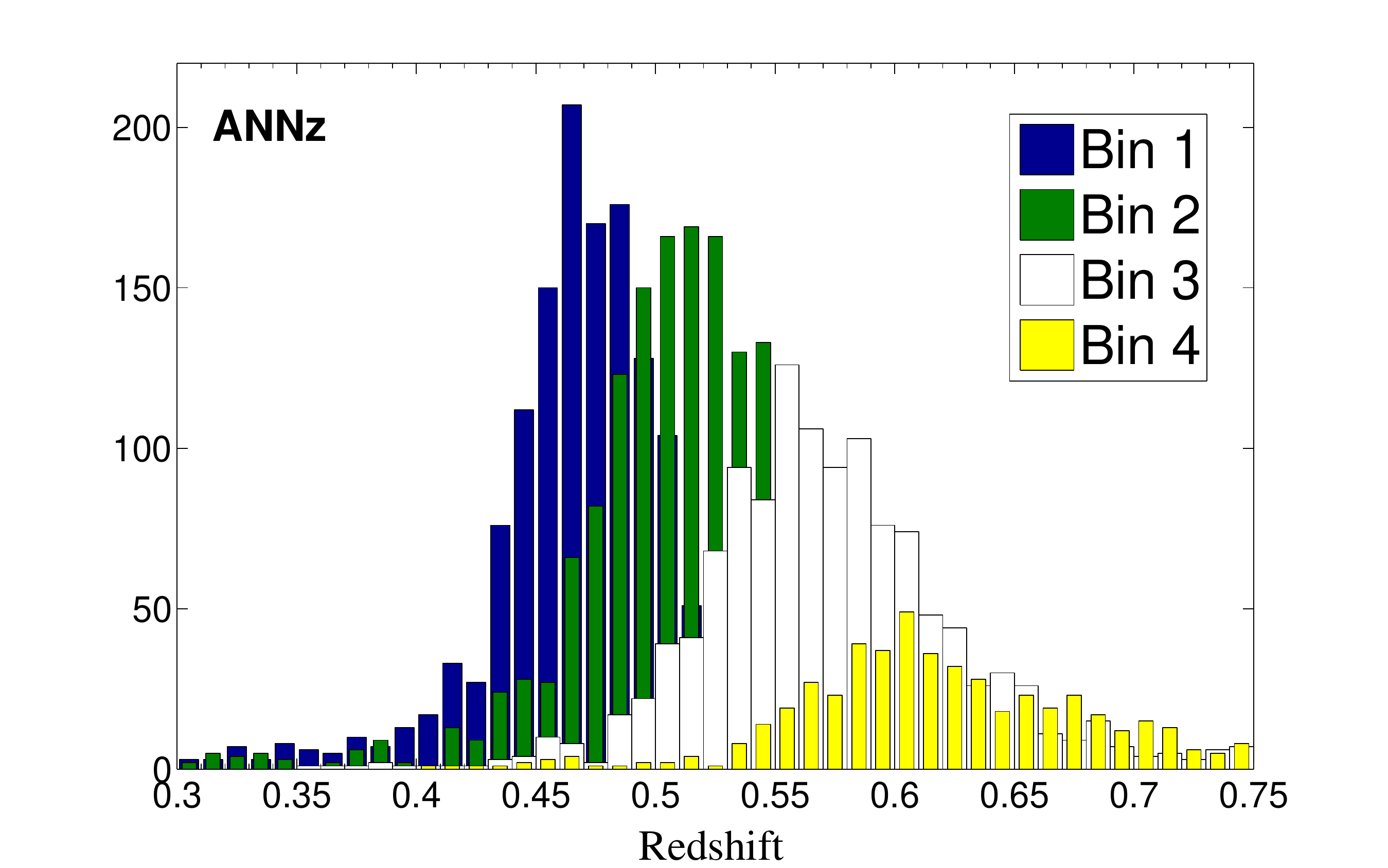}
      \includegraphics[width=4.36cm,height=4.36cm]{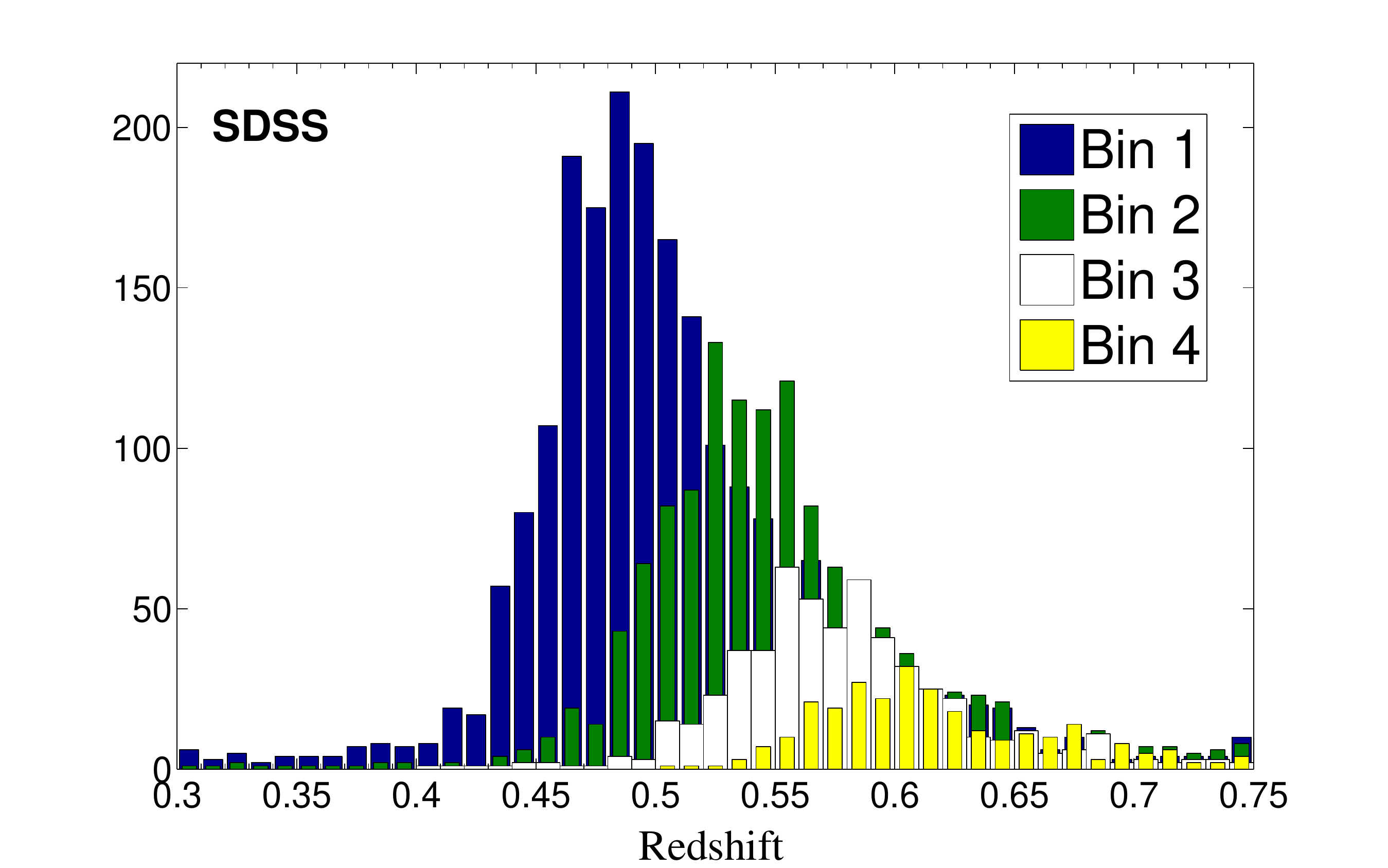}
      \includegraphics[width=4.36cm,height=4.36cm]{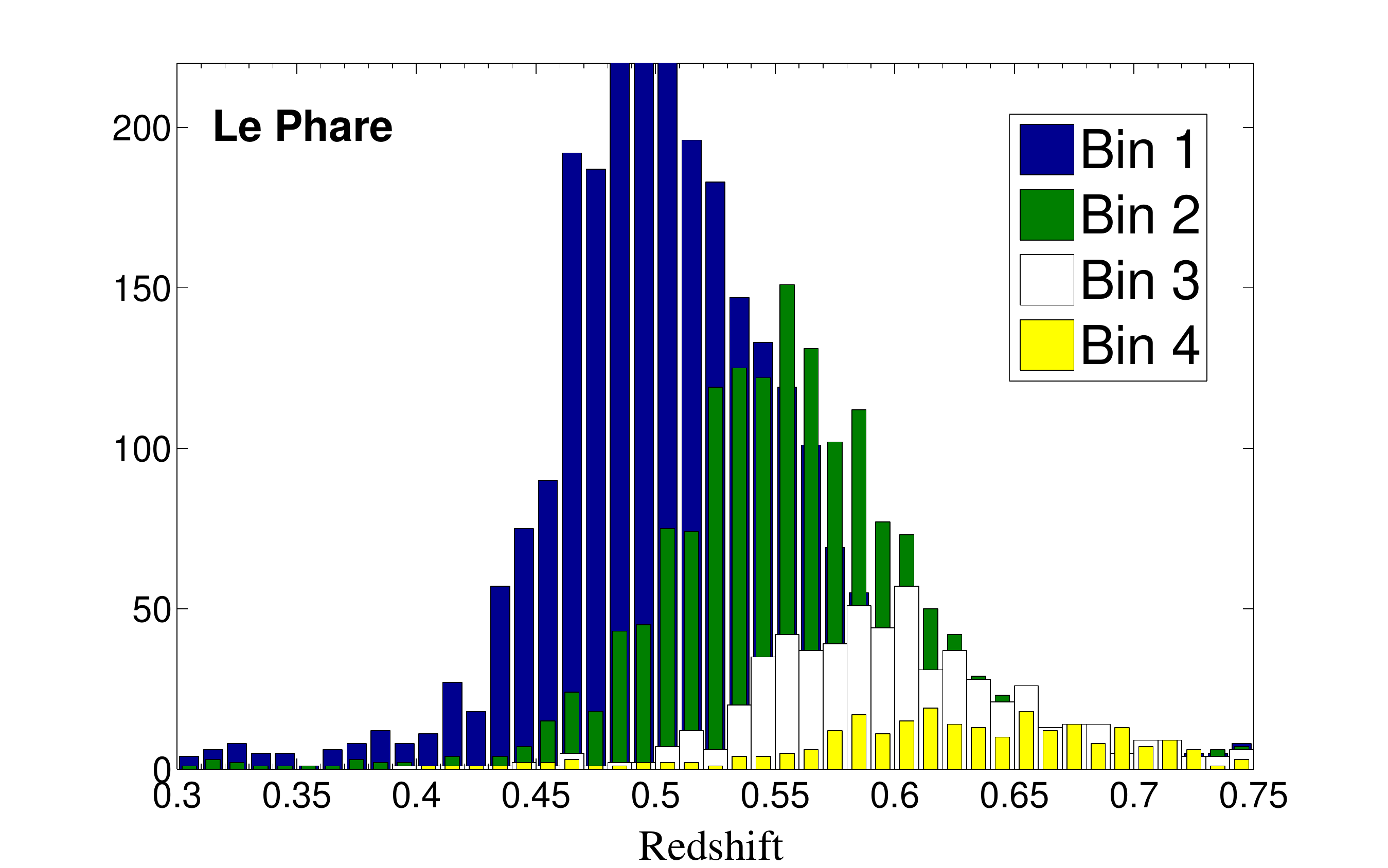}
      \includegraphics[width=4.36cm,height=4.36cm]{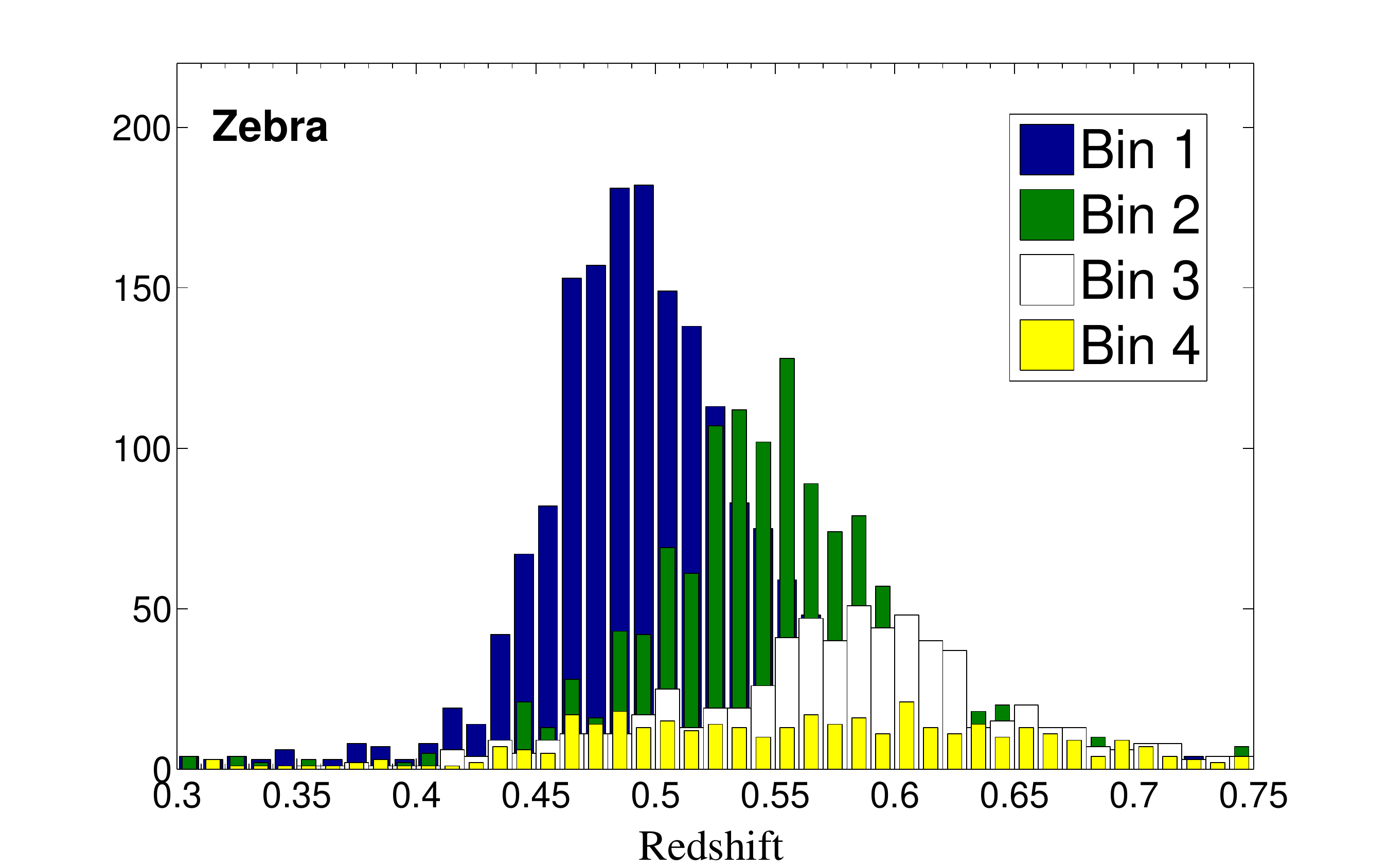}\\
      \includegraphics[width=4.36cm,height=4.36cm]{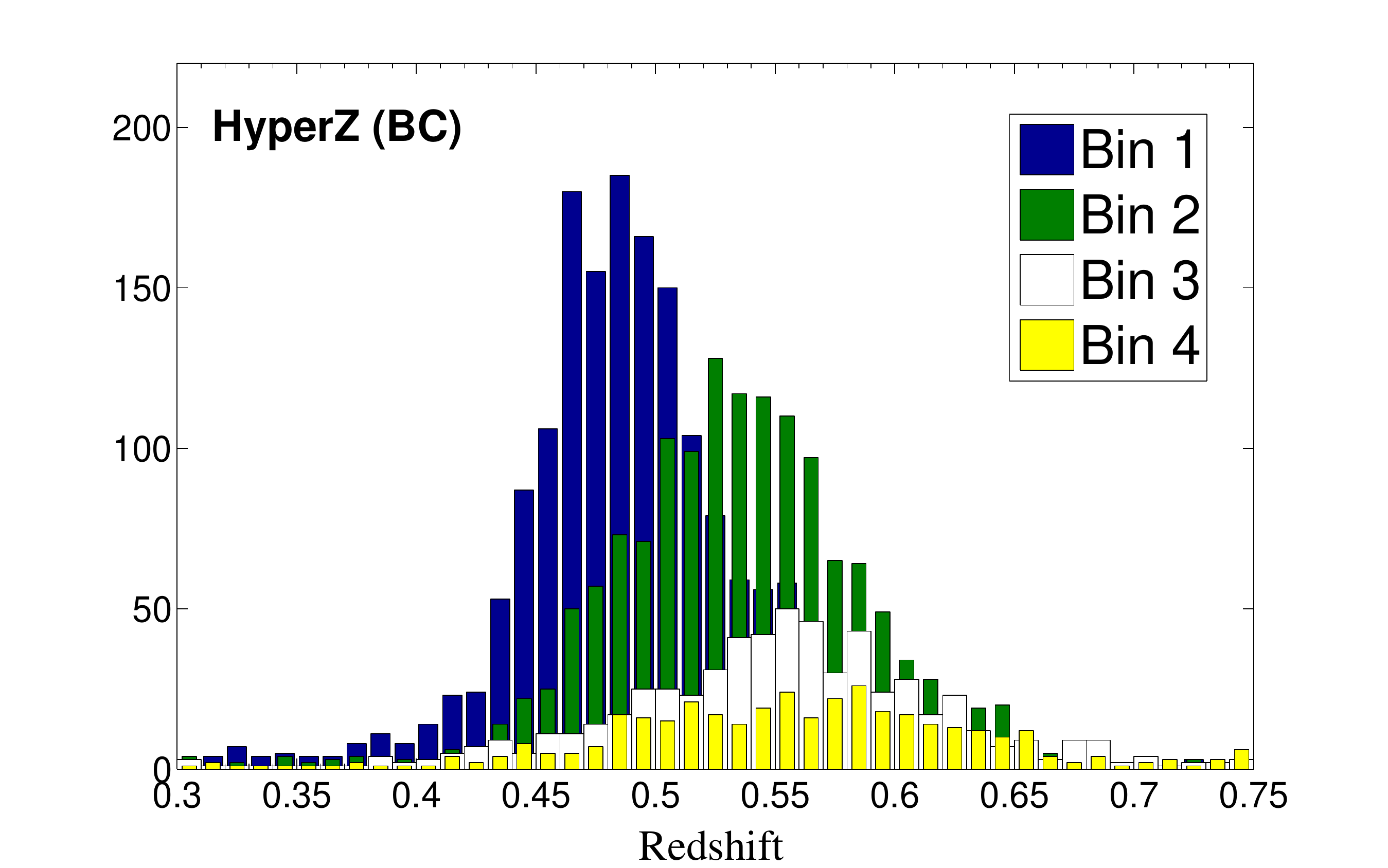}
      \includegraphics[width=4.36cm,height=4.36cm]{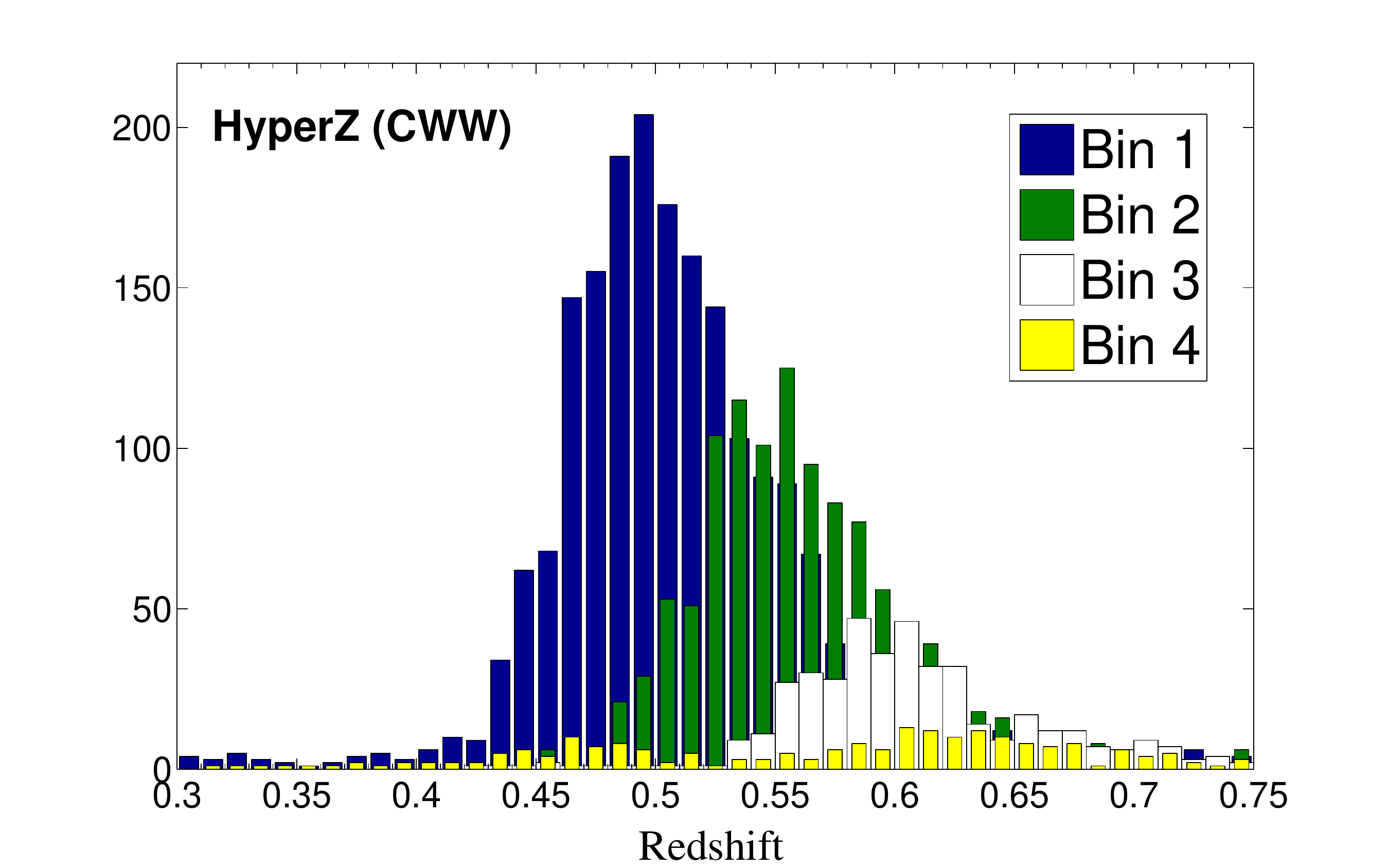}
      \includegraphics[width=4.36cm,height=4.36cm]{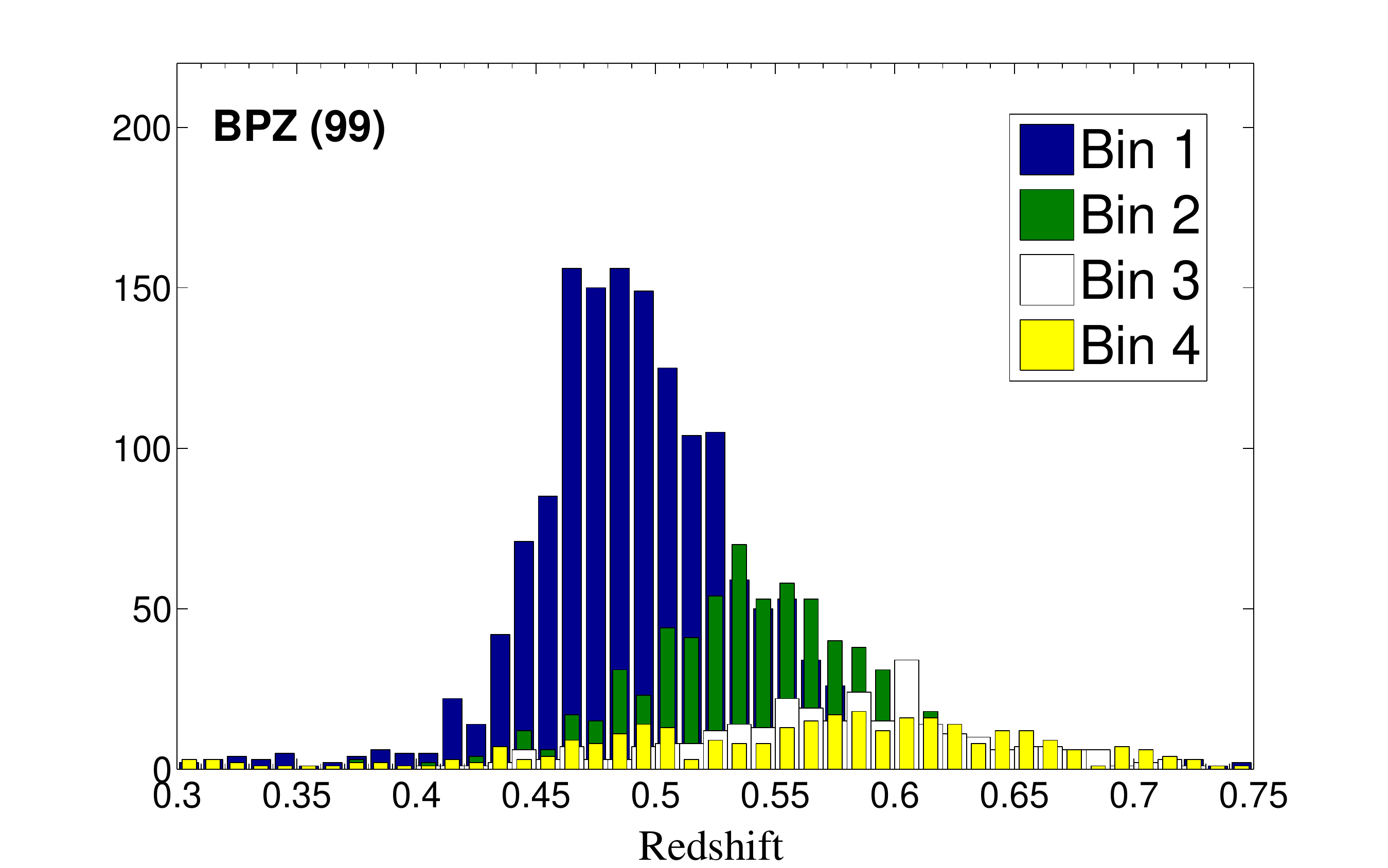}
      \includegraphics[width=4.36cm,height=4.36cm]{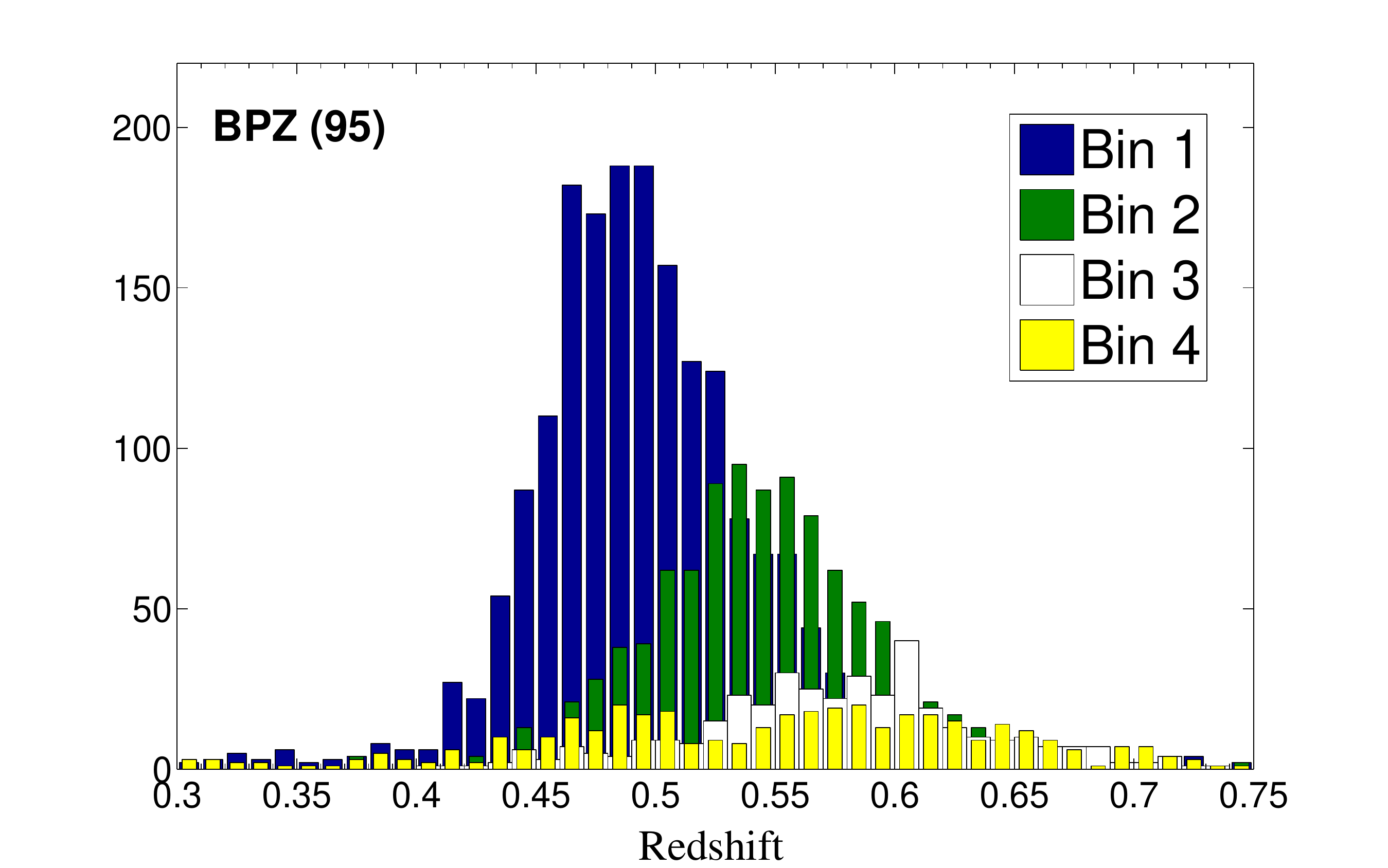}
    \end{minipage}
    \caption{\small{The spectroscopic redshift distributions n(z) in four photometric redshift bins for different photometric codes and templates. The y-axis denotes the number of galaxies in some small redshift interval $\delta z$. This is based on the DR6 catalogues provided by \citealt{Abdalla08}. In order to derive the combined constraint for each code each differing redshift distribution was taken into account. For the HyperZ code CWW and BC refer to observed and synthetic templates given by \citealt{Coleman80} and \citealt{Bruzual03}, respectively. For BPZ we use the `Bayesian' output for the redshift and therefore (99) and (95) refer to cuts of $\mathrm{{\it odds}} > 0.99$ and $\mathrm{{\it  odds}} > 0.95$, respectively (see \citealt{Abdalla08} and references therein for details).}}
    \label{fig:DR6n_z_distributions}
  \end{flushleft}
\end{figure*}
\noindent

\begin{table*}
\centering

\begin{tabular}{ccccccccc} 
\hline
Code & $\mu_{1}$ & $\sigma_{1}$ & $\mu_{2}$ & $\sigma_{2}$ & $\mu_{3}$ & $\sigma_{3}$ & $\mu_{4}$ & $\sigma_{4}$\\
\hline
ANNz & $0.475$ & $0.0323$ & $0.523$ & $0.0428$ & $0.571$ & $0.0429$ & $0.627$ & $0.0537$ \\
SDSS & $0.499$ & $0.0431$ & $0.544$ & $0.0420$ & $0.573$ & $0.0383$ & $0.618$ & $0.0421$ \\
Le PHARE & $0.509$ & $0.0463$ & $0.557$ & $0.047$ & $0.599$ & $0.0489$ & $0.632$ & $0.0572$ \\
ZEBRA & $0.499$ & $0.0403$ & $0.549$ & $0.047$ & $0.578$ & $0.069$ & $0.571$ & $0.085$ \\
HYPERZ (BC) & $0.491$ & $0.0417$ & $0.535$ & $0.0497$ & $0.551$ & $0.061$ & $0.558$ & $0.0664$ \\
HYPERZ (CWW) & $0.507$ & $0.0407$ & $0.557$ & $0.040$ & $0.598$ & $0.033$ & $0.6036$ & $0.177$ \\
BPZ (99) & $0.495$ & $0.0393$ & $0.544$ & $0.0466$ & $0.577$ & $0.0595$ & $0.570$ & $0.0849$ \\
BPZ (95) & $0.495$ & $0.0405$ & $0.543$ & $0.044$ & $0.577$ & $0.0576$ & $0.557$ & $0.0870$ \\
\hline
\hline
\\
\end{tabular}
\caption{\small{The mean $\mu_{i}$ and deviation $\sigma_{i}$ of the Gaussian fitting to the spectroscopic redshift distribution $n(z)$ in photometric bin $i$ for different codes in DR6. The spectroscopic n(z) are illustrated in Figure~\ref{fig:DR6n_z_distributions}.}}
\label{table:gaussianredshiftfitsDR6}
\end{table*}

\end{document}